\documentclass[10pt,twocolumn]{article}

\usepackage[utf8]{inputenc}
\usepackage[T1]{fontenc}
\usepackage{amsmath,amssymb,amsfonts,mathtools}
\usepackage{graphicx}

\usepackage{dblfloatfix}

\usepackage{booktabs}
\usepackage{siunitx}
\usepackage{xcolor}
\colorlet{deepred}{black}
\colorlet{pendingorange}{black}
\colorlet{magenta}{black}
\colorlet{cyan}{black}
\colorlet{red}{black}
\colorlet{blue}{black}
\colorlet{purple}{black}
\renewcommand{\textcolor}[2]{#2}
\renewcommand{\color}[1]{}
\usepackage{caption}
\usepackage{subcaption}
\usepackage{needspace}
\usepackage[margin=0.75in,columnsep=0.3in]{geometry}

\usepackage[super,sort&compress,comma]{natbib}

\usepackage{hyperref}
\hypersetup{colorlinks=true,linkcolor=black,urlcolor=black,citecolor=black}

\usepackage{authblk}

\usepackage{enumitem}
\usepackage{placeins}

\usepackage{bm}
\usepackage{multirow} 

\newcommand{\code}[1]{\texttt{\small #1}}
\newcommand{\cmi}{\ensuremath{\mathrm{cm}^{-1}}}

\newcommand{\ket}[1]{\ensuremath{\lvert #1 \rangle}}
\newcommand{\bra}[1]{\ensuremath{\langle #1 \rvert}}

\title{Vibrational Sample-Based Quantum Diagonalization with Unitary-Cluster-Jastrow Ansätze on IBM QPUs}

\author[a]{Kunal Kumar}
\author[b,c,d]{Erik L\"otstedt\thanks{Corresponding author: \href{mailto:loetstedte@riken.jp}{loetstedte@riken.jp}}}
\author[a]{Yuri Kobayashi}
\author[e,f,g,d]{Seiji Yunoki}

\affil[a]{IBM Quantum, IBM Japan, Tokyo 103-8510, Japan}
\affil[b]{RIKEN Center for Interdisciplinary Theoretical and Mathematical Sciences (iTHEMS), Wako, Saitama 351-0198, Japan}
\affil[c]{Trapped Ion Quantum Computer Team, TRIP Headquarters, RIKEN, Wako, Saitama 351-0198, Japan}
\affil[d]{Computational Condensed Matter Physics Laboratory, RIKEN Pioneering Research Institute (PRI), Wako, Saitama 351-0198, Japan}
\affil[e]{RIKEN Center for Emergent Matter Science (CEMS), Wako 351-0198, Japan}
\affil[f]{RIKEN Center for Computational Science (R-CCS), Kobe 650-0047, Japan}
\affil[g]{RIKEN Center for Quantum Computing (RQC), Wako 351-0198, Japan}

\date{\today}

\begin{document}
\maketitle

\begin{abstract}
We extend sample-based quantum diagonalization (SQD) to anharmonic vibrational
spectra and use it to benchmark how vibrational ans\"atze scale on present-day
hardware under a direct one-hot encoding (up to 72 qubits). Five particle-conserving circuits are warm-started through vibrational
coupled-cluster (VCC) $t_1,t_2$ amplitudes: UVCCSD, the compact heuristic
circuit (CHC), and three unitary-cluster--Jastrow ans\"atze introduced here,
VLUCJ, Vg-uCJ, and VIm-uCJ. A self-consistent configuration-recovery loop
reconstructs correlated states from hardware samples in the low-retention
regime where na\"ive post-selection breaks down, enabling the calculation of ground-state energies for H$_2$O, CH$_2$O, and CH$_2$ClF (3--9 modes).

We compare harmonic-oscillator and vibrational self-consistent-field (VSCF,
``modal'') reference bases. In the small-modal-basis regime relevant for near-term
devices, the modal basis is typically
more accurate
than the
harmonic-oscillator representation; empirically, it also yields substantially
better-converged hardware sampling for deep circuits at these qubit counts.
Across ans\"atze, we find a clear retention/accuracy tradeoff when scaling: the
new VLUCJ circuit is the most hardware-robust, retaining the largest fraction of
physically valid one-hot outcomes, whereas the more expressive uCJ variants and
UVCCSD can recover more correlation when retention is sufficient.

Finally, we extend the recovery protocol to excited states by targeting a chosen
per-mode VSCF occupation and warm-starting it with a state-specific VCC
calculation, and we demonstrate
the calculation of one representative fundamental vibrational frequency per
molecule on IBM QPUs.
\end{abstract}

\section{Introduction}
\label{sec:intro}

Infrared (IR) and Raman vibrational spectroscopies are fundamental
tools for chemical identification and analysis. In particular, the
mid-infrared region is traditionally interpreted by assigning
characteristic functional-group vibrations to individual bands.
\cite{Fuhrer1972ChemRev,LinVien1991HandbookIRRaman} Vibrational
spectroscopy has long been a tool for chemical fingerprinting, and
experimental advances have enabled increasingly high-resolution
IR and near-infrared (NIR) spectra of large biological and material
systems.\cite{Baiz2020ChemRev,Burns2007NIRHandbook,Bokobza2019Polymers,Ito2014JACS,Freier2011PNAS,Muroda2012Biochemistry} To achieve
comparably high resolution, computational methods must be developed that
accurately treat vibrational anharmonicity.\cite{Bowman1978JCP,Bowman1986AccChemRes,TranBerkelbach2023JCP,AietaCeottoAnharmonicDensities}

From a theoretical perspective, anharmonic vibrational structure theory
closely parallels electronic structure theory. In a modal basis, each
vibrational mode is represented by occupation of one among several modal
states, and anharmonic couplings between modes generate a many-body
Hamiltonian that can be formulated in a second-quantized Fock-space
framework analogous to that used for correlated electrons.
\cite{Christiansen2004,Ollitrault2020VibrationalQC,SeidlerChristiansenVCC,Glaser2023vDMRG}
Classical vibrational-structure methods inherit much of the
computational complexity of their electronic-structure counterparts.
Full vibrational configuration interaction (VFCI) scales exponentially
with the number of vibrational modes and the size of the one-mode basis.
To overcome this bottleneck, several vibrational selected configuration
interaction (SCI) methods have been developed to systematically prune
the exponentially large configuration space. Methods such as
configuration selective vibrational CI \cite{NeffRauhut2009} and vibrational adaptive sampling
CI \cite{Lesko2019VibASCI} laid the groundwork for this approach, whereas vibrational heat-bath
configuration interaction (VHCI) \cite{FetherolfBerkelbach2021VHCI,TranBerkelbach2023JCP} reduces this cost by selecting only the
configurations expected to contribute significantly to the target
eigenstates.
Vibrational
coupled-cluster (VCC) theory instead employs a size-extensive, truncated
excitation hierarchy with more favorable scaling, but the accuracy of a
particular truncation cannot generally be guaranteed in advance.
\cite{SeidlerChristiansenVCC} As illustrated by the results in
Table~\ref{tab:classicalrefs}, low-order VCC or vibrational
self-consistent-field (VSCF; often abbreviated ``VibSCF'')
approximations can
deviate by tens of cm$^{-1}$ from the converged VFCI energies, even for the small systems considered here.

Sample-based quantum diagonalization (SQD) was developed to reduce the
dependence of quantum electronic-structure calculations on the accuracy
of a single variational ansatz. In common electronic-SQD workflows, a
classical coupled-cluster calculation is used to warm-start a low-depth
LUCJ sampling circuit, and SQD then draws bitstrings from the prepared
state rather than using the circuit only for expectation-value
estimation. SQD samples electronic configurations from its measured
bitstring distribution and classically diagonalizes the physical
Hamiltonian in the subspace spanned by selected and recovered
configurations. This procedure can recover correlation effects that are
absent from the original circuit ansatz, while a self-consistent
configuration-recovery protocol mitigates sampling errors and restores
configurations satisfying the required particle-number and spin
constraints.\cite{RobledoMoreno2025SQD,Barison2025ExtendedSQD} Large-scale IBM--RIKEN
demonstrations have subsequently combined SQD on IBM Heron processors
with the Fugaku supercomputer, extending electronic-structure
calculations to active spaces beyond the reach of exact diagonalization.
\cite{RobledoMoreno2025SQD,Shirakawa2025ClosedLoop}
The original idea of a quantum sampling-based approach was introduced by Kanno et al.~\cite{Kannoetal} under the name of quantum-selected configuration interaction (QSCI).

In this work, we adapt SQD to the vibrational Fock-space representation
and adopt the same warm-start idea by using vibrational coupled-cluster
(VCC) information to initialize the sampling circuits. This extension is
not simply a change of notation: the qubit encoding, warm-start
amplitudes, quantum-circuit ans\"atze, and configuration-recovery rule
are all designed specifically for the vibrational problem, as described
in Section~\ref{sec:methods}. The central empirical objective of this
study
is to determine which of several physically motivated ans\"atze provides
sufficient improvement in accuracy to justify its additional circuit
cost when self-consistent configuration recovery is used to identify
the correlated vibrational subspace.
Quantum sample-based approach to vibrational energy evaluation was pioneered in
Ref.~\cite{LotstedtSzidarovszky2026Rovibrational}, where a triatomic molecule (H$_2$O) was investigated. In Ref.~\cite{LotstedtSzidarovszky2026Rovibrational}, the binary mapping combined with an ansatz quantum circuit constructed by approximate time evolution \cite{Mikkelsen2025a,Sugisaki2025,Yu2025} was employed.

This paper takes the next step by focusing specifically on a
vibrational-SQD workflow: we (i) formulate SQD directly in the
vibrational Fock space with a direct one-hot encoding, (ii) warm-start a
family of particle-conserving vibrational ans\"atze from classical
force-field information, and (iii) evaluate those ans\"atze under a
self-consistent configuration-recovery loop that robustly reconstructs
the correlated vibrational subspace at qubit counts where na\"ive
post-selection alone fails. Our scope is purely vibrational. The inclusion of rovibrational coupling and the calculation of transition probabilities are left to future work.

The contribution of this paper is threefold. First, we present a
hardware-realistic, ansatz-agnostic vibrational-SQD pipeline that starts
from classical force-field information and requires no
molecule-specific circuit engineering. In particular, we start from
anharmonic force-field information in normal-mode coordinates, truncated
at quartic order, from which we construct the second-quantized
vibrational Hamiltonian used for downstream quantum simulation
(Section~\ref{sec:mop}). We demonstrate this end-to-end workflow on
three molecules spanning 3 to 9 modes (H$_2$O, CH$_2$O, CH$_2$ClF;
Table~\ref{tab:molecules}) and a controlled 2/4/6/8-modal sweep spanning
6 to 72 qubits under a direct one-hot encoding, executed on two superconducting quantum computers, \code{ibm\_kingston} and \code{ibm\_phoenix}.
These three molecules are selected because they are small enough to enable a thorough comparison with reference classical methods, while CH$_2$ClF's 9 modes still reach the paper's largest qubit counts (up to 72).

Second, we provide a head-to-head comparison of five distinct
particle-number-conserving ans\"atze (UVCCSD, VLUCJ, VIm-uCJ, Vg-uCJ,
CHC), evaluated on equal footing across the full set of modal
truncations (2/4/6/8 modals per mode), using matched qubit counts in the
harmonic and modal bases, on IBM hardware rather than a noiseless
simulator. This directly tests whether additional
circuit structure -- including hardware-efficient designs previously
proposed for vibrational calculations\cite{Ollitrault2020VibrationalQC}
-- yields a measurable accuracy advantage once self-consistent
configuration recovery is used to identify the correlated vibrational
subspace, and finds that the answer depends on which regime is asked
about: in the noiseless limit, VLUCJ's diagonal correlator provably
cannot move weight off the reference configuration by itself, while
the other four ansätze -- including CHC, whose cross-modal-index phase
is not subject to the same obstruction despite sharing VLUCJ's
singles--phase--singles skeleton -- can and do; on IBM hardware, the
retention ordering inverts this, because how much raw signal survives
decoherence dominates over how much correlation an ideal circuit could
reach (Section~\ref{sec:results}, Section~\ref{sec:discussion}).

Third, we extend the recovery procedure to excited states by injecting an
arbitrary per-mode vibrational self-consistent-field (VSCF; ``modal'')
reference occupation and running a \emph{separate} classical vibrational
coupled-cluster (VCC) calculation to obtain that state's warm-start
amplitudes (Section~\ref{sec:excited}), reporting one excited-state energy
per molecule (Table~\ref{tab:excited}). Finally, we execute
both ground- and excited-state workflows on IBM hardware
(\code{ibm\_kingston}) up to 72 qubits (Section~\ref{sec:hardware}), reaching
problem sizes beyond the noiseless statevector-tractable
regime.

\section{Methods}
\label{sec:methods}

\subsection{Vibrational Hamiltonian}
\label{sec:mop}

For detailed vibrational-structure theory and derivations, we refer the
reader to Ref.~\cite{WilsonDeciusCross1980}. Connections between these
vibrational formulations and quantum-computing implementations have
been discussed in the context of hardware-efficient vibrational
algorithms and rovibrational energy-level calculations.
\cite{Ollitrault2020VibrationalQC,LotstedtSzidarovszky2026Rovibrational}
Here, we focus on the pipeline-specific ingredients required to make
sample-based quantum diagonalization practical for anharmonic
vibrational Hamiltonians.

The starting point of the workflow is a molecular vibrational
Hamiltonian expressed in dimensionless normal coordinates $q_i$. The
harmonic contribution describes each normal mode as an independent
oscillator, whereas cubic and quartic terms account for anharmonicity
and coupling between the modes. Retaining terms through fourth order
gives a computationally tractable Taylor approximation that captures
the leading anharmonic and mode-coupling contributions near the
equilibrium geometry; its accuracy for a given molecule depends on the
underlying potential, the coordinate choice, and the displacement
range over which the expansion is evaluated. The resulting quartic
force-field Hamiltonian can be
written as
\begin{equation}
  \hat H \;=\; \sum_i \omega_i \left(\hat n_i + \tfrac12\right)
  \;+\; \sum_{\boldsymbol{p}\in\mathcal{A}} f_{\boldsymbol{p}} \prod_i \hat q_i^{p_i},
  \label{eq:mop}
\end{equation}
where $\omega_i$ and $\hat n_i$ are the harmonic frequency and the number operator, respectively, of mode $i$,
$\boldsymbol{p}=(p_1,p_2,\ldots)$ specifies the powers of the normal
coordinates in a given anharmonic term, and $f_{\boldsymbol{p}}$ is the
corresponding force-field coefficient. Here, $\mathcal{A}$ denotes the
set of force-field terms retained in the molecular-operator
representation, with a maximum total polynomial order
$|\boldsymbol{p}|=\sum_i p_i$ of four. 
In this study, we obtain $f_{\boldsymbol{p}}$
from density-functional theory (DFT) electronic-structure calculations (B3LYP functional \cite{Lee1988,Becke1993,Stephens1994},
aug-cc-pVQZ basis set \cite{Kendall1992} for H$_2$O and aug-cc-pVTZ \cite{Kendall1992,Woon1993} for CH$_2$O and CH$_2$ClF) implemented in {\sc gaussian}\cite{g16} (\code{freq=anharmonic} keyword). The anharmonic coupling coefficients $f_{\boldsymbol{p}}$ are evaluated by numerical differentiation of the potential energy surface obtained by DFT
\cite{Barone2004a,Barone2005}.
The {\sc gaussian} output files are supplied in the Supplementary material. The numerical values of $f_{\boldsymbol{p}}$ can be obtained by multiplying the values found in the {\sc gaussian} output files by suitable combinatorial prefactors.

The three systems provide a progression from three to nine vibrational
modes and from 6 to 72 qubits under the modal truncations considered
here (Table~\ref{tab:molecules}).

\begin{table}[t]
\caption{\textbf{Molecular systems and modal-space sizes.} We consider a
modal-truncation sequence $M\in\{2,4,6,8\}$ (modals per mode) under a direct
one-hot encoding, in which each retained modal is represented by one
qubit. The table reports each molecule's symmetry, number of vibrational
modes $N$, the maximum truncation \textcolor{purple}{$M_{\max}$,} and the corresponding
maximum qubit count $N M_{\max}$.}
\label{tab:molecules}
\small
\setlength{\tabcolsep}{3pt}
\renewcommand{\arraystretch}{1.05}
\begin{tabular}{@{}l c c c c c@{}}
\toprule
Molecule & Atoms & Sym. & $N$ & $M_{\max}$ & Qubits ($N M_{\max}$) \\
\midrule
H$_2$O     & 3 & $C_{2v}$        & 3 & 8 & 24 \\
CH$_2$O    & 4 & $C_{2v}$        & 6 & 8 & 48 \\
CH$_2$ClF  & 5 & $C_s$           & 9 & 8 & 72 \\
\bottomrule
\end{tabular}
\end{table}
\normalsize
\renewcommand{\arraystretch}{1.0}

\subsection{Modal representation and qubit encoding}
\label{sec:encoding}

To make \textcolor{blue}{the continuous-coordinate Hamiltonian \eqref{eq:mop}} accessible to a quantum
processor, we first represent the motion along each normal mode $i$ in a
finite set of $M$ one-dimensional basis functions, referred to as
\textcolor{cyan}{modals.\cite{Christiansen2004}} Specifically, the $k$-th modal for mode $i$, denoted
$\phi_{i k}(q_i)$, is expanded in a harmonic-oscillator (HO) basis
$\{\chi_v(q_i)\}_{v=0}^{K-1}$ as
\begin{equation}
  \phi_{i k}(q_i) = \sum_{v=0}^{K-1} C_{k v}^{(i)}\,\chi_v(q_i),
  \label{eq:modal_expansion}
\end{equation}
where $v$ is the HO quantum number, $C_{k v}^{(i)}$ are expansion
coefficients, and $K$ is the HO cutoff. These modals may be the bare HO functions
(in which case $C_{k v}^{(i)}=\delta_{k v}$) or mode-specific functions
obtained from a vibrational self-consistent-field (VSCF) calculation.\cite{Bowman1978JCP,Bowman1986AccChemRes,SeidlerChristiansenVCC}
A multidimensional vibrational configuration is then a direct product
containing one occupied modal for each of the $N$ vibrational modes.
Increasing $M$ systematically enlarges the local basis and improves its
ability to represent anharmonic motion, but also increases the quantum
resources required.

We encode this modal representation using a direct, or one-hot, mapping,
\textcolor{blue}{using $M$ qubits for each vibrational mode,}
following established quantum formulations of molecular vibrations.
\cite{McArdle2019DigitalVibrations,Ollitrault2020VibrationalQC} Each
modal is assigned one qubit, and the occupied modal of a given
vibrational mode is identified by the single qubit in state
$\lvert 1\rangle$ within its $M$-qubit register. We label the qubit
assigned to modal $k$ ($k=0,\dots,M-1$) of mode $i$ as qubit $(i,k)$.
For example, for $N=2$
modes and $M=3$ modals per mode (6 qubits total), we order the qubits as
$(i,k) = (1,0),(1,1),(1,2),(2,0),(2,1),(2,2)$. The computational
basis state $\ket{001010}$ then corresponds to the vibrational
configuration in which \textcolor{magenta}{the $k=2$ modal is occupied in mode~1 and
the $k=1$ modal is occupied in mode~2.} Here and throughout, ket labels
and bitstrings are written left to right in this same $(i,k)$ order.
The full calculation therefore requires $NM$ qubits. This mapping uses more
qubits than compact binary encodings, but retains a transparent
one-to-one correspondence between modal occupations and measured
bitstrings. It also maps modal-transfer operators to
excitation-preserving qubit operations, which is particularly
convenient for the configuration sampling and recovery steps of
vibrational SQD.

\textcolor{cyan}{The encoding  determines} the relevant structure of the
state-preparation circuit: an admissible ansatz should generate
correlations between modes while preserving one occupied modal within
each mode register. We consequently compare five
modal-number-conserving ans\"atze that implement different trade-offs
between circuit depth, parameter count and the ability to transfer
amplitude among correlated vibrational configurations.

\subsection{Modal-number-conserving state preparation}
\label{sec:ansatz_construction}\label{sec:ansatze}

Within \textcolor{blue}{the one-hot} encoding, state preparation amounts to redistributing
amplitude among the allowed modal configurations without leaving the
one-excitation-per-mode physical subspace. The circuit must also
introduce correlations between different vibrational modes, because
independent intramode rotations alone cannot represent the correlated
response induced by the anharmonic coupling terms in
Eq.~\eqref{eq:mop}. Circuit expressivity must therefore be balanced
against hardware cost: richer excitation operators can represent
correlation more directly, but generally require deeper circuits and
more entangling gates.

To examine the trade-off between circuit cost and variational
expressibility systematically, we compare five modal-number-conserving
ans\"atze, UVCCSD, VLUCJ, VIm-uCJ, Vg-uCJ and CHC, under identical
conditions for every combination of molecule, modal truncation and
one-mode basis considered here. UVCCSD provides the systematically
motivated unitary vibrational coupled-cluster construction developed in
Refs.~\cite{McArdle2019DigitalVibrations,Ollitrault2020VibrationalQC,JensenChristiansen2025UVCC},
whereas VLUCJ, VIm-uCJ and Vg-uCJ adapt shallow Jastrow-inspired
molecular electronic-structure circuits to the mode-resolved occupation
constraints of the vibrational problem.\cite{Motta2023LUCJ,Tkachenko2025BeyondReal}
CHC is included as a compact hardware-oriented comparison inspired by
Ref.~\cite{Ollitrault2020VibrationalQC}.

\begin{figure}[!t]
  \centering
  \includegraphics[width=\columnwidth]{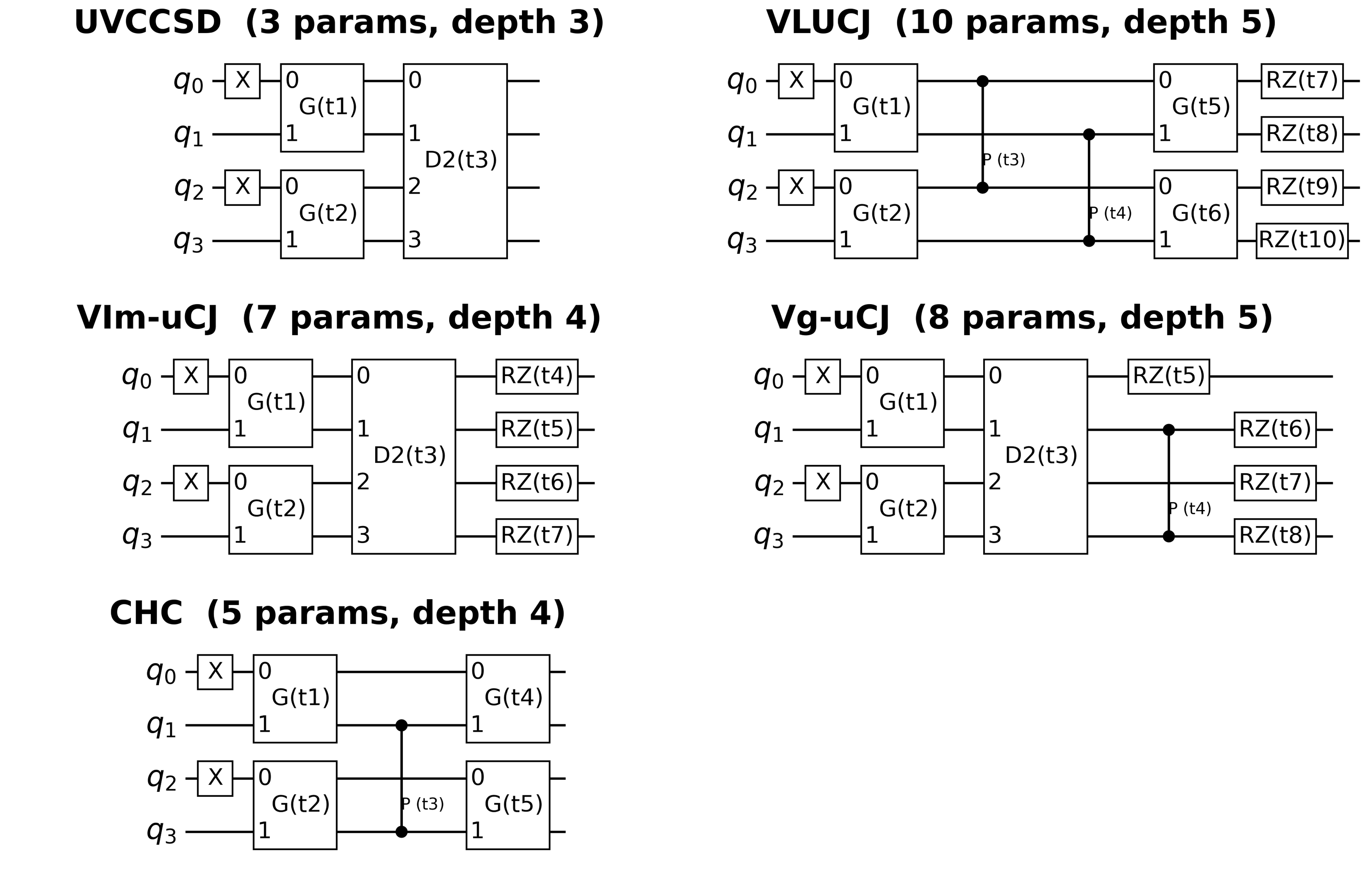}
  \caption{All five ansätze at the toy $N=2,M=2$ (4-qubit) system, one
  layer, $P=1$ pair (VIm-uCJ, Vg-uCJ) -- $P$ is the number of
  intermode modal pairs $(i,k),(j,l)$ chosen to receive a diagonal
  Jastrow correlator/$D_2$ primitive, out of the $\binom{N}{2}(M-1)^2$
  possible pairs (Table~\ref{tab:ansatzlayout}), so $P=1$ here is the
  minimal, non-empty choice at this toy system's own small pair count.
  Givens (``G'') and paired double excitation (``D2'') are drawn as opaque boxes;
  the diagonal Jastrow (``P'') and $R_Z$ tail are drawn as their native gates.
  The true native-gate compilation of a single $D_2(\theta)$ on this system is
  depth 66, 25 CZ gates once transpiled at \code{optimization\_level=3} to generic
  Heron basis gates -- illustrative of relative primitive cost, not a universal
  per-backend number. Table~\ref{tab:ansatzlayout} and the gate-cost figures in
  Section~\ref{sec:results} report the measured costs for the production circuits.
  For this $N=2,M=2,P=1$ toy example, UVCCSD and VIm-uCJ display the same
  Givens-plus-$D_2$ connectivity because the complete double-excitation set
  contains only one pair at this size; their constructions differ more generally
  in pair selection, parameterization, and the optional $R_Z$ tail
  (Table~\ref{tab:ansatzlayout}). VLUCJ and CHC share the
  ``singles--diagonal phase--singles'' sandwich structure and differ in which
  qubits the diagonal phase couples (same modal index across modes for VLUCJ,
  different excited modal indices for CHC) -- the difference analyzed at the
  operator level in Appendix~\ref{app:ansatz_structure}.}
  \label{fig:ansatzcircuits}
\end{figure}

\FloatBarrier
\begin{table*}[!t]
\centering
\caption{Per-layer primitive count for each ansatz, $N$ modes, $M$
modals/mode, $P$ chosen pairs (VIm-uCJ, Vg-uCJ only). The table reports
primitive \emph{counts}, not their two-qubit gate cost, which differs
sharply by primitive type and is relative, not absolute: a Givens rotation is $\mathcal{O}(1)$ native
two-qubit gates, a diagonal Jastrow phase contributes one logical
two-qubit primitive before transpilation, and a paired
double excitation is $\mathcal{O}(10)$ (25 CZ gates on the 4-qubit toy
system above, \code{optimization\_level=3}, generic Heron basis gates,
seed unfixed -- illustrative of relative primitive cost, not a
universal per-backend number)
-- so an ansatz's total \emph{parameter} count understates its
relative circuit cost whenever it includes even one $D_2$ term.}
\label{tab:ansatzlayout}
\begin{tabular}{@{}lccccc@{}}
\toprule
Ansatz & Givens & Diag.\ Jastrow ($\mathrm{cp}$) & $D_2$ & Params/layer & Uses $D_2$? \\
\midrule
UVCCSD  & $N(M-1)$ & --                     & $\binom{N}{2}(M-1)^2$ & $N(M-1)+\binom{N}{2}(M-1)^2$ & yes \\
VLUCJ   & $2N(M-1)$ & $(N-1)M$              & --                    & $2N(M-1)+(N-1)M+NM$          & no \\
VIm-uCJ & $N(M-1)$  & --                     & $P$                   & $N(M-1)+P+NM$                & yes \\
Vg-uCJ  & $N(M-1)$  & $P$ (excited-modal)    & $P$                   & $N(M-1)+2P+NM$               & yes \\
CHC     & $2N(M-1)$ & $\binom{N}{2}(M-1)^2$ & --                    & $2N(M-1)+\binom{N}{2}(M-1)^2$ & no \\
\bottomrule
\end{tabular}
\end{table*}

The five constructions are assembled from a common set of
modal-number-preserving primitives but differ in which primitives are
included, how modes and modals are connected, and how the operations
are ordered. These distinctions determine whether a circuit merely
changes phases, independently rotates individual modes, or introduces
an explicit correlated two-mode excitation. \textcolor{magenta}{Further details are provided in
Appendix~\ref{app:ansatz_structure}.}
Throughout, $N$ denotes the number
of vibrational modes, $M$ the number of modals per mode and $L$
the number of ansatz layers.

\paragraph{Primitive 1: intramode Givens rotation.}
Let $r_i$ denote the reference modal of mode $i$ -- the modal occupied
by the initial (VSCF or HO) product state, not necessarily modal $0$ --
encoded on qubit $(i,r_i)$, and let $k\neq r_i$ denote an excited modal
of the same mode, encoded on qubit $(i,k)$. The primitive is implemented
directly as the pair of commuting Pauli-string rotations
\begin{equation}
  G_{i,k}(\theta) = \exp\!\left[ 
  \textcolor{cyan}{i}\frac{\theta}{2} X_{i,k}Y_{i,r_i} \right]
  \exp\!\left[ -i\frac{\theta}{2} Y_{i,k}X_{i,r_i} \right].
  \label{eq:givens_pauli}
\end{equation}
Here, $X_{i,k}$ denotes a Pauli $X$ acting on qubit $(i,k)$, and
similarly for the other Pauli operators. Because $X_{i,k}Y_{i,r_i}$ and
$Y_{i,k}X_{i,r_i}$ commute, Eq.~\eqref{eq:givens_pauli} is equivalently
\begin{equation}
  G_{i,k}(\theta) = \exp\!\left[ i\frac{\theta}{2}
  \left( X_{i,k}Y_{i,r_i} - Y_{i,k}X_{i,r_i} \right) \right].
  \label{eq:givens_pauli_compact}
\end{equation}
On the local encoded basis, restricted to the one-hot subspace of mode
$i$, the action is
\begin{align}
  G_{i,k}(\theta)\ket{1_{i,r_i}0_{i,k}} &=
  \cos(\theta)\ket{1_{i,r_i}0_{i,k}} + \sin(\theta)\ket{0_{i,r_i}1_{i,k}},
  \\
  G_{i,k}(\theta)\ket{0_{i,r_i}1_{i,k}} &=
  \cos(\theta)\ket{0_{i,r_i}1_{i,k}} - \sin(\theta)\ket{1_{i,r_i}0_{i,k}}.
\end{align}
Thus, for a nontrivial angle, this primitive transfers amplitude between
the reference modal and one excited modal of a \emph{single} mode. It is
the ansatz's only single-excitation-like primitive. In our
implementation, it is passed to the 
\textcolor{magenta}{transpiler} as the two symbolic Pauli
rotations in Eq.~\eqref{eq:givens_pauli}, rather than decomposed as a
textbook two-qubit Givens gate. Its \textcolor{magenta}{transpiled} cost is consequently
reported with respect to the target hardware's native gate set.
\cite{Ollitrault2020VibrationalQC}

\paragraph{Primitive 2: diagonal Jastrow phase.}
The second primitive is the controlled-phase gate
\begin{equation}
  \operatorname{cp}(\varphi) = \operatorname{diag}(1,1,1,e^{i\varphi}),
\end{equation}
applied between two modal qubits belonging to different vibrational
modes. In VLUCJ, it is applied between qubits carrying the same modal
index $k$, namely $(i,k)$ and $(j,k)$. In CHC, it is applied between
$(i,k)$ and $(j,l)$ for every pair of excited-modal indices $k,l\in\{1,\dots,M-1\}$
of the two modes, including $k=l$. In Vg-uCJ, it is applied between
$(i,k)$ and $(j,l)$, where $k$ and $l$ are the excited-modal indices of
the specific mode-pair coupling selected for the accompanying $D_2$
primitive (Section~\ref{sec:warmstart}, Eq.~\eqref{eq:paired_double_pauli}'s
own $k,l$ notation); because the selected $D_{ij}^{kl}$ coupling itself
can have $k=l$, equality of the two excited-modal indices is not
excluded here, just as it is not excluded for CHC above. The controlled phase is treated
here as one logical two-qubit primitive; its native-gate and CZ cost
after transpilation depends on the backend target and the angle
$\varphi$, and is reported for the actual submitted circuits in
Section~\ref{app:transpilation}. Unlike the Givens and paired double-excitation
primitives, however, it is diagonal in the computational basis. It
therefore cannot change computational-basis populations by itself.
Whether it can affect subsequent amplitude transfer depends entirely on
the qubit pair on which it acts and on whether the input state already
has coherent support on basis states that acquire different phases.

\paragraph{Primitive 3: paired two-mode double excitation.}
Let $r_i$ and $r_j$ denote the reference modals of two distinct modes
$i$ and $j$, encoded on qubits $(i,r_i)$ and $(j,r_j)$, and let $k\neq
r_i$ and $l\neq r_j$ denote excited modals of those respective modes,
encoded on qubits $(i,k)$ and $(j,l)$. Under the modal-to-qubit
encoding, the paired double-excitation primitive is implemented directly
as the sum of eight commuting four-qubit Pauli-string rotations
\begin{equation}
\begin{aligned}
  D_{ij}^{kl}(\theta)
  &= \exp\Bigl[-i\frac{\theta}{4}\Bigl(
  X_{i,k}Y_{j,l}X_{i,r_i}X_{j,r_j}\\
  &\quad+ Y_{i,k}X_{j,l}X_{i,r_i}X_{j,r_j}
  + Y_{i,k}Y_{j,l}X_{i,r_i}Y_{j,r_j}\\
  &\quad+ Y_{i,k}Y_{j,l}Y_{i,r_i}X_{j,r_j}
  - X_{i,k}X_{j,l}X_{i,r_i}Y_{j,r_j}\\
  &\quad- X_{i,k}X_{j,l}Y_{i,r_i}X_{j,r_j}
  - X_{i,k}Y_{j,l}Y_{i,r_i}Y_{j,r_j}\\
  &\quad- Y_{i,k}X_{j,l}Y_{i,r_i}Y_{j,r_j}
  \Bigr)\Bigr].
\end{aligned}
  \label{eq:paired_double_pauli}
\end{equation}
following the standard eight-term Pauli-string decomposition of the UVCCSD
paired double-excitation operator (see Refs.~\cite{McArdle2019DigitalVibrations,Ollitrault2020VibrationalQC,JensenChristiansen2025UVCC}), adapted here from
electronic-structure spin-orbitals to vibrational modals. All eight
Pauli strings in Eq.~\eqref{eq:paired_double_pauli} mutually commute,
so the exponential of their sum is implemented as the product of eight
individual Pauli-string rotations, each a native two-qubit-gate-based
primitive under the target basis gate set. Writing
$\ket{\mathrm{ref}}\equiv\ket{1_{i,r_i}0_{i,k}}\ket{1_{j,r_j}0_{j,l}}$
and $\ket{kl}\equiv\ket{0_{i,r_i}1_{i,k}}\ket{0_{j,r_j}1_{j,l}}$,
restricted to this two-dimensional subspace the action of
Eq.~\eqref{eq:paired_double_pauli} is
\begin{align}
  D_{ij}^{kl}(\theta)\ket{\mathrm{ref}}
  &=
  \cos(2\theta)\,\ket{\mathrm{ref}}
  +
  \sin(2\theta)\,\ket{kl},
  \\
  D_{ij}^{kl}(\theta)\ket{kl}
  &=
  -\sin(2\theta)\,\ket{\mathrm{ref}}
  +
  \cos(2\theta)\,\ket{kl},
  \label{eq:paired_double_configs}
\end{align}
i.e.\ it simultaneously transfers one modal occupation $r_i\to k$ in
mode $i$ and one modal occupation $r_j\to l$ in mode $j$; the two
transfers occur together or not at all, never separately, and the
physical rotation angle in this subspace is $2\theta$ rather than
$\theta$, a consequence of the $\theta/4$ prefactor in
Eq.~\eqref{eq:paired_double_pauli}. This is the
only primitive considered here that is both off-diagonal and capable of
transferring amplitude across two vibrational modes in a single
operation. It is also by far the most expensive primitive in the
compiled circuits. On the four-qubit toy system used throughout this
section, one instance of $D_{ij}^{kl}(\theta)$ transpiled in isolation
(\code{optimization\_level=3}, a generic Heron-basis-gate target) to
25 CZ gates and depth 66, versus depth approximately $1$--$2$ for one
intramode Givens rotation under the same convention; because no
transpiler seed was fixed for this specific illustrative check and the
result can vary with the chosen compilation target and the Qiskit\cite{Qiskit2024} release --
the software environment specified for this work provides only a minimum
requirement \textcolor{purple}{(Qiskit version at least $1.0$}, Section~\ref{app:transpilation}) rather than a
pinned version, and the exact release used for this illustrative check
was not separately documented --
these
two numbers should be read as illustrative of the qualitative cost gap
between the two primitives, not as reproducible, universal figures --
the CZ counts and depths reported for every actual production circuit
in this paper (Section~\ref{app:transpilation}) are measured directly
on the submitted circuit rather than estimated from this toy
comparison. The large cost of
$D_{ij}^{kl}(\theta)$ arises because each of its eight Pauli-string
rotations requires a separate entangling ladder, and the transpiler
does not cancel sufficient structure across the eight individual terms.

Figure~\ref{fig:ansatzcircuits} draws all five ansätze, one layer, $P=1$
pair where applicable, at the same minimal $N=2,M=2$ (4-qubit) toy
system used throughout this section -- with Givens and $D_2$
(shorthand used throughout this paper for the paired double-excitation
primitive $D_{ij}^{kl}(\theta)$ of
Eqs.~\eqref{eq:paired_double_pauli}--\eqref{eq:paired_double_configs})
drawn as
labeled boxes rather than their native-gate compilation, since the
latter (depth 66 for a single $D_2$, as above) would be unreadable as a
structure diagram; the caption states the depth alongside the
schematic. The five parameter counts (3, 10, 7, 8, 5) match
Table~\ref{tab:ansatzlayout} exactly at $N=2,M=2,L=1,P=1$.

\paragraph{Per-ansatz layout.} Table~\ref{tab:ansatzlayout} gives each
ansatz's per-layer primitive count as a closed-form function of $N,M$
(and, for VIm-uCJ/Vg-uCJ, a chosen pair count $P\le\binom{N}{2}(M-1)^2$,
the number of largest-$|t_2|$ mode-pair/excited-modal couplings kept --
see Section~\ref{sec:warmstart}). All of this paper's main hardware
results use \textcolor{cyan}{one layer ($L=1$),} for every ansatz. UVCCSD has no repeated-layer
structure at all (it is always $L=1$); VLUCJ, VIm-uCJ, Vg-uCJ, and CHC
each support layer repetition, and a separate \textcolor{cyan}{simulation with two layers} confirms
this paper's conclusions are unchanged at $L=2$ for these four
ansätze, across the same 24 parameter settings.

VLUCJ, VIm-uCJ and Vg-uCJ each include a single-qubit
$R_Z(\theta)$ tail containing $NM$ parameters per layer in every
calculation reported in this paper.
These parameters
are included in the ``Params/layer'' column of
Table~\ref{tab:ansatzlayout}. Because the $R_Z$ tail is diagonal and
contains only one-qubit operations, it does not introduce the
$\binom{N}{2}(M-1)^2$ two-mode scaling associated with the complete set
of modal pairs.

The central structural distinction is therefore not the parameter count
alone, but whether the circuit contains an explicit off-diagonal two-mode
correlator. UVCCSD, VIm-uCJ and Vg-uCJ contain $D_2$ operations, while
VLUCJ and CHC generate correlations indirectly through single-mode
rotations and diagonal phase operations. The consequences of this
distinction, including the role of diagonal phases and the additional
excited-modal phase in Vg-uCJ, are analyzed in
Appendix~\ref{app:ansatz_structure}, and their empirical impact on
recovered energies is reported in Section~\ref{sec:results} and
discussed in Section~\ref{sec:discussion}.

\subsection{Warm start and direct bind}
\label{sec:warmstart}

Every ansatz above is warm-started from classical VCC $t_1,t_2$
amplitudes, obtained from a vibrational
coupled-cluster calculation\cite{SeidlerChristiansenVCC} on the
Hamiltonian of Section~\ref{sec:mop} (Appendix~\ref{app:ho-basis} gives
the full amplitude-to-angle mapping), rather than
from a random or all-zero initial point. In UVCCSD, VLUCJ, and CHC,
the intramode Givens angles are initialized directly from the
corresponding classical $t_1$ amplitudes; in UVCCSD specifically, the
paired double-excitation angles are additionally initialized directly
from the corresponding $t_2$ amplitudes, since UVCCSD's non-diagonal
primitives are literally the exponentiated single/double excitation
operators the VCC amplitudes were fit to in the first place. In
VIm-uCJ and Vg-uCJ, the selected $D_2$ angles are instead initialized
from the retained $t_2$ amplitudes through a bounded map, because the
two-level rotation these primitives implement
(Eq.~\eqref{eq:paired_double_configs}) is periodic in $\theta$:
directly identifying an unbounded cluster amplitude with this rotation
angle can place the initialization on an unintended angular branch of
the periodic rotation, producing a state on the wrong side of the
intended correlation entirely, not merely a less-accurate one. To avoid
such branch-selection failures, we apply a bounded arctangent map
(Eq.~\eqref{eq:t2toangle},
Appendix~\ref{app:ho-basis}) applied to the $D_2$-type paired-rotation
angle only, which agrees with the direct copy for small amplitudes
while never overshooting; VIm-uCJ's and Vg-uCJ's own intramode Givens
block is, by contrast, left at zero initial value and warm-started by
the optimizer alone (Appendix~\ref{app:ho-basis}) -- a deliberate
implementation choice, not a mathematical necessity, made because the
paired correlator's own $D_2$-angle warm start already supplies the
dominant correction. VLUCJ's and CHC's diagonal Jastrow phases, and
every ansatz's optional $R_Z$ tail, have no unique direct counterpart
in a VCC amplitude set and are likewise left at zero initial
value. For VIm-uCJ and Vg-uCJ, only the $P$
largest-amplitude mode-pair couplings are kept per layer, out of
$\binom{N}{2}(M-1)^2$ possible pairs, since every pair kept costs one
more paired-excitation primitive at $\mathcal{O}(10)$ two-qubit gates
(Section~\ref{sec:ansatze}); $P$ is therefore this paper's primary control parameter for
these two ansätze's circuit cost, independent of layer count $L$. Every
VIm-uCJ/Vg-uCJ result reported in this paper uses the same fixed $P=8$
per layer, at every molecule, modal count, and basis; $P$ itself is not
swept.
\textcolor{cyan}{The numerical values of the VCC $t_1$ and $t_2$ amplitudes are obtained using the {\sc MidasCpp} program. \cite{MidasCpp}}
The
full amplitude-to-angle mapping and pair-selection procedure are given
in Appendix~\ref{app:ho-basis}.

\subsection{Harmonic-oscillator (HO) basis vs.\ modal (VSCF) basis}
\label{sec:basis}

The same state can be expanded in harmonic-oscillator (HO) products or
VSCF-modal products, related by a separable rotation $C^{(i)}$ per
mode; the modal Hamiltonian is $H_{\text{modal}} = U H_{\text{HO}}
U^\dagger$ with $U = \bigotimes_i \big(C^{(i)}[{:}M]\big)^\dagger$,
where $[{:}M]$ denotes keeping only the lowest $M$ rows (the $M$
retained modals, ranked by VSCF energy) of the full per-mode VSCF
rotation matrix returned by {\sc MidasCpp}, discarding the higher,
untruncated modals -- the same truncation notationally introduced in
Eq.~\eqref{eq:modal_expansion} (i.e.,
$U$ maps HO $\to$ modal). Single-mode
anharmonicity is absorbed into the modal ground state $\psi_0$, so the
modal reference $\ket{0\cdots0}$ \emph{is} the VSCF ground state -- a
much better starting point than the bare HO product. We therefore report
results in both bases throughout, not as redundant bookkeeping, but
because doing so separates two independently verifiable contributions to
the overall accuracy of the workflow. The modal basis's reference-config
energy is closer to the converged/exact value than the HO basis's at
most, though not all, tested (molecule, modals, ansatz) settings
(Section~\ref{sec:results}); this is a property of which basis the
qubits represent, fixed before any ansatz or warm-start is chosen; the
ansatz-dependent gaps between ``E before'' and ``E after''
(Appendix~\ref{app:fulltables-artifacts}) are a property of
what the circuit does inside that basis. Reporting both bases
side by side is how this paper avoids conflating the two. The HO basis
carries a second, purely practical justification: its one-mode
operator factors retain a sparse band structure that the corresponding
modal-basis operators lose (Appendix~\ref{app:ho-basis}), so an
HO-representation Hamiltonian truncated to the same $M$ retained
functions per mode can be assembled and applied via sparse Kronecker
products without ever materializing a dense matrix -- for CH$_2$ClF,
this paper's own pipeline builds this $M{=}8$-truncated HO-representation
operator, of dimension $M^N=8^9=134{,}217{,}728$, in seconds
this way, whereas an explicitly stored dense matrix at this same
dimension would require $(8^9)^2\approx1.8\times10^{16}$ 
\textcolor{cyan}{entries} -- not a slower calculation to materialize densely, an
impossible one on any conceivable classical hardware. (The primitive,
untruncated HO basis used for the classical VFCI/VHCI references,
Table~\ref{tab:classicalrefs}, retains $K{=}11$ functions per mode,
giving a formal dimension of $K^9\approx2.36\times10^9$ for CH$_2$ClF,
itself well beyond dense construction and part of why no VFCI
reference exists for this molecule.) The relevant distinction is
therefore not simply the formal Hilbert-space dimension, but whether
sparse, factorized operator structure survives the basis choice: it
does for the HO representation and does not for the modal one, which
is often more compact for representing the VSCF-adapted low-energy
states but loses this sparse band structure once transformed
(Appendix~\ref{app:ho-basis}). That contrast, not just the smaller
accuracy difference at fixed problem settings (molecule, basis size
$M$, and ansatz choice), is the justification for reporting the
HO basis at all in this paper.

Figure~\ref{fig:ho-vs-modal} in Appendix~\ref{app:ho-basis} visualizes
this contrast directly, plotting both bases' full compact Hamiltonians
side by side for H$_2$O.

The full derivation of the VSCF rotation matrix $C^{(i)}$, its
per-mode heatmaps, the modal-count density sweep, and the ground-state
probability-distribution comparison across molecules are given in
Appendix~\ref{app:ho-basis}.

\subsection{Self-consistent configuration recovery}
\label{sec:recovery}
\label{sec:cliff}

Each mode's $M$-qubit block must return exactly one set bit to be
physical, i.e.
\begin{equation}
  \sum_{k=0}^{M-1} \hat n_{i,k} = 1,\qquad i=1,\dots,N,
  \label{eq:physical}
\end{equation}
where $\hat n_{i,k}$ denotes the occupation (measurement outcome) of
qubit $(i,k)$. If mode blocks fail independently and each survives
noise with approximately equal probability $\bar p$, the probability
that all $N$ blocks survive together in a given shot scales as
$P_{\text{valid}} \approx \bar p^{\,N}$ -- exponential in mode count.
Hardware errors can be correlated across qubits and blocks, so
this is a heuristic scaling argument rather than an exact model; the
retention values actually measured on IBM hardware (Section~\ref{sec:results})
are used throughout this paper rather than this approximation. At the
larger CH$_2$O and CH$_2$ClF settings considered here, simple
post-selection on physical shots alone would discard most or all
measured shots for several ansätze.
We therefore introduce a recovery procedure
that repairs broken samples instead of discarding them.

\subsubsection{Motivation and background}
\label{sec:recoveryprovenance}

The recovery method used throughout this pipeline is the vibrational
adaptation of the self-consistent configuration recovery
technique developed for electronic-structure SQD by Robledo-Moreno
et al.\cite{RobledoMoreno2025SQD} The core idea -- project
each noise-corrupted measurement outcome back onto the physical
(particle- or occupation-number-conserving) subspace, biased by the
current best estimate of the ground state's own occupation statistics,
then diagonalize in the resulting compact subspace and iterate to
self-consistency -- is identical here; only the notion of ``physical
subspace'' changes, from electronic particle-number conservation to
the vibrational one-hot constraint of Eq.~\eqref{eq:physical}. Three
further results from the broader SQD literature motivate specific
choices made throughout this paper: Reinholdt et al.\cite{ReinholdtQSCILimitations2025}
establish that the operative metric under sample~$\to$~project~$\to$~diagonalize is
configuration coverage, not raw sampled-state fidelity, which is
why we report one-hot retention and recovered energy together
(Section~\ref{sec:onehotexplains}) rather than fidelity alone.
Vaquero-Sabater et al.\cite{Broers2026NoiseConfigRecovery}
document that ansätze which concentrate amplitude on very few
configurations (e.g.\ a diagonal-Jastrow correlator such as VLUCJ) may
provide insufficient configuration-space coverage even when their
state fidelity is comparatively high. This motivates reporting
configuration retention, recovered-subspace size, and energy accuracy
as distinct diagnostics throughout this paper
(Section~\ref{sec:onehotexplains}), rather than inferring one from the
others. Consistent with this concern, VLUCJ's diagonal-Jastrow
correlator is, acting alone, unable to change computational-basis
populations (Section~\ref{app:diagonal_phases}); whether this
translates into VLUCJ's recovered energy remaining at or near the
reference-configuration energy in a specific calculation is an
empirical question examined in Section~\ref{sec:results} and
Section~\ref{sec:ansatzdiscuss}, not a general property established by
this circuit-structure argument alone.
Motta et al.\cite{Motta2023LUCJ} introduced the
electronic-structure LUCJ ansatz, whose vibrational analogue (VLUCJ) is
one of the five ansätze compared here
(Section~\ref{sec:ansatze}).

\subsubsection{Average modal occupancy}
\label{sec:occ}

The central self-consistent quantity is the average modal
occupancy, a per-mode probability distribution over which modal level
$k$ is occupied ($k=0,\dots,M-1$, consistent with the modal index of
Section~\ref{sec:ansatze}), estimated from the current target-state
Ritz vector\cite{Saad2011EigenvalueProblems} (the lowest eigenvector for
ground-state recovery; the overlap-selected eigenvector of
Section~\ref{sec:excited} for excited-state recovery). Given a set of physical
configurations
$\{\bm{v}^{(p)}\}_{p=1}^{|\mathcal S|}$ (each
$\bm{v}^{(p)} = (v_1^{(p)},\dots,v_N^{(p)})$) with amplitudes $c_p$
from a diagonalization in the subspace $\mathcal S$
(Section~\ref{sec:subspace}), the average modal occupancy of mode $i$
is
\begin{equation}
  \bar n_{i,k} \;=\; \sum_{p=1}^{|\mathcal S|} |c_p|^2\,
    \delta_{v_i^{(p)},\,k}
  \,, \qquad i=1,\dots,N,\;\; k=0,\dots,M-1,
  \label{eq:occ-main}
\end{equation}
i.e.\ the fraction of the diagonalized state's own probability weight
sitting on each modal level $k$ across all sampled configurations,
with $\sum_k \bar n_{i,k}=1$ for a normalized subspace eigenvector. In
the noiseless and sufficiently complete-subspace limit, this
distribution approaches the true one-mode marginal modal-occupation
distribution of the target eigenstate; it reduces to a delta function
at that mode's own reference modal $r_i$ only for an uncorrelated
product state occupying a single modal in each mode, not for a
correlated many-body eigenstate in general. The recovery loop's
purpose is to estimate this marginal distribution reliably from a
necessarily incomplete sampled subspace, starting from a uniform prior
across all $M$ modals before any diagonalization has happened
(Eq.~\eqref{eq:occ0}, Appendix~\ref{app:recovery}).

\subsubsection{Repairing a single measurement outcome}
\label{sec:repair}

\textcolor{blue}{Each mode's $M$-qubit block of a raw, possibly non-physical bitstring
is repaired independently by a simple rule: if the block already has
exactly one set bit, it is a physical, well-formed observation and is
kept unchanged; otherwise -- zero set bits, or more than one -- the
invalid block (not the entire shot) is replaced by a modal index drawn
from the current average-occupancy distribution
(Section~\ref{sec:occ}) rather than assigned arbitrarily or dropped.
Applying this rule to every mode
block of a bitstring and re-encoding one-hot yields a
guaranteed-physical repaired bitstring. In the numerical implementation,
this repair draw is performed exactly once per distinct raw bitstring
key present in the raw counts dictionary (not once per individual
shot, and not an analytical enumeration of every possible repair
outcome), and that bitstring's full observed weight is then
accumulated onto the single repaired outcome drawn; renormalizing the
resulting weights over all distinct repaired outcomes gives the
repaired probability distribution over physical bitstrings only, which
is what every subsequent step of the pipeline actually operates on.} The full formal
repair rule (Eq.~\eqref{eq:repair}) and recovered-distribution formula
(Eq.~\eqref{eq:recdist}) are given in Appendix~\ref{app:recovery}.

\subsubsection{Subspace selection and diagonalization}
\label{sec:subspace}

\textcolor{blue}{Every distinct physical configuration in the repaired
distribution is used: the vibrational Hamiltonian is restricted to
exactly this set of configurations -- never the full
$M^N$-dimensional one-hot-sector operator ($M^N=2^3=8$ for our smallest parameter
setting, H$_2$O at $M{=}2$, but $8^9\approx1.3\times10^8$ for the
largest, CH$_2$ClF at $M{=}8$), and still less the full $2^{NM}$-dimensional
qubit Hilbert space (up to $2^{72}\approx4.7\times10^{21}$ for that
same largest parameter setting), of which only the $M^N$-dimensional
one-hot sector is physically relevant -- both far beyond dense
diagonalization at the qubit counts reached here, which is exactly the
dimensional blowup SQD is designed to avoid. The
Hamiltonian matrix elements within this subspace are built directly
from the quartic force field: in the HO calculations, the required
matrix elements are evaluated directly in the truncated HO basis; in
the modal calculations, the one-mode operators are transformed with
the per-mode VSCF rotation $C^{(i)}$ introduced in
Section~\ref{sec:basis} (Appendix~\ref{app:ho-basis}).
Diagonalizing this small subspace Hamiltonian gives the current best
energy estimate and the expansion coefficients of the selected
target-state Ritz vector (the lowest eigenvector for ground-state
recovery, the overlap-selected eigenvector of Section~\ref{sec:excited}
for excited-state recovery), which feed back into the average
modal occupancy (Section~\ref{sec:occ}) to produce the next
iteration's occupancy prior.} The full formal subspace-Hamiltonian
construction (Eq.~\eqref{eq:hsub}, Eq.~\eqref{eq:modalop}) and
diagonalization (Eq.~\eqref{eq:diag}) are given in
Appendix~\ref{app:recovery}.

\subsubsection{The self-consistent loop}
\label{sec:loop}

\textcolor{blue}{Repair (Section~\ref{sec:repair}), subspace selection and
 diagonalization (Section~\ref{sec:subspace}), and the occupancy update
 (Section~\ref{sec:occ}) together form a genuine fixed-point iteration:
 the occupancy estimate from one iteration's diagonalization biases
 which broken bitstrings get repaired into which configurations at the
 next iteration, which produces a new diagonalization and a new
 occupancy estimate, and so on. Iteration continues, using a fixed
 seed per (molecule, modals, basis, ansatz) parameter setting, until
 two consecutive energy estimates agree to within \SI{0.1}{\cmi} or
 until a maximum of 30 iterations is reached. Every ground-state
 parameter setting reported in this paper (Appendix~\ref{app:fulltables-artifacts})
 converged by this criterion, in a range of 2--11 iterations; no
 non-converged (max-iteration-limited) case was observed among them.}
 The full iteration diagram
(Eq.~\eqref{eq:loop}) is given in Appendix~\ref{app:recovery}.

\subsubsection{Before- and after-recovery energies}
\label{sec:beforeafter}

One-hot retention (Section~\ref{sec:onehotexplains}) is the fraction of
raw shots satisfying the physical constraint of Eq.~\eqref{eq:physical},
evaluated before any Hamiltonian diagonalization; it should be
distinguished from the before- and after-recovery \emph{energies}
defined here. The before-recovery energy is obtained by diagonalizing
the Hamiltonian in the subspace spanned by the distinct, already-valid
one-hot configurations empirically observed among the raw shots,
augmented by the guaranteed reference configuration
(Section~\ref{sec:mainconfig}). The after-recovery energy is obtained
from the subspace sampled from the final, converged repaired
distribution of the self-consistent loop
(Eq.~\eqref{eq:recdist}, Eq.~\eqref{eq:loop}); the production
calculations diagonalize the full set of distinct configurations in
this final repaired distribution, with no fixed subspace-size cap.

The difference between the two energies measures the effect of
configuration recovery, whereas retention measures only the fraction
of raw valid shots; a parameter setting can have low retention and
still show a large before/after gap (recovery still does real work,
e.g.\ UVCCSD in Figure~\ref{fig:beforeafterbarsmodal}), or low
retention with almost no gap left to close, once retention is at the
noise floor (Section~\ref{sec:onehotexplains}) -- so knowing one
quantity does not determine the other. The uniform-random-sampler
control of Section~\ref{sec:uniform} shows that applying this same
recovery procedure to a structureless valid-configuration baseline
does not reproduce the hardware-derived energies at the larger modal
counts; this control rejects a structureless uniform-sampling
explanation, but does not by itself separate chemically meaningful
circuit structure from structured hardware bias.

\subsubsection{Ensuring inclusion of the reference configuration}
\label{sec:mainconfig}

One refinement over the recovery procedure as originally described:
the sampled subspace $\mathcal S$ always includes the target state's
own reference configuration, whether or not it was actually measured.
For the ground state, every mode's reference modal is $r_i=0$, so this
is the direct-encoded bitstring $\ket{v_1\cdots v_N} = \ket{r_1\cdots
r_N} = \ket{0\cdots0}$ with modal $0$ occupied in every mode's block
(e.g.\ \ket{100\,100\,\cdots\,100} for $M=3$); Section~\ref{sec:excited}
below generalizes this to $r_i\ne0$ for excited-state targets. This is
a guaranteed-inclusion rule layered on top of the existing sampling and
recovery (Section~\ref{sec:cliff}--\ref{sec:recovery}), not a
replacement for it: it protects against the case where the reference
configuration itself is under-sampled or entirely repaired away before
the subspace is drawn, which would otherwise bias the diagonalization
away from the state's own dominant configuration.

\subsubsection{Excited states via VSCF-reference preparation}
\label{sec:excited}

Excited-state calculations use the same self-consistent configuration-recovery
workflow, but with an alternative reference configuration used to seed the
selected subspace. For a fundamental excitation of mode $i$ into modal level
$k\ne 0$, the guaranteed-included configuration (Section~\ref{sec:mainconfig}) is
chosen as $r_i=k$ for the excited mode and $r_j=0$ for all other modes $j$.
Consistent with this choice of reference, the ansatz warm start is obtained from
a separate classical vibrational coupled-cluster calculation\cite{SeidlerChristiansenVCC}
performed for the same excited occupation, rather than by reusing the
ground-state amplitudes with a substituted reference.
\textcolor{magenta}{We report results for  one excited state per molecule: the lowest-frequency
fundamental (Table~\ref{tab:excited}).}
Infrared intensities, which require \textcolor{magenta}{dipole-moment-surfaces} in
addition to the potential-energy surface provided by the force field, are beyond
the scope of this work and are not considered here.

All ansatz-construction and warm-start conventions used for the ground state
extend to excited references via a simple relabeling of the reference modal
indices $\{r_i\}$ (Appendix~\ref{app:recovery}). This relabeling leaves the
remainder of the workflow unchanged and reduces to the standard ground-state
case when $r_i=0$ for all modes.

The one place the excited-state loop's diagonalization step genuinely
differs from the ground-state one (Section~\ref{sec:subspace}) is which
eigenvector of the sampled subspace is taken as the current iteration's
state estimate. The ground-state loop always selects the lowest
eigenvalue. By the Rayleigh--Ritz principle\cite{MacDonald1933}, the
lowest eigenvalue obtained in any trial subspace is an upper bound to
the exact ground-state energy within the same Hamiltonian and finite
basis;
convergence toward that exact energy is expected as the trial
subspaces become systematically more complete, but inclusion of the
reference product configuration alone does not by itself guarantee
this. For an excited
state, selecting the lowest eigenvalue would only track the intended
state if the sampled subspace never
happened to contain a lower-energy configuration than the target
excited state -- a condition that recovery starting from a uniform occupancy
prior cannot guarantee once there is sampling noise. The excited-state
loop therefore instead selects, at each iteration, the eigenvector with
maximum overlap onto the guaranteed-included excited reference
configuration (Section~\ref{sec:mainconfig}). This criterion is used to
track the Ritz state most strongly connected to the target reference,
even when lower-energy Ritz states are present in the sampled subspace;
it improves state continuity across iterations but is not an absolute
guarantee if the target state becomes strongly mixed with other states
or loses reference overlap. A controlled synthetic test on a 4-mode
model system (99\% of raw counts on the exact correct
excited-reference bitstring, 1\% uniform noise, single fixed seed)
confirms the two rules
agree once the sampled subspace is clean (both recover the excited
reference to within \SI{5}{\cmi} of the model system's own exact
answer, a check on internal consistency of the selection rule rather
than a demonstration of high absolute accuracy), and that the
max-overlap rule alone
tracks the intended reference once noise introduces a lower-energy configuration into
the subspace, which is the regime IBM hardware execution reaches for
this paper's own $M{=}6/8$ excited-state parameter settings.
A practical consequence for interpreting Table~\ref{tab:excited}: the
classical excited-state warm-start amplitudes (obtained from a separate
VCC calculation for the excited occupation, Section~\ref{sec:excited})
are identically zero for a raising operator that would need to exceed
the highest modal retained at a given truncation $M$ -- e.g.\ at the
minimal truncation $M=2$, if the excited modal is already the highest
retained level, that mode's warm-start channel from the excited
reference is zero at that specific truncation. This is a property of
the classical warm-start amplitudes only; it does not restrict which
absolute modal levels the quantum circuit itself can reach, since the
logical relabeling of Appendix~\ref{app:recovery} maps every retained
modal, including those below the excited reference, into the ansatz's
standard connectivity.

\subsection{Classical reference methods}
\label{sec:ClassicalRef}
\textcolor{blue}{
We compare the results of the SQD method with the vibrational energies obtained using several classical methods, (i) the HO approximation, (ii) vibrational self-consistent field
(VSCF)\cite{Bowman1978JCP,Bowman1986AccChemRes}, (iii) vibrational coupled-cluster (VCC) 
\cite{Christiansen2004a, SeidlerChristiansenVCC}, (iv) vibrational heat-bath configuration interaction
(VHCI) \cite{FetherolfBerkelbach2021VHCI,Bhatty2021,TranBerkelbach2023JCP}, and vibrational full configuration interaction (VFCI).
}

\textcolor{blue}{
In the HO approximation, the vibrational energy is defined as 
\begin{equation}
    E_{\rm HO} = \sum_{i=1}^N \omega_i(v_i+1/2), 
\end{equation}
where $v_i$ is the vibrational quantum number of  mode $i$.
 In the case of the vibrational ground state, 
$\bm{v}=(v_1,v_2,\ldots,v_N)=(0,0,\ldots,0)$. The HO energy is a reference energy which does not include any effects of the anharmonic coupling. The VSCF energy $E_{\rm VSCF}$ is defined  using a product of VSCF modals $\phi_{i k}$ [see Eq.~\eqref{eq:modal_expansion}] 
\begin{equation}
    E_{\rm VSCF} = \langle \Phi_{\bm{k}}^{\rm VSCF} |\hat{H} |\Phi_{\bm{k}}^{\rm VSCF}\rangle,
\end{equation}
where 
\begin{equation}
    |\Phi_{\bm{k}}^{\rm VSCF}\rangle = |\phi_{1 k_1}\cdots \phi_{N k_N}\rangle
\end{equation}
and $\bm{k}=(k_1,\ldots,k_N)=(0,\ldots,0)$ for the ground state.
VSCF treats anharmonicity and intermode coupling at the mean-field product-state level, optimizing each mode's own one-mode functions self-consistently in the effective potential of all other modes, but does not include correlation beyond a single product of these optimized one-mode functions.
}

\textcolor{blue}{
The VCC energy is defined in terms of the reference VSCF wave function $\Phi_{\bm{k}}^{\rm VSCF}$ as \cite{Christiansen2004a}
\begin{equation}
    E_{\rm VCC} = \langle \Phi_{\bm{k}}^{\rm VSCF} | e^{-T} \hat{H} e^T |\Phi_{\bm{k}}^{\rm VSCF}\rangle,
\end{equation}
where $T$ is a coupled-cluster operator. We employ the {\sc MidasCpp} program \cite{MidasCpp} for evaluating $E_{\rm VCC}$.
}

Similarly to the SQD method, the VHCI method is defined by a subset $\mathcal{S}_{\rm VHCI}$ of the total set of vibrational configurations,
\begin{equation}
    E_{\rm VHCI} = \langle \Phi^{\rm VHCI} |\hat{H} |\Phi^{\rm VHCI}\rangle,
\end{equation}
where 
\begin{equation}
    |\Phi^{\rm VHCI}\rangle = \sum_{\bm{v}\in \mathcal{S}_{\rm VHCI}} 
    c_{\bm{v}} |\bm{v}\rangle ,
\end{equation}
and $|\bm{v}\rangle=|\chi_{v_1} \chi_{v_2} \cdots \chi_{v_N}\rangle$ is a single, uncorrelated product of HO basis functions specified by the vibrational quantum numbers $\bm{v}=(v_1,v_2,\ldots,v_N)$.
We select the set of VHCI configurations according to the method presented in Ref.~\cite{FetherolfBerkelbach2021VHCI}. Briefly, starting from an initial set $\mathcal{S}^0_{\rm VHCI}$ of configurations defined as the configurations $|\bm{v}\rangle$ satisfying
$\sum_{i=1}^N v_i\le 2$,
we solve for the ground state wave function  using $\mathcal{S}^0_{\rm VHCI}$ and obtain the expansion coefficients $c_{\bm{v}}^0$ for every selected configuration $|\bm{v}\rangle\in\mathcal{S}^0_{\rm VHCI}$. In the next step, for every selected configuration $|\bm{v}\rangle \in \mathcal{S}^0_{\rm VHCI}$ and every connected candidate configuration $|\bm{v}'\rangle$ not yet in $\mathcal{S}^0_{\rm VHCI}$, we add $|\bm{v}'\rangle$ to form a new set $\mathcal{S}^1_{\rm VHCI}$ whenever $|W_{\bm{v}\bm{v}'} c_{\bm{v}}|>\epsilon_{\rm VHCI}$ is satisfied, using the already-known coefficient $c_{\bm v}$ of the selected configuration $|\bm v\rangle$, not a coefficient of the not-yet-added candidate $|\bm v'\rangle$.
Here $W_{\bm{v}\bm{v}'}$ is a \textcolor{purple}{certain average matrix element} (see Ref.~\cite{FetherolfBerkelbach2021VHCI} for the detailed definition), and
$\epsilon_{\rm VHCI}$ is a small parameter controlling the convergence. The Hamiltonian is then diagonalized in the expanded set $\mathcal{S}^1_{\rm VHCI}$, giving both a new energy and new coefficients $c_{\bm v}^1$ for the selection screening of the next cycle. The selection--diagonalization cycle is repeated until the energy is converged to the chosen threshold, or the set $\mathcal{S}^\ell_{\rm VHCI}$ becomes too large to diagonalize. We have written our own implementation of the VHCI method using {\sc matlab} \cite{MATLAB_2026a}.

\textcolor{blue}{The VFCI energy is defined  as
\begin{equation}
    E_{\rm VFCI} = \langle \Phi^{\rm VFCI} |\hat{H} |\Phi^{\rm VFCI}\rangle
\end{equation}
with 
\begin{equation}
    |\Phi^{\rm VFCI}\rangle = \sum_{v_1,\ldots,v_N=0}^{K-1} 
    c^{\rm VFCI}_{\bm{v}} |\bm{v}\rangle ,
\end{equation}
meaning that all possible \textcolor{magenta}{HO} configurations are included in the wave function expansion. The expansion coefficients
$c^{\rm VFCI}_{\bm{v}}$ are obtained by solving the eigenvalue problem defined by the full $K^N\times K^N$ Hamiltonian matrix.
The VFCI energy $E_{\rm VFCI}$ is the exact variational eigenvalue within the chosen quartic force field and finite $K^N$ HO product basis, but is only possible to compute for small molecules due to the rapid increase of the size of the Hamiltonian matrix with an increasing number of modes.
}

\textcolor{blue}{The vibrational ground state energies $E_0$ (zero-point energy, ZPE), and the excitation energy $\Delta E_1$ of the lowest excited state, defined as
\begin{equation}
    \Delta E_1 = E_1-E_0,
\end{equation}
obtained with the classical methods described above are shown in Table~\ref{tab:classicalrefs}. $\Delta E_1$ corresponds to the fundamental frequency $\omega_i$ of the vibrational mode with smallest $\omega_i$. We include the experimental measurements of $\Delta E_1$ for reference. Table~\ref{tab:classicalrefs} shows that anharmonic treatment lowers the ZPE, relative to the HO approximation, by \SI{55.81}{\cmi} for H$_2$O and \SI{76.35}{\cmi} for CH$_2$O; the corresponding changes in the targeted fundamental are \SI{77.13}{\cmi} for H$_2$O and \SI{36.62}{\cmi} for CH$_2$O -- the anharmonic correction to the fundamental is therefore substantial for both molecules.
For CH$_2$ClF, comparison with the largest VHCI calculation gives changes of \SI{86.68}{\cmi} in the ZPE and \SI{5.26}{\cmi} in the fundamental. The VFCI energies for H$_2$O and CH$_2$O are reproduced to within \SI{1}{\cmi} both by the VCC method with $M=4$ modals and the VHCI method using a small fraction of the total set of configurations (234 of $11^3=1331$ configurations for H$_2$O, about \SI{17.6}{\percent}; 4175 of $11^6=1{,}771{,}561$ for CH$_2$O, about \SI{0.24}{\percent}) -- the fraction required for comparable accuracy depends strongly on the molecule. 
We have not calculated the VFCI energy for CH$_2$ClF because of the large size (dimension $11^9\approx\num{2.36e9}$) of the Hamiltonian matrix. The largest VHCI calculation for CH$_2$ClF, containing \num{1006515} configurations, gives a ZPE of \SI{6681.59}{\cmi} and a fundamental of \SI{371.20}{\cmi}; these values are essentially unchanged from the \num{190167}-configuration calculation (ZPE unchanged at \SI{6681.59}{\cmi}, fundamental changing from \SI{371.21}{} to \SI{371.20}{\cmi}), suggesting they are close to the converged VHCI limit, though they are not independently confirmed against an exact VFCI value.
While the purpose of this paper is not to obtain vibrational energies agreeing with experiment, we comment that the VFCI (or, for CH$_2$ClF, largest-VHCI) results for $\Delta E_1$ differ from the experimentally measured excitation energies by \SI{42.54}{\cmi} for H$_2$O, \SI{5.72}{\cmi} for CH$_2$O, and \SI{11.80}{\cmi} for CH$_2$ClF. In order to obtain a better \textcolor{magenta}{agreement} with experiment, a more accurate  potential energy surface \cite{Charmet2013} and/or more sophisticated basis sets \cite{Matyus2007} would need to be employed.}
\textcolor{cyan}{We also comment that the VFCI energies for H$_2$O do not agree with those in Ref.\cite{LotstedtSzidarovszky2026Rovibrational} because of the difference in the set of anharmonic coupling constants employed.}
\begin{table}[!t]
\centering
\captionsetup{skip=4pt}
\caption{\textcolor{blue}{
Vibrational ground state energies (ZPE) and lowest excitation energy ($\Delta E_1$) in units of cm$^{-1}$.
We target one fundamental per molecule, chosen as the lowest-frequency fundamental listed here: the $\nu_2$ bending mode for H$_2$O, the $\nu_4$ CH$_2$ wagging mode for CH$_2$O, and  the $\nu_6$  ClCF bending mode for CH$_2$ClF.
The number of HO basis function is $K=11$ for all molecules. The VCC energies include contributions from single, double, and triple excitations (denoted VCC[3] in {\sc MidasCpp}'s own excitation-rank notation, corresponding to a VCCSDT-level truncation). In the VHCI calculations, $\epsilon_{\rm VHCI}=10$ cm$^{-1}$, and the total number of configurations included is denoted by $|\mathcal{S}|$. 
The experimental $\Delta E_1$ values are quoted to the precision reported by their respective sources; the integer value for CH$_2$ClF reflects the precision of that source, not a lower confidence in the calculated values.}
}
\label{tab:classicalrefs}
\footnotesize
\begin{tabular}{l S S} 
\toprule
\multicolumn{1}{l}{Method} & \multicolumn{1}{c}{ZPE / \si{\cmi}} & \multicolumn{1}{c}{$\Delta E_1$ / \si{\cmi}} \\
\midrule
\multicolumn{3}{c}{H$_2$O}\\
HO                                & 4670.06    & 1629.34      \\
VSCF                              & 4630.42    & 1558.52      \\
VCC ($M=2$)                       & 4630.40    & 1558.49      \\
VCC ($M=4$)                       & 4614.43    & 1552.49      \\
VCC ($M=6$)                       & 4614.25    & 1552.22      \\
VCC ($M=8$)                       & 4614.25    & 1552.21      \\
VHCI ($|\mathcal{S}|=10$)         & 4636.00    & 1564.53      \\
VHCI ($|\mathcal{S}|= 72$)        & 4614.29    & 1552.64      \\
VHCI ($|\mathcal{S}|= 234$)       & 4614.25    & 1552.22      \\
VHCI ($|\mathcal{S}|=542$)        & 4614.25    & 1552.21      \\
VFCI                              & 4614.25    & 1552.21      \\
Expt.\cite{Furtenbacher2020a}     &            & 1594.75 \\
&&\\
\multicolumn{3}{c}{CH$_2$O}\\
HO                                & 5814.12    & 1198.16      \\
VSCF                              & 5755.21    & 1166.14      \\
VCC ($M=2$)                       & 5753.85    & 1166.09      \\
VCC ($M=4$)                       & 5738.01    & 1161.66      \\
VCC ($M=6$)                       & 5737.81    & 1161.55      \\
VCC ($M=8$)                       & 5737.80    & 1161.55      \\
VHCI ($|\mathcal{S}|=28$)         & 5761.29    & 1169.20      \\
VHCI ($|\mathcal{S}|= 547$)       & 5737.86    & 1161.88      \\
VHCI ($|\mathcal{S}|= 4175$)      & 5737.77    & 1161.54      \\
VFCI                              & 5737.77    & 1161.54      \\
Expt.\cite{Germann2024}           &            & 1167.26 \\
&&\\
\multicolumn{3}{c}{CH$_2$ClF}\\
HO                                & 6768.27    & 376.46       \\
VSCF                              & 6708.12    & 372.27       \\
VCC ($M=2$)                       & 6701.56    & 372.19      \\
VCC ($M=4$)                       & 6682.03    & 371.23      \\
VCC ($M=6$)                       & 6681.77    & 371.22      \\
VCC ($M=8$)                       & 6681.77    & 371.22      \\
VHCI ($|\mathcal{S}|=55$)         & 6714.83    & 373.60      \\
VHCI ($|\mathcal{S}|= 1840$)      & 6681.94    & 372.46       \\
VHCI ($|\mathcal{S}|= 24749$)     & 6681.60    & 371.29      \\
VHCI ($|\mathcal{S}|= 190167$)    & 6681.59    & 371.21      \\
VHCI ($|\mathcal{S}|= 1006515$)   & 6681.59    & 371.20       \\
Expt.\cite{Charmet2013}           &            & 383 \\
\bottomrule
\end{tabular}
\end{table}

\section{Results}
\label{sec:results}

This paper's primary results are executions on IBM hardware on
\emph{two} backends, \code{ibm\_kingston} (IBM Heron r2, heavy-hex
coupling map) and \code{ibm\_phoenix} (IBM Nighthawk r2, square
coupling map). The ground-state campaign uses the same logical set of
24 molecule--modal-count--basis \emph{parameter
settings} (the term used throughout this section for this
quadruple, excluding ansatz choice; two bases (HO and modal), four
modal counts 2/4/6/8, three molecules from Table~\ref{tab:molecules}:
H$_2$O, CH$_2$O, CH$_2$ClF) on both backends, and, with five
ansätze, this yields 120 (molecule, modals, basis, ansatz) \emph{circuit
settings} for the ground state per backend, all at layers=1 (a separate
layers=2 sweep, Section~\ref{sec:ansatze}, confirms this paper's
conclusions are unchanged one layer deeper), with the
identical qubit-count-tiered shot budget on both backends
(\num{100000} shots for $N_q\le24$, \num{500000} for $24<N_q\le56$,
and \num{1000000} for $56<N_q\le72$). Excited-state coverage is not
identical across the two backends: it extends to 72 circuit settings on
\code{ibm\_kingston} (VLUCJ, VIm-uCJ, and Vg-uCJ; UVCCSD and CHC were not
run there) and the full 120 circuit settings on \code{ibm\_phoenix} (all five
ansätze) -- 192 circuit settings on \code{ibm\_kingston} and 240 on
\code{ibm\_phoenix}, 432 hardware circuit settings in total. For every executed
circuit setting, the
raw hardware counts, before-recovery energy (an explicit
diagonalization of the Hamiltonian restricted to the one-hot subspace
spanned by all empirically observed valid configurations, augmented by
the analytically known reference configuration (included even if it was
not observed; Section~\ref{sec:mainconfig})) and after-recovery energy
(the self-consistent loop's own final diagonalization) are reported for
every executed circuit setting; nothing here is a
noiseless-simulator placeholder. \textcolor{deepred}{The complete
ground- and excited-state numbers, both backends side by side, every
per-parameter-setting physical qubit mapping and job identifier
included, are available as four interactive online ledgers
(Appendix~\ref{app:fulltables-artifacts}) rather than as printed
tables.}

\begin{figure*}[t]
  \centering
  \includegraphics[width=\textwidth]{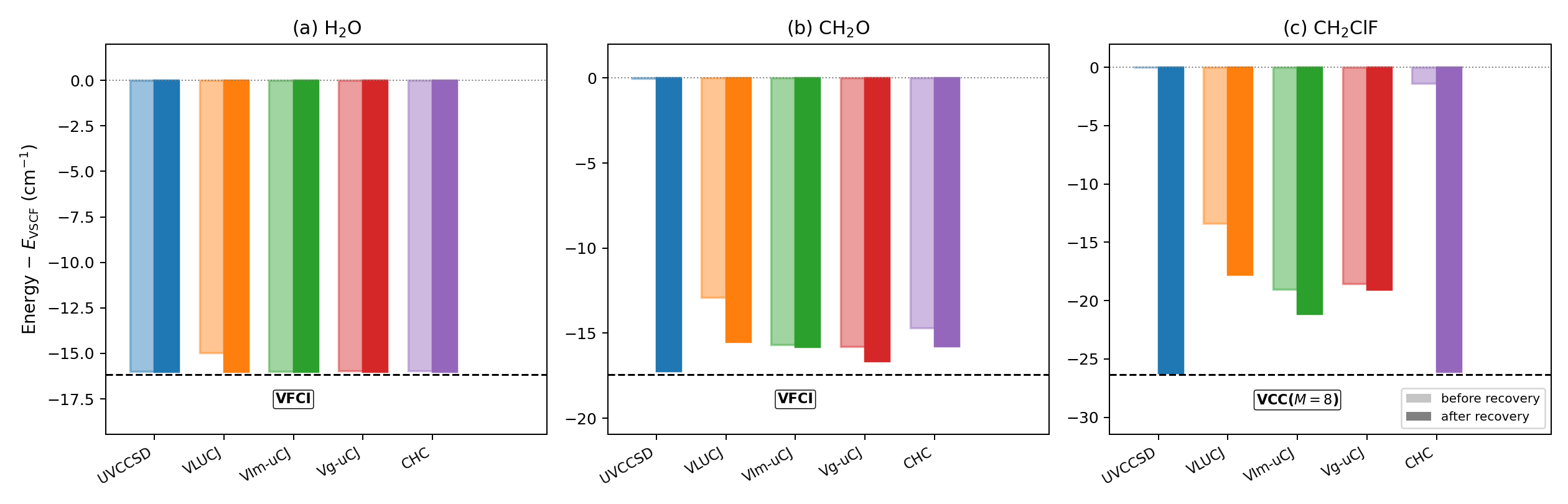}
  \caption{\code{ibm\_phoenix} hardware, layers=1,
  vibrational ground-state energy relative
  to the VSCF reference ($E-E_{\rm VSCF}$) for each of the five
  ansätze, modal basis, $M{=}4$ (12/24/36 qubits for
  H$_2$O/CH$_2$O/CH$_2$ClF respectively) -- faded bars are before
  self-consistent recovery (an explicit subspace diagonalization over
  whatever the raw sample actually contained), filled bars are after.
  The dashed line (labeled in-panel) is the converged classical target
  (VFCI for H$_2$O/CH$_2$O; VCC($M{=}8$) for CH$_2$ClF, whose
  Hamiltonian is too large for VFCI, Table~\ref{tab:classicalrefs}'s
  footnote). We show $M{=}4$ because several ansätze retain sufficient
  pre-recovery signal for the effect of recovery to remain visible. The
  corresponding \code{ibm\_kingston} comparison is given in
  Appendix~\ref{app:beforeafterkingston}
  (Figure~\ref{fig:beforeafterbarsmodalkingston}).}
  \label{fig:beforeafterbarsmodal}
\end{figure*}

\begin{figure*}[t]
  \centering
  \includegraphics[width=\textwidth]{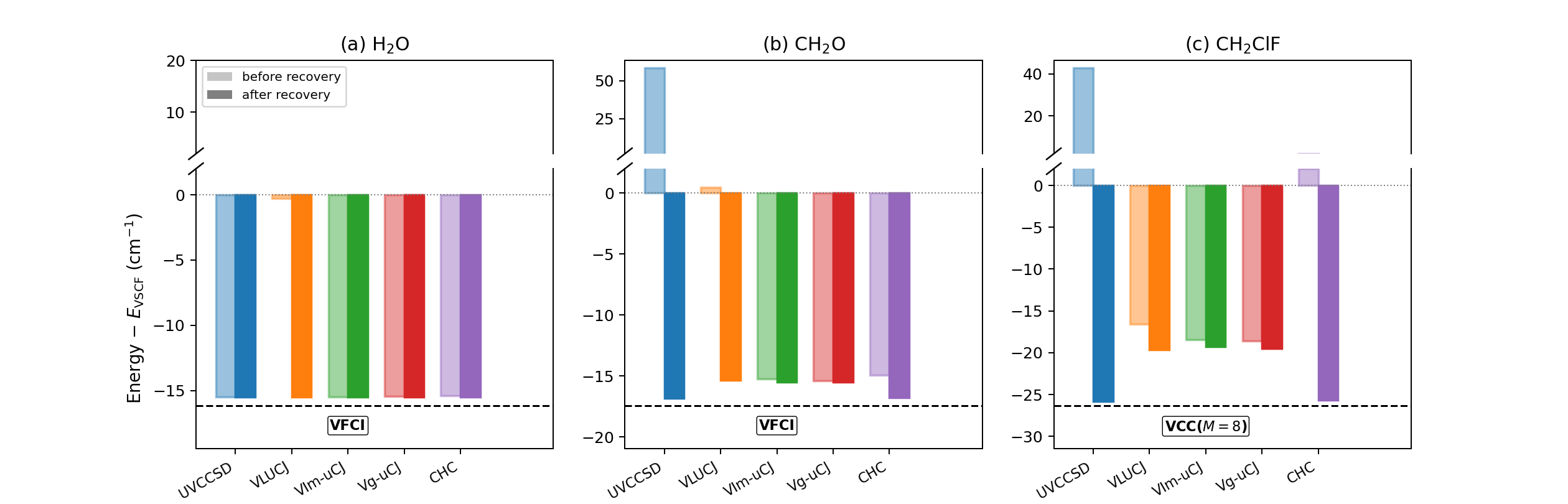}
  \caption{Same as Figure~\ref{fig:beforeafterbarsmodal}, but in the HO
  basis. The comparison shows that the relative quality of the raw
  modal- and HO-basis samples depends on the ansatz and its retained
  one-hot signal. The corresponding \code{ibm\_kingston} comparison is
  shown in Appendix~\ref{app:beforeafterkingston},
  Figure~\ref{fig:beforeafterbarshokingston}.}
  \label{fig:beforeafterbarsho}
\end{figure*}

\textcolor{cyan}{The main results of this investigation are shown in Figures~\ref{fig:beforeafterbarsmodal} and \ref{fig:beforeafterbarsho}, where we show \textcolor{purple}{the vibrational ground state energy} obtained by the SQD method using \code{ibm\_phoenix}, for each molecule, circuit type, and basis type (modal basis in Fig.~\ref{fig:beforeafterbarsmodal} and 
HO basis in Fig.~\ref{fig:beforeafterbarsho}).}
Table~\ref{tab:configcounts} reports the numbers of distinct
configurations underlying the bars in
Figures~\ref{fig:beforeafterbarsmodal} and \ref{fig:beforeafterbarsho}.
Here, $|\mathcal{S}_{\rm before}|$ is the number of empirically
observed one-hot configurations, augmented by the guaranteed reference
configuration (Section~\ref{sec:mainconfig}), and
$|\mathcal{S}_{\rm after}|$ is the size of the final repaired
distribution.

\begin{figure*}[!t]
  \centering
  \includegraphics[width=\textwidth]{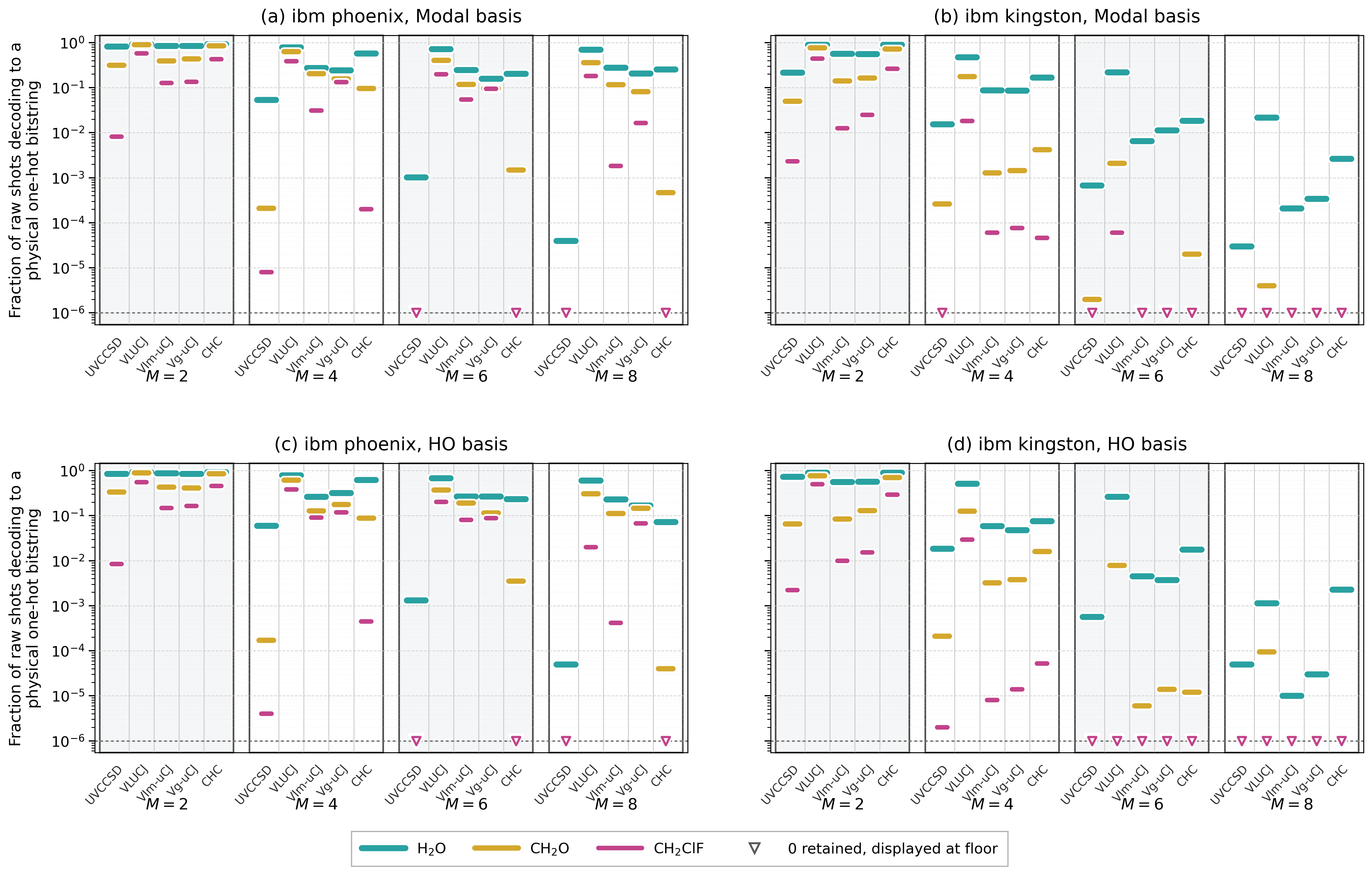}
  \caption{One-hot retention (fraction of raw shots decoding to a
  physical bitstring, before recovery) vs.\ modal count $M$, log scale,
  in four panels: \code{ibm\_phoenix} (left) vs.\ \code{ibm\_kingston}
  (right), modal basis (top) vs.\ HO basis (bottom). Each outlined,
  alternately-shaded group corresponds to one modal count $M$ and
  contains five ansatz columns (in the order UVCCSD, VLUCJ, VIm-uCJ,
  Vg-uCJ, CHC, left to right, separated by thin vertical rules). Within
  one ansatz column, each molecule's individually measured retention
  (H$_2$O, CH$_2$O, CH$_2$ClF, each run once at that parameter setting
  -- \emph{not} repeated trials of the same system) is drawn as its own
  short horizontal line at that value, colored by molecule (teal =
  H$_2$O, gold = CH$_2$O, magenta = CH$_2$ClF); the three lines are
  drawn at deliberately different lengths (H$_2$O longest, CH$_2$O
  medium, CH$_2$ClF shortest) so that two or three molecules landing on
  the same, or nearly the same, retention value all remain individually
  visible rather than overdrawing each other. 
  A hollow downward triangle, in that molecule's color, replaces the
  horizontal line whenever a molecule's raw retention at that parameter
  setting was \emph{exactly zero} -- plotted at the $10^{-6}$ floor
  since a true zero cannot be placed on a log axis, but visually
  distinct from an ordinary line so it is never mistaken for a real,
  small-but-nonzero measurement; when more than one molecule is
  floor-censored at the same parameter setting, their triangles are
  nested at slightly different sizes so each remains individually
  visible. The retention trends and ansatz comparisons are discussed in
  Section~\ref{sec:onehotexplains}.}
  \label{fig:onehotcliff}
\end{figure*}

\begin{figure*}[t]
  \centering
  \includegraphics[width=\textwidth]{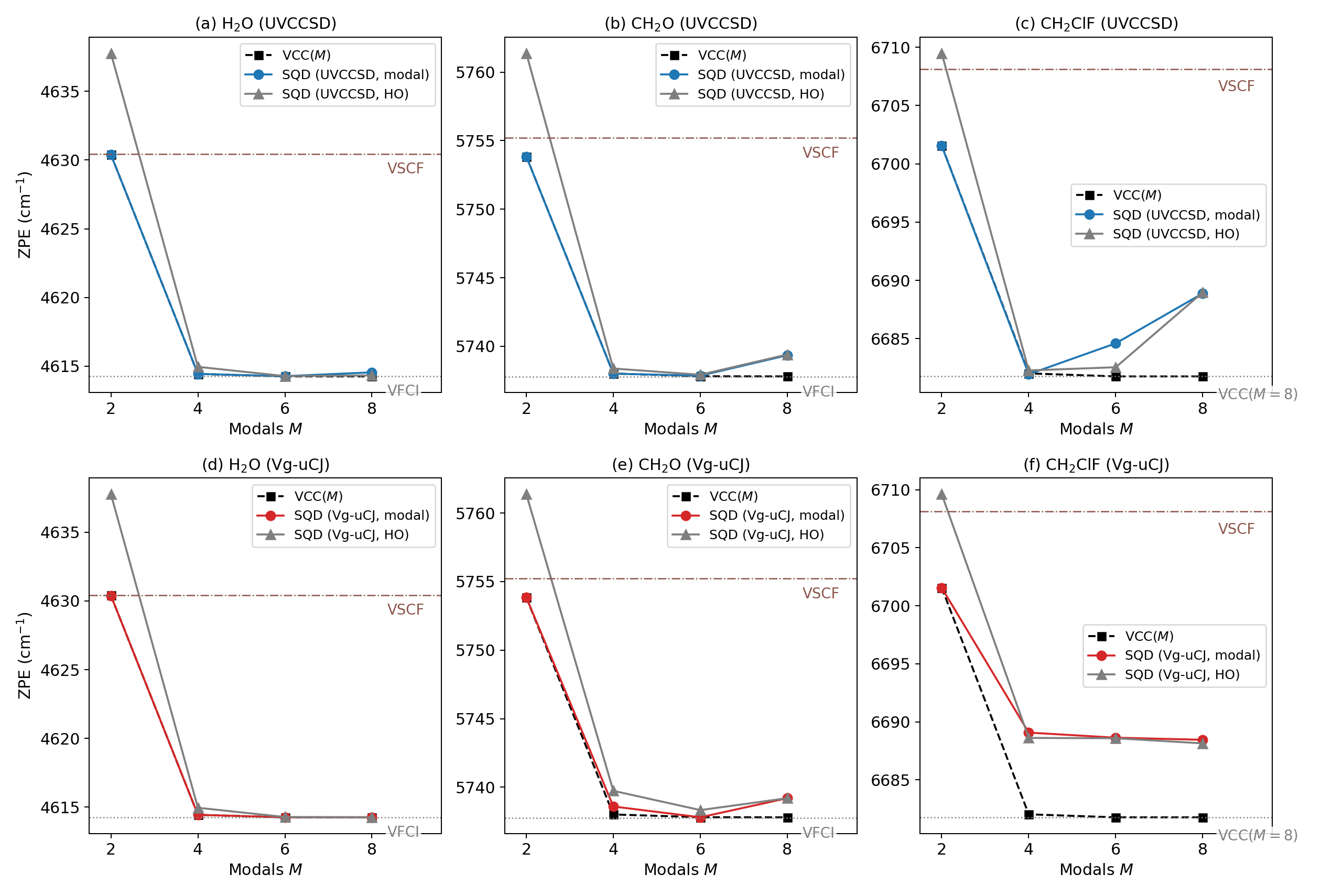}
  \caption{Ground-state energy vs.\ modal count $M$ on \code{ibm\_phoenix} for UVCCSD (top row, panels a--c) --
  the deepest, highest-CZ-count ansatz tested (Table~\ref{tab:ansatzlayout})
  -- and Vg-uCJ (bottom row, panels d--f), a representative shallow
  Jastrow-family circuit, in both the modal and HO bases. Results are
  compared with the classical VCC($M$) hierarchy and with the converged
  VFCI reference for H$_2$O and CH$_2$O, or the VCC($M{=}8$) reference
  for CH$_2$ClF (Table~\ref{tab:classicalrefs}). Numerical errors and
  retention-dependent trends are discussed in the molecule-specific
  subsections below.}
  \label{fig:energyvsMground}
\end{figure*}

\textcolor{cyan}{The fraction of raw shots decoding to a
  physical bitstring, before recovery as a function of the number of modals $M$ is shown in Fig.~\ref{fig:onehotcliff}, and the ground and excited state energies as a function of $M$ are displayed in Figs.~\ref{fig:energyvsMground} and \ref{fig:energyvsMexcited}, respectively}.

\begin{table}[t]
\centering
\caption{Number of distinct one-hot configurations diagonalized for
each bar in Figure~\ref{fig:beforeafterbarsmodal} (modal basis) and
Figure~\ref{fig:beforeafterbarsho} (HO basis), both \code{ibm\_phoenix} hardware, $M{=}4$. $|\mathcal{S}_{\rm before}|$ is
the raw one-hot subspace size (faded bars); $|\mathcal{S}_{\rm after}|$
is the repaired distribution size (filled bars) -- both counts are of
\emph{distinct configurations}, not raw shots, and
$|\mathcal{S}_{\rm after}|\ge|\mathcal{S}_{\rm before}|$ by construction
since recovery only adds decoded configurations to the raw set.
Section~\ref{sec:recovery} describes the recovery loop that produces
$\mathcal{S}_{\rm after}$. In the production calculations, we
diagonalize in the complete set of distinct configurations appearing in
the final repaired distribution shown here; no \emph{a priori}
\code{basis\_size} cutoff is imposed.}
\label{tab:configcounts}
\footnotesize
\setlength{\tabcolsep}{4pt}
\begin{tabular}{@{}llrrrr@{}}
\toprule
 & & \multicolumn{2}{c}{Modal, Phoenix} & \multicolumn{2}{c}{HO, Phoenix} \\
Molecule & Ansatz & $|\mathcal{S}_{\rm before}|$ & $|\mathcal{S}_{\rm after}|$ & $|\mathcal{S}_{\rm before}|$ & $|\mathcal{S}_{\rm after}|$ \\
\midrule
H$_2$O & UVCCSD & 64 & 64 & 64 & 64 \\
 & VLUCJ & 18 & 23 & 16 & 17 \\
 & VIm-uCJ & 54 & 55 & 53 & 53 \\
 & Vg-uCJ & 28 & 32 & 31 & 35 \\
 & CHC & 28 & 32 & 30 & 31 \\
\addlinespace
CH$_2$O & UVCCSD & 22 & 1613 & 18 & 1601 \\
 & VLUCJ & 27 & 40 & 30 & 39 \\
 & VIm-uCJ & 148 & 292 & 181 & 417 \\
 & Vg-uCJ & 184 & 376 & 161 & 343 \\
 & CHC & 131 & 317 & 148 & 457 \\
\addlinespace
CH$_2$ClF & UVCCSD & 5 & 18866 & 3 & 19292 \\
 & VLUCJ & 116 & 251 & 120 & 265 \\
 & VIm-uCJ & 540 & 3069 & 367 & 1260 \\
 & Vg-uCJ & 276 & 745 & 458 & 1441 \\
 & CHC & 74 & 6855 & 168 & 9390 \\
\bottomrule
\end{tabular}
\end{table}

Two conclusions follow directly from Figures~\ref{fig:beforeafterbarsmodal}--\ref{fig:beforeafterbarsho} and guide the interpretation of the remaining results. First, \textbf{the relative quality of modal- and HO-basis raw samples depends on whether the circuit retains enough physical signal for basis-dependent structure to remain observable}. For shallow, high-retention circuits such as VLUCJ, both bases yield an informative raw subspace, but modal basis is the more accurate of the two against the converged VFCI target (H$_2$O: before-recovery error of \SI{1.19}{\cmi} in modal basis vs.\ \SI{15.90}{\cmi} in HO basis), consistent with Section~\ref{sec:basis}'s a priori argument for modal basis. The two bars plotted in
Figures~\ref{fig:beforeafterbarsmodal}--\ref{fig:beforeafterbarsho} are
measured relative to $E_{\rm VSCF}$, not to this converged target, so
they can be misleading if read in isolation: HO's raw sample happens to sit
almost exactly on $E_{\rm VSCF}$ itself ($-0.27\,\mathrm{cm}^{-1}$ in
that plot's units) while modal's sits \SI{14.98}{\cmi} below it, but
$E_{\rm VSCF}$ is a mean-field, pre-correlation energy, not the
anharmonic ground state -- HO's proximity to it is a coincidence of
where that basis's uncorrelated raw sample happens to land, not
evidence that HO basis is more accurate. For deep circuits such as UVCCSD, retention can approach the noise floor, at which point large differences between the two raw energies should not be interpreted as a systematic basis advantage: on \code{ibm\_phoenix} hardware, UVCCSD's ``before'' bar is far from converged in HO basis specifically ($+58.6\,\mathrm{cm}^{-1}$ CH$_2$O, $+43.0\,\mathrm{cm}^{-1}$ CH$_2$ClF, both near \SI{0}{\cmi} in the modal basis at the same molecule/modals/ansatz/backend), and this pattern is not the universal one seen on \code{ibm\_kingston} (Appendix~\ref{app:beforeafterkingston}, Figures~\ref{fig:beforeafterbarsmodalkingston}--\ref{fig:beforeafterbarshokingston}). The observed difference correlates more strongly with circuit depth and one-hot retention than with basis choice alone: UVCCSD's circuit is 30--100$\times$ more CZ gates than VLUCJ's at the corresponding setting in the other basis (Table~\ref{tab:ansatzlayout}), and its one-hot retention has already collapsed below \SI{0.02}{\percent} for CH$_2$O/CH$_2$ClF at this modal count, while VLUCJ's much shallower circuit retains \SI{38}{}--\SI{79}{\percent} at the same settings (Appendix~\ref{app:fulltables-artifacts}) -- consistent with Section~\ref{sec:basis}'s a priori argument for modal basis, but only once circuit depth is shallow enough for that argument's premise (a real, basis-dependent sample) to hold at all. The before/after bars visualize this retention dependence directly: where retention is high, the faded (before) and filled (after) bars remain close because the raw sample already spans a physically meaningful subspace, and the gap grows sharply as retention collapses.

Second, \textbf{VLUCJ yields the highest one-hot retention on both backends}, quantified as the fraction of raw shots decoding to a physically valid configuration. Averaged over all (molecule, modals) parameter settings, VLUCJ retains \SI{25.2}{\percent} (modal) and \SI{26.0}{\percent} (HO) on \code{ibm\_kingston}, 
 versus \SI{8.4}{}/\SI{9.0}{\percent} for the other four ansätze combined; on \code{ibm\_phoenix} the same ordering holds (\SI{56.4}{}/\SI{53.0}{\percent} for VLUCJ versus \SI{20.5}{}/\SI{21.3}{\percent}). Per molecule (Figure~\ref{fig:onehotcliff}), VLUCJ sits above every other ansatz at almost every modal count, in both bases and on both backends -- by roughly one to three orders of magnitude on \code{ibm\_kingston} ($M{\ge}6$), and by a more modest factor of $5$--$40\times$ on the lower-noise \code{ibm\_phoenix}, where the other ansätze's own retention has not yet collapsed as far; the ranking itself is not backend-specific even though its margin is (Section~\ref{app:transpilation}). Retention alone, however, does not fully determine the final post-recovery accuracy: in the less severe noise regime associated with \code{ibm\_phoenix}'s shallower transpiled circuits (Section~\ref{app:transpilation}), VIm-uCJ and Vg-uCJ can achieve smaller final errors against the classical VCC($M$) hierarchy than VLUCJ (Figure~\ref{fig:energyvsMground}), despite retaining less raw signal. Accordingly, the apparent ``best'' ansatz depends on whether the objective is maximizing retained signal (where VLUCJ is consistently strongest) or minimizing the post-recovery energy error (where the ranking can change with backend and transpilation).

Figure~\ref{fig:onehotcliff} shows that one-hot retention falls by
approximately one to two orders of magnitude per doubling of qubit
count across the tested ansätze and bases, consistent with the
rapid-retention-decay mechanism discussed in Section~\ref{sec:cliff}.
Although VLUCJ delays this decline, its retention also decreases with
system size.
Extension to substantially larger active spaces will therefore require
stronger error mitigation, larger shot budgets, more retention-efficient
circuits, or a combination of these improvements.

UVCCSD's post-recovery accuracy (median
$|E_{\rm SQD}-E_{{\rm VCC}(M)}|=\SI{0.27}{\cmi}$, worst case
\SI{7.92}{\cmi}, across all 24 parameter settings) is at least as good
as Vg-uCJ's (median \SI{0.99}{\cmi}, worst case \SI{8.06}{\cmi}) --
concrete evidence that self-consistent recovery is not restricted to
shallow circuits: a circuit with $\binom{N}{2}(M-1)^2$-scaling phase
cost (UVCCSD, and the phase block CHC shares its algebraic structure
with -- Table~\ref{tab:ansatzlayout}) can still yield an accurate
post-recovery energy when the observed and decoded configurations span
a sufficiently informative subspace. Per-molecule trends and numerical
errors are discussed in the molecule-specific subsections below.

\subsection{H$_2$O (water)}
H$_2$O's small size, with only three vibrational modes, is a major
reason for its high accuracy: at $M{=}8$ the repaired distribution's
$|\mathcal{S}_{\rm after}|$ already covers roughly \SI{30}{}--\SI{38}{\percent}
of the full $8^3=512$-configuration one-hot space (UVCCSD/Vg-uCJ; see
Table~\ref{tab:configcounts} for the analogous $M{=}4$ counts), so the
recovery loop is diagonalizing
close to the whole physically relevant subspace rather than a small
fraction of it -- unlike CH$_2$O and CH$_2$ClF below, where the same
fraction is under 1\% and under 0.01\% respectively at the same modal
count. Across all five ansätze at $M{=}8$, ground-state modal-basis results
sit within \SI{0.00}{}--\SI{0.30}{\cmi} of VCC($M{=}8$)
(\SI{4614.25}{\cmi}), and HO-basis results within
\SI{0.00}{}--\SI{0.08}{\cmi} (Appendix~\ref{app:fulltables-artifacts})
-- an order of magnitude tighter than either larger molecule below,
consistent with H$_2$O's retention staying high at every modal count
tested (Figure~\ref{fig:onehotcliff}). Excited-state agreement across
all five ansätze is similarly tight, \SI{0.01}{}--\SI{3.32}{\cmi} in
modal basis (VLUCJ the one outlier; every other ansatz within
\SI{0.04}{\cmi}) and \SI{0.04}{}--\SI{1.52}{\cmi} in HO basis
(Appendix~\ref{app:fulltables-artifacts}). After recovery, Vg-uCJ on
\code{ibm\_phoenix} reaches \SI{4614.26}{\cmi} at $M{=}8$, within
\SI{0.01}{\cmi} of the converged VFCI ground-state energy
(\SI{4614.25}{\cmi}, Table~\ref{tab:classicalrefs}), and stays within
\SI{0.01}{\cmi} of the classical VCC($M$) hierarchy at every modal count 2
through 8 (Figure~\ref{fig:energyvsMground}a). The other four ansätze
give similarly small errors, all within \SI{0.30}{\cmi} of VCC($M{=}8$)
(Appendix~\ref{app:fulltables-artifacts}).
At $M{=}8$, modal-basis VIm-uCJ gives \SI{1552.20}{\cmi} above ground,
only \SI{0.01}{\cmi} below the VCC($M{=}8$) value
(\SI{1552.21}{\cmi}, Table~\ref{tab:classicalrefs}) -- among the most
accurate results in this paper, consistent with H$_2$O's uniquely high
retention throughout (Figure~\ref{fig:onehotcliff}). UVCCSD, Vg-uCJ,
and CHC are also within \SI{0.04}{\cmi}, whereas VLUCJ is the sole
outlier, lying \SI{3.32}{\cmi} above the reference
(Appendix~\ref{app:fulltables-artifacts}).

Thus, H$_2$O shows no systematic ground-state separation by ansatz
family; only the modal-basis VLUCJ excited-state result is clearly
distinct, in contrast to the more ansatz-dependent excited-state
behavior found for CH$_2$ClF. Already at the smaller, classically
tractable benchmark points $M{=}2$ and $M{=}4$, the recovered hardware
energies follow the corresponding classical VCC($M$) results and can
also be compared directly with VFCI
(Figures~\ref{fig:beforeafterbarsmodal}--\ref{fig:energyvsMground}),
calibrating the hardware-plus-recovery workflow against an independent
classical benchmark that remains available at every modal count
studied for this molecule.

\subsection{CH$_2$O (formaldehyde)}
CH$_2$O's 6 modes make the full one-hot space at $M{=}8$
($8^6=262{,}144$ configurations) far too large to cover even a small
fraction of: the repaired distribution's $|\mathcal{S}_{\rm after}|$
reaches only \SI{0.4}{}--\SI{0.8}{\percent} of it (UVCCSD/Vg-uCJ,
Appendix~\ref{app:fulltables-artifacts}) -- two orders of magnitude below H$_2$O's
own coverage above, which is consistent with the larger, though still
modest, ground-state deviations reported next. At $M{=}8$, the five ansätze yield ground-state energies that are both
more accurate and less dispersed than for CH$_2$ClF below. In the modal
basis, all ansätze lie within \SI{1.40}{}--\SI{2.04}{\cmi} of
VCC($M{=}8$) (\SI{5737.80}{\cmi}); in the HO basis, all results fall
within \SI{1.38}{}--\SI{2.00}{\cmi} (Appendix~\ref{app:fulltables-artifacts}).
As for H$_2$O above, modal basis is the more accurate of the two for
CH$_2$O's raw (before-recovery), shallow-circuit sample when measured
against the converged target rather than against $E_{\rm VSCF}$.
This behavior is consistent with CH$_2$O's intermediate one-hot retention
(Figure~\ref{fig:onehotcliff}). The excited-state deviations remain
below \SI{1.5}{\cmi} overall
(Appendix~\ref{app:fulltables-artifacts}), although VIm-uCJ
and VLUCJ form a less accurate pair in the modal basis at $M{=}8$, as
detailed below. Modal-basis Vg-uCJ on \code{ibm\_phoenix} tracks the classical
VCC($M$) ground-state hierarchy closely through $M{=}6$ before drifting upward by
\SI{1.40}{\cmi} at $M{=}8$
(Figure~\ref{fig:energyvsMground}b) -- the same pattern H$_2$O does not
show, consistent with CH$_2$O's lower one-hot retention at
these qubit counts (Figure~\ref{fig:onehotcliff}). The remaining
ansätze show a comparable $M{=}8$ shift, with modal-basis errors
confined to \SI{1.40}{}--\SI{2.04}{\cmi} (Appendix~\ref{app:fulltables-artifacts}).
CH$_2$O's
excited-state fundamental ($\nu_4$, CH$_2$ wagging) recovers to
\SI{1160.61}{\cmi} above ground at $M{=}6$ (VIm-uCJ, modal basis,
\code{ibm\_phoenix}), within \SI{0.94}{\cmi}, and settles to
\SI{1160.14}{\cmi} at $M{=}8$, within \SI{1.41}{\cmi} of the converged
classical value (\SI{1161.55}{\cmi}, Table~\ref{tab:classicalrefs}).
UVCCSD (\SI{1161.68}{\cmi}, \SI{0.13}{\cmi} above), Vg-uCJ
(\SI{1161.31}{\cmi}, \SI{0.24}{\cmi} below), and CHC (\SI{1161.78}{\cmi},
\SI{0.23}{\cmi} above) are all closer to the classical value at $M{=}8$ than
either Jastrow-based VIm-uCJ or VLUCJ (\SI{1160.14}{\cmi},
\SI{1.41}{\cmi} below, Appendix~\ref{app:fulltables-artifacts}).

For CH$_2$O, no clean Jastrow-versus-non-Jastrow separation is evident
at $M{=}8$. VIm-uCJ and VLUCJ give the two largest excited-state
undershoots, but Vg-uCJ, which is also a Jastrow-family ansatz, remains
much closer to the classical reference. The ground-state results
likewise show no ansatz-family separation, with all five ansätze
producing similarly sized positive deviations. At the classically
tractable benchmark point $M{=}4$, the recovered hardware energies
follow the corresponding classical VCC($M$) results and can also be
compared directly with VFCI
(Figures~\ref{fig:beforeafterbarsmodal}--\ref{fig:energyvsMground}),
calibrating the workflow before the $M{=}6/8$ results above, for which
CH$_2$O's own VFCI calculation is no longer tractable.

\subsection{CH$_2$ClF (fluorochloromethane)}
\label{sec:ch2fcl}
Unlike H$_2$O and CH$_2$O above, CH$_2$ClF has no classically tractable
VFCI calculation at any modal count (Table~\ref{tab:classicalrefs}'s
footnote). Converged VCC($M{=}8$) and large-$|\mathcal{S}|$ VHCI values
remain available as classical references instead
(Table~\ref{tab:classicalrefs}), so the results below are not devoid of
classical comparison, even though they cannot be calibrated against an
exact VFCI benchmark the way H$_2$O's and CH$_2$O's smaller-$M$ results
are above.
CH$_2$ClF's 9 modes push the full one-hot space at $M{=}8$ to
$8^9\approx1.3\times10^8$ configurations, of which the repaired
distribution's $|\mathcal{S}_{\rm after}|$ covers only
\SI{0.002}{}--\SI{0.007}{\percent} (UVCCSD/Vg-uCJ,
Appendix~\ref{app:fulltables-artifacts}) -- another order of magnitude below
CH$_2$O's own coverage, and the clear continuation of the same
trend (a smaller fraction of a rapidly growing space as mode count
increases) that already separates H$_2$O from CH$_2$O above; this is
consistent with CH$_2$ClF showing this section's largest ground-state
deviations from VCC($M{=}8$), reported next. Across the five ansätze
evaluated in each basis at $M{=}8$, absolute ZPE errors relative to
VCC($M{=}8$), whose ZPE is \SI{6681.77}{\cmi}, range from
\SI{4.51}{\cmi} (CHC, modal) to \SI{7.21}{\cmi} (UVCCSD, HO) -- all
five ansätze land within a narrow \SI{4.5}{}--\SI{7.2}{\cmi} band in
modal basis, and \SI{3.7}{}--\SI{7.2}{\cmi} in HO basis, with
VIm-uCJ's HO-basis result (\SI{3.72}{\cmi}) the smallest $M{=}8$ ZPE
error for this molecule (Appendix~\ref{app:fulltables-artifacts}). Unlike H$_2$O and CH$_2$O above,
HO basis is marginally more accurate than modal basis for CH$_2$ClF's
raw, shallow-circuit sample against the converged target. One-hot retention
at $M{=}8$ is uniformly low across the five ansätze, ranging from
\SI{0}{}--\SI{18}{\percent} in the modal basis and \SI{0}{}--\SI{7}{\percent}
in the HO basis, and represents the sharpest retention decline among
the three molecules (Figure~\ref{fig:onehotcliff},
Appendix~\ref{app:fulltables-artifacts}). Excited-state
accuracy against $\Delta E_1^{\rm VCC(8)}$ (\SI{371.22}{\cmi}) is more
scattered across the five ansätze: modal-basis errors range from
\SI{0.26}{\cmi} (CHC) to \SI{4.42}{\cmi} in magnitude (Vg-uCJ, an
undershoot -- Appendix~\ref{app:fulltables-artifacts}), and HO-basis errors range
from \SI{0.17}{\cmi} (Vg-uCJ) to \SI{1.07}{\cmi} (VIm-uCJ,
Appendix~\ref{app:fulltables-artifacts}); as discussed above (Section~\ref{sec:results}),
$\Delta E_1$ inherits no one-sided bound from the ground- and
excited-state energies' own individual variational character, so both
over- and under-shoots of this size are expected, not anomalous.

CH$_2$ClF is the largest vibrational system considered here, with nine
modes and up to 72 qubits at $M{=}8$. It also exhibits the largest
sustained hardware-to-VCC deviation among the three molecules: its
modal-basis Vg-uCJ result on \code{ibm\_phoenix} settles into a
persistent \SI{6.7}{}--\SI{7.0}{\cmi} offset above the classical VCC($M$)
ladder from $M{=}4$ onward and does not reconverge toward the
VCC($M$) sequence (Figure~\ref{fig:energyvsMground}c).
\textcolor{magenta}{We have not calculated the  VFCI reference energy for this
molecule because of the large size ($11^9\times11^9$) of the Hamiltonian matrix.} The largest VHCI calculation
($|\mathcal{S}|\approx 10^6$ configurations, Table~\ref{tab:classicalrefs})
is judged close enough to VFCI to serve in its place, giving
\SI{6681.59}{\cmi} (ZPE) and \SI{371.20}{\cmi} ($\Delta E_1$) as the
tightest available classical references -- VCC($M{=}8$), the target
used throughout this section for consistency with the other two
molecules, sits just \SI{0.18}{\cmi} (ZPE) and \SI{0.02}{\cmi}
($\Delta E_1$) above these VHCI values, so the two classical methods
agree to well within this paper's own hardware-vs-classical
discrepancies. 
\textcolor{purple}{As can be seen in Fig.~\ref{fig:energyvsMexcited},} CH$_2$ClF's excited-state fundamental ($\nu_6$, ClCF bending)
recovers to \SI{369.08}{\cmi} above ground at $M{=}6$ (VIm-uCJ, modal
basis, \code{ibm\_phoenix}, \SI{0.16}{\percent} one-hot retention) --
\SI{2.14}{\cmi} below the converged classical VCC($M{=}6$) value
(\SI{371.22}{\cmi}) -- and to \SI{367.25}{\cmi} at $M{=}8$
(\SI{0.01}{\percent} retention), \SI{3.97}{\cmi} below the same
molecule's own VCC($M{=}8$) target -- CH$_2$ClF's excited-state error
for VIm-uCJ grows with modal count here, unlike the other two
molecules. This increase appears to arise because the reduction in the
recovered ground-state energy with increasing $M$ is not matched by a
comparable reduction in the excited-state 
\textcolor{purple}{energy.} The
other four ansätze do not show a single consistent split at either
modal count: at $M{=}6$, VIm-uCJ (\SI{2.14}{\cmi} below) and CHC
(\SI{3.14}{\cmi} below) both undershoot, while UVCCSD
(\SI{0.28}{\cmi} below), Vg-uCJ (\SI{0.25}{\cmi} below), and VLUCJ
(\SI{0.54}{\cmi} above) stay close to the ladder; at $M{=}8$ the
pattern shifts -- UVCCSD (\SI{0.46}{\cmi} above) and VLUCJ
(\SI{0.72}{\cmi} above) now overshoot, CHC crosses over to a small
undershoot (\SI{0.26}{\cmi} below), and VIm-uCJ and Vg-uCJ remain the
two largest errors, both undershooting by \SI{3.97}{}--\SI{4.42}{\cmi}
(Appendix~\ref{app:fulltables-artifacts}). Only VIm-uCJ and Vg-uCJ -- the two
Jastrow-family ansätze with an excited-modal correlator
(Table~\ref{tab:ansatzlayout}) -- undershoot at both modal counts;
CHC's direction flips between them, and UVCCSD/VLUCJ stay small and
mostly on the overshoot side. Thus, the ansatz grouping observed for
CH$_2$O does not persist consistently across modal counts for
CH$_2$ClF. CH$_2$ClF's repaired set can be large even at modest $M$,
and we always diagonalize in the complete repaired subspace
$\mathcal{S}_{\rm after}$ (Table~\ref{tab:configcounts}) rather than a
smaller truncation of it, since a smaller fixed-dimensional subspace
would retain only the highest-probability configurations while
discarding low-probability contributions that can be collectively
important: at $M{=}4$ (VLUCJ, modal basis, \code{ibm\_kingston}), using
the complete repaired subspace ($|\mathcal{S}_{\rm after}|{=}1334$)
rather than its own highest-probability 200 configurations reduces the
discrepancy with VCC($M{=}4$) from \SI{8.6}{\cmi} to \SI{3.23}{\cmi}.

\subsection{Validation controls and the role of one-hot retention}
\label{sec:hardware}

\subsubsection{Uniform valid-configuration control}
\label{sec:uniform}

To determine whether the recovery procedure alone can reproduce the
reported energies, we applied the same repair-and-resampling workflow
to samples drawn uniformly from the valid one-hot configuration space,
at every modal count for every molecule. The uniform control agrees
with the hardware-derived results at the smallest modal count for
every molecule, and throughout the full modal-count range for H$_2$O
specifically, where the valid-configuration space is small enough to
be sampled extensively either way. For CH$_2$O and CH$_2$ClF, the
uniform control diverges from hardware once the configuration
space grows large enough that uniform sampling can no longer cover it
completely (from $M{=}6$ and $M{=}4$ respectively); at $M{=}8$ the
uniform-control energy differs from hardware's best ansatz by
\SI{63.8}{\cmi} to \SI{74.5}{\cmi}, depending on the molecule. The
hardware-derived results for these two molecules therefore cannot be
reproduced, at the same accuracy, by uniformly sampling valid
configurations and applying the same recovery procedure -- while for
H$_2$O, at the modal counts studied here, blind uniform sampling
remains sufficient. This control rejects a structureless
valid-sampling baseline once the configuration space is large enough
for one to be meaningfully tested, although it does not by itself
distinguish chemically meaningful circuit structure from structured
hardware bias (e.g.\ initialization asymmetry, readout bias, or the
guaranteed-included reference configuration). The complete protocol
and molecule-resolved results are reported in
Appendix~\ref{app:uniformcontrol}.

\subsubsection{One-hot retention as a limiting factor for SQD recovery}
\label{sec:onehotexplains}

UVCCSD has by far the deepest circuit of the five ansätze
(Table~\ref{tab:ansatzlayout}, Section~\ref{sec:ansatze}) but is not
uniformly the worst performer in
Figures~\ref{fig:beforeafterbarsmodal}--\ref{fig:beforeafterbarsho}.
Figure~\ref{fig:onehotcliff} helps explain this by plotting one-hot
retention -- the fraction of raw shots that decode to a physical
one-hot bitstring at all, before any recovery is applied -- against
modal count, for both backends and both bases.

Every ansatz's retention drops by roughly one to two orders of
magnitude per doubling of qubit count, for every molecule, in both
bases -- this sharp one-hot retention drop-off (Section~\ref{sec:cliff})
is the default outcome at these qubit counts, not a theoretical
concern. VLUCJ's retention survives meaningfully past 24 qubits on
both backends, remaining above \SI{1}{\percent} at 36 qubits in both
bases, whereas the other four ansätze fall to values ranging from
exactly zero valid one-hot shots at the applied shot budget to a few
$\times10^{-3}\,\%$ at the same parameter settings
(Appendix~\ref{app:fulltables-artifacts}). This retention advantage holds
independent of raw circuit cost: VLUCJ is also the cheapest of the
five ansätze by CZ count at essentially every parameter setting
(Table~\ref{tab:ansatzlayout}), so its retention advantage is
consistent with, not contradicted by, its low gate count. On the
lower-noise \code{ibm\_phoenix} backend specifically, VLUCJ's
retention remains well above this floor even past 48 qubits (30--40\%
at 48 qubits, 19--20\% at 54 qubits); on \code{ibm\_kingston} backend,
VLUCJ's own retention has likewise fallen below \SI{2}{\percent} by 48
qubits, so this extended survival past 48 qubits is a
\code{ibm\_phoenix}-specific effect, not a property of VLUCJ
independent of backend.

Once retention approaches this empirical floor for a given ansatz,
circuit-depth comparisons become less informative, because the
recovered subspace is increasingly determined by a very small number
of valid observations and by the recovery procedure rather than by
which configurations the ansatz's ideal circuit could in principle
reach. This regime is reached earlier for UVCCSD and CHC than for
VLUCJ, while VIm-uCJ and Vg-uCJ exhibit intermediate behavior
(Section~\ref{sec:ansatzdiscuss}). This is the regime
Figure~\ref{fig:energyvsMground} shows for CH$_2$O and CH$_2$ClF at
$M{=}6/8$: the recovered energy drifts upward from the classical
VCC($M$) target as retention collapses, while H$_2$O -- whose
retention stays highest of the three molecules throughout
(Figure~\ref{fig:onehotcliff}, shown earlier in
Section~\ref{sec:results}) -- remains close to the classical hierarchy
over the tested modal counts.

Low retention limits the interpretation of recovered energies. When
only a small number of valid configurations are observed, the final
energy becomes increasingly sensitive to the composition of the
retained sample, the guaranteed reference configuration, and the
repair-and-resampling procedure, so a small deviation from a classical
reference is not, by itself, sufficient evidence of an accurate
hardware state in this regime. Conversely, retention alone does not
determine accuracy, because a high-retention sample may still omit
configurations that are important to the target eigenstate
(Section~\ref{sec:ansatzdiscuss}). The uniform control
(Section~\ref{sec:uniform}) rejects uniform valid-space sampling as an
explanation for the hardware-derived results, but it does not
independently certify each recovered energy when retention is very
low.

\FloatBarrier
\subsubsection{Excited-state fundamentals: definition, state selection, and interpretation}
\label{sec:excited-results-overview}
Excited-state calculations were performed at $M{=}2,4,6,8$ in both
bases, for all five ansätze on \code{ibm\_phoenix}, and for
VLUCJ, VIm-uCJ, and Vg-uCJ on \code{ibm\_kingston}
(Section~\ref{sec:results}). The complete results are available as
interactive online ledgers (Appendix~\ref{app:fulltables-artifacts}).
Figure~\ref{fig:energyvsMexcited} below shows the \code{ibm\_phoenix}
results for UVCCSD and Vg-uCJ; Figure~\ref{fig:energyvsMexcitedothers}
(Appendix~\ref{app:fulltables-ground-others}) shows the remaining
three ansätze.

The excited-state fundamental frequency reported throughout this
section is $\Delta E_1^{\rm SQD} = E_{\rm ex}^{\rm SQD}-E_0^{\rm SQD}$.
Here $E_{\rm ex}^{\rm SQD}$ is the Ritz eigenvalue of the eigenvector
selected by maximum overlap with the guaranteed-included excited
reference configuration (Section~\ref{sec:excited}).\cite{Saad2011EigenvalueProblems}
This selected root need not coincide with the second-lowest Ritz
eigenvalue of the sampled subspace. $E_0^{\rm SQD}$ is the corresponding SQD ground-state energy
from the same workflow, so $\Delta E_1^{\rm SQD}$ is an excitation
energy internal to the SQD calculation.

\begin{figure*}[!t]
  \centering
  \includegraphics[width=\textwidth]{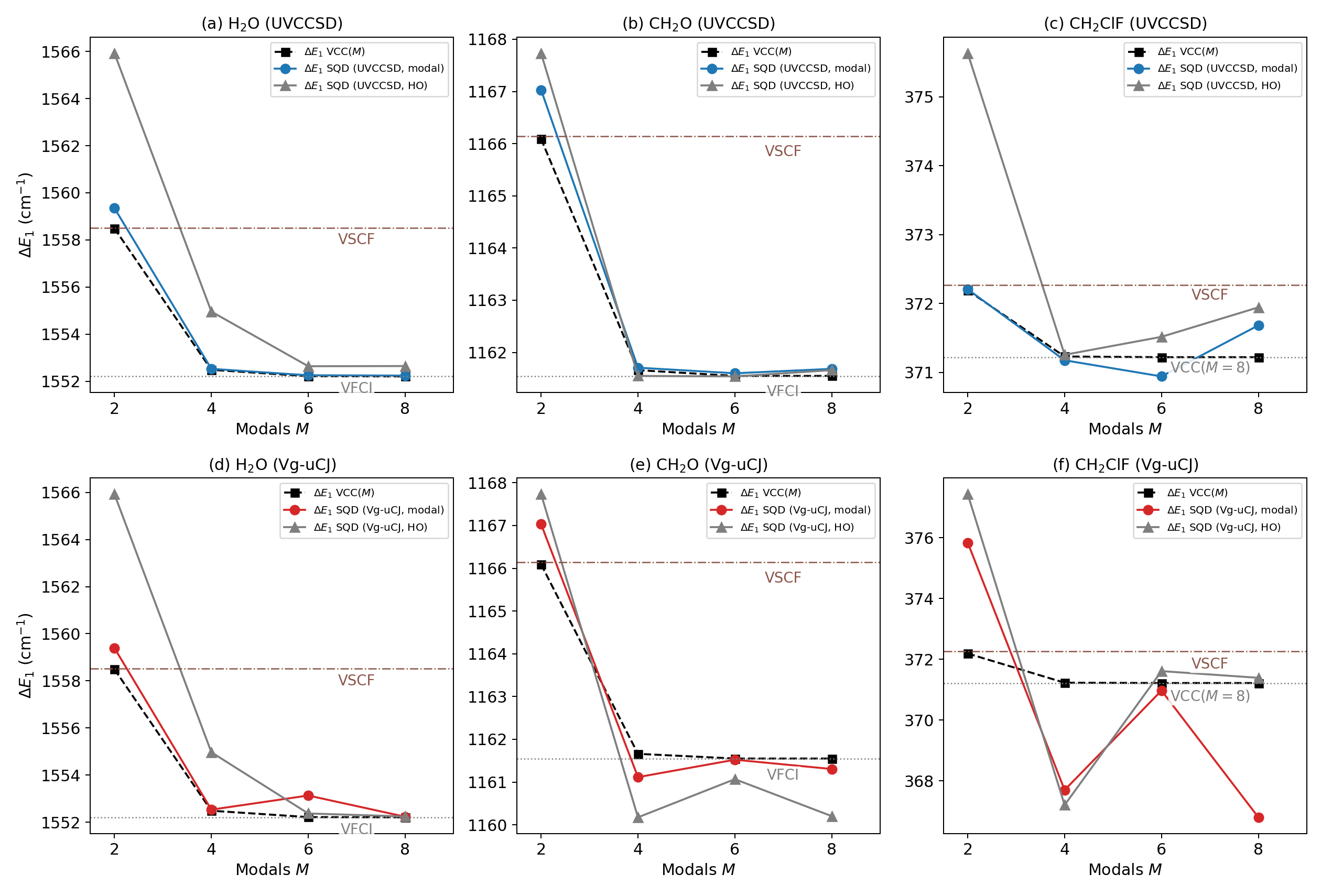}
  \caption{SQD fundamental frequency
  $\Delta E_1^{\rm SQD}=E_{\rm ex}^{\rm SQD}-E_0^{\rm SQD}$ as a
  function of modal count $M$ on \code{ibm\_phoenix} hardware, for
  UVCCSD (top row, panels a--c) and Vg-uCJ (bottom row, panels d--f) in
  both the modal and HO bases. The recovered values are compared with
  the VCC($M$) sequence and with the converged VFCI reference where
  available, or the VCC($M{=}8$) reference for CH$_2$ClF
  (Table~\ref{tab:classicalrefs}). Because $\Delta E_1^{\rm SQD}$ is an
  energy difference, deviations may occur on either side of the
  classical reference (Section~\ref{sec:results}). Medians, worst
  cases, and molecule-specific ranges are given in the text below.}
  \label{fig:energyvsMexcited}
\end{figure*}

\begin{table*}[t]
\centering
\caption{Excited-state targets and recovered fundamental frequencies
(lowest-frequency fundamental per molecule), this paper's current
H$_2$O/CH$_2$O/CH$_2$ClF molecule set, one column per molecule so the
same quantity can be compared directly down each row. All values are
in \si{\cmi}. ``Classical VCC($M{=}8$)'' is the converged
VCC($M$)-hierarchy excitation energy from Table~\ref{tab:classicalrefs}
-- the corresponding VCC($M{=}8$) state-energy difference, not a
difference from the VSCF zero-point energy. ``SQD (\code{ibm\_phoenix})'' is this paper's own recovered
$\Delta E_1^{\rm SQD}=E_{\rm ex}^{\rm SQD}-E_0^{\rm SQD}$
(Section~\ref{sec:results}, Figure~\ref{fig:energyvsMexcited}), with
both energies obtained from the SQD procedure on \code{ibm\_phoenix} execution (UVCCSD, modal basis) at the
same modal count $M$ as the row, not a noiseless-simulator placeholder.}
\label{tab:excited}
\footnotesize
\setlength{\tabcolsep}{4pt}
\begin{tabular}{@{}lccc@{}}
\toprule
 & H$_2$O ($\nu_2$, HOH bend) & CH$_2$O ($\nu_4$, CH$_2$ wagging) & CH$_2$ClF ($\nu_6$, ClCF bending) \\
\midrule
Classical VCC($M{=}8$) & 1552.21 & 1161.55 & 371.22 \\
SQD (\code{ibm\_phoenix}), $M{=}6$ & 1552.26 & 1161.60 & 370.94 \\
SQD (\code{ibm\_phoenix}), $M{=}8$ & 1552.25 & 1161.68 & 371.68 \\
\bottomrule
\end{tabular}
\end{table*}

UVCCSD's accuracy (median $|\Delta E_1^{\rm SQD}-\Delta E_1^{\rm
VCC(M)}|$ of \SI{0.21}{\cmi}, worst case \SI{7.43}{\cmi}, across all 24
parameter settings) is substantially tighter than Vg-uCJ's (median
\SI{0.91}{\cmi}, worst case \SI{7.43}{\cmi}) -- the same
deep-circuit-still-recovers pattern as the ground-state comparison
above (Figure~\ref{fig:energyvsMground}), now for the excitation
energy specifically. UVCCSD's deviations split 20 above/4 below zero
across the 24 cells, and Vg-uCJ's split 14 above/10 below -- both
ansätze legitimately undershoot the classical ladder at some cells
without that indicating an error (Section~\ref{sec:results}). H$_2$O
and CH$_2$O converge to within \SI{1.7}{\cmi} of the ladder by
$M{=}6$--$8$ for both ansätze; the worst outlier for both is H$_2$O's
HO basis at $M{=}2$ (\SI{7.43}{\cmi} above the ladder, the same
parameter setting where the ground-state figure shows its own largest
HO-basis excursion). CH$_2$ClF, the same molecule with the persistent
ground-state offset, shows UVCCSD tracking the ladder within
$-0.28$--\SI{3.44}{\cmi} at every $M$ (its best-behaved modal-basis
case is $M{=}4$--$8$, all under \SI{0.5}{\cmi}), while Vg-uCJ ranges
more widely and less consistently, $-4.42$--\SI{5.24}{\cmi}, including
a \SI{4.42}{\cmi} undershoot at $M{=}8$ (modal basis).

The same comparison for the three ansätze not already shown above --
VIm-uCJ, VLUCJ, and CHC, one row each -- is shown in
Figure~\ref{fig:energyvsMexcitedothers}, given in
Appendix~\ref{app:fulltables-ground-others} alongside its ground-state
counterpart.

VIm-uCJ's accuracy (median \SI{1.20}{\cmi}, worst case \SI{7.43}{\cmi},
across all 24 parameter settings) is looser than UVCCSD's
(\SI{0.21}{\cmi}/\SI{7.43}{\cmi}, Figure~\ref{fig:energyvsMexcited})
and undershoots more often than it overshoots (13 of 24 cells below
zero); its CH$_2$ClF deviations run \SI{-4.67}{}--\SI{2.98}{\cmi},
undershooting at every modal count from $M{=}4$ onward. VLUCJ's
accuracy (median \SI{1.38}{\cmi}, worst case \SI{16.87}{\cmi}) is the
loosest of the five ansätze in this section overall. Its single worst
cell is CH$_2$O, modal basis, $M{=}4$ (\SI{16.87}{\cmi} above, panel
e), despite a one-hot retention of \SI{43.5}{\percent}
(Appendix~\ref{app:fulltables-artifacts}) -- high relative to most cells in this
study; this example shows that high retention alone does not guarantee
accurate recovery, and that the composition of the retained and
repaired subspace also matters. CH$_2$O's other three modal counts all
stay under \SI{2}{\cmi}, and VLUCJ's CH$_2$ClF range,
\SI{0.34}{}--\SI{8.91}{\cmi}, overshoots at every modal count. CHC's
accuracy (median \SI{0.44}{\cmi}, worst case \SI{9.18}{\cmi}) is the
tightest of the three ansätze shown here, close to UVCCSD's own; its
CH$_2$ClF deviations split sign with modal count (\SI{4.88}{},
$-0.12$, $-3.14$, $-0.26\,{\rm cm}^{-1}$ at $M{=}2$/$4$/$6$/$8$ modal
basis respectively) rather than settling on one side. H$_2$O's HO
basis at $M{=}2$ (\SI{7.4}{\cmi} above, the same worst outlier as
UVCCSD and Vg-uCJ above) is VIm-uCJ's and CHC's own worst cell as
well; only VLUCJ's single CH$_2$O outlier above exceeds it. This
repeated \SI{7.4}{\cmi} worst case across four of the five ansätze
traces to the same underlying cell (H$_2$O, HO basis, $M{=}2$), where
retention is high enough for every ansatz to return a similar,
reference-dominated recovered distribution.
With these error trends established, we now examine whether the SQD
fundamental frequency admits a variational interpretation, and where such
arguments may break down.

\paragraph{Variational interpretation of the fundamental frequency.}
Diagonalization in a fixed SQD-selected subspace produces an ordered
sequence of Ritz eigenvalues, and by the min--max principle, the $k$th
ordered Ritz eigenvalue is an upper bound on the corresponding exact
eigenvalue within the same symmetry sector~\cite{MacDonald1933}. 
This statement applies to eigenvalues identified by their ordering: when the
ground state is the lowest Ritz eigenvalue, $E_0^{\rm SQD}\ge E_0^{\rm
VFCI}$. The excited state reported throughout this paper, however, is
instead selected by maximum overlap with the guaranteed-included
excited reference (Section~\ref{sec:excited}), so its state identity
is tracked separately from the eigenvalue ordering; a one-to-one
upper-bound statement relative to the corresponding physical excitation
does not follow solely from the overlap criterion. Consequently, the
fundamental frequency $\Delta E_1^{\rm SQD}=E_{\rm ex}^{\rm SQD}-E_0^{\rm
SQD}$ has no fixed sign relative to the exact fundamental
$\Delta E_1^{\rm VFCI}$: even granting a one-sided ground-state bound,
$\Delta E_1^{\rm SQD}$ can legitimately fall above \emph{or} below
$\Delta E_1^{\rm VFCI}$ without contradiction, since the two absolute
errors need not agree in magnitude. Where VFCI is unavailable,
deviations from $\Delta E_1^{\rm VCC(M)}$ may likewise fall on either
side, but that comparison is between two approximate methods rather
than an application of a variational bound. This is a general property
of excitation energies, not a defect specific to this recovery
procedure -- excited-state calculations that lack a lower-bound
guarantee (e.g.\ folded-spectrum approaches that target interior
roots by minimizing \textcolor{purple}{$\langle \psi|(\hat H-\sigma I)^2|\psi\rangle$
\cite{WangZunger1994FoldedSpectrum}}
or other non-ground-state variational methods) face the same asymmetry).
Because $E_{\rm ex}^{\rm
SQD}$ and $E_0^{\rm SQD}$ are obtained from the same diagonalization
and therefore tend to share correlated errors, $\Delta E_1^{\rm SQD}$
can also be more accurate than either absolute energy individually
would suggest
(Figures~\ref{fig:energyvsMexcited} \textcolor{cyan}{and} \ref{fig:energyvsMexcitedothers}
show deviations of both signs at every ansatz). The relevant
diagnostics are therefore the magnitude and stability of the deviation
as the basis size grows (Section~\ref{sec:results}) and the continuity
of the selected state across modal counts, not the sign of the
deviation, nor whether the ground-state-accuracy grouping of
Section~\ref{sec:ansatzdiscuss} persists unchanged.

\section{Discussion}
\label{sec:discussion}

\subsection{Ansatz structure, measured excitation content, and
practical performance}
\label{sec:ansatzdiscuss}

VLUCJ's diagonal Jastrow correlator is implemented as a controlled-phase
gate,
\begin{equation}
  \mathrm{cp}(\varphi) = \operatorname{diag}(1,1,1,e^{i\varphi})
  = \exp\!\left(i\varphi\,\ket{11}\!\bra{11}\right),
  \label{eq:cpdiscuss}
\end{equation}
diagonal in the computational basis, with Hermitian generator
$G_{\rm cp}=\ket{11}\!\bra{11}=\operatorname{diag}(0,0,0,1)$ on the
physical $\{\ket{00},\ket{01},\ket{10},\ket{11}\}$ subspace
(Appendix~\ref{app:diagonal_phases}). A controlled-phase gate acting
alone cannot change computational-basis populations; population
transfer arises only through its combination with the surrounding
mode-mixing (Givens) rotations, and $U^\dagger\mathrm{cp}(\varphi)U$ is
not diagonal, nor does it reduce to the identity, merely because
$\mathrm{cp}(\varphi)$ is diagonal -- that depends on the specific
qubit pairing and parameter sharing between the phase and the
surrounding rotations. Consequently, the expressivity of the complete
VLUCJ block cannot be inferred from the diagonality of the
controlled-phase operator alone. Appendix~\ref{app:diagonal_phases}
analyzes the specific coupling pattern used by VLUCJ (the \code{cp}
phase couples the \emph{same} modal-index qubit its own surrounding
Givens blocks rotate) and contrasts it with the broader
cross-modal-index coupling pattern CHC's own \code{cp} block uses
instead, despite both ansätze sharing the same ``singles -- diagonal
phase -- singles'' sandwich (Figure~\ref{fig:ansatzcircuits}).

Measured excitation-level decomposition of each ansatz's \code{ibm\_phoenix} recovered distribution for CH$_2$ClF, the largest
molecule in this study, provides empirical evidence consistent with
this structural distinction, though it does not by itself constitute a
rigorous proof for the complete circuit. At $M{=}8$ (72 qubits, modal
basis), CHC and VIm-uCJ/Vg-uCJ all place non-zero, comparable weight on
double and triple excitations (CHC \SI{8.9}{}/\SI{1.4}{\percent},
VIm-uCJ \SI{10.0}{}/\SI{2.4}{\percent}, Vg-uCJ
\SI{8.5}{}/\SI{1.8}{\percent}), while VLUCJ's own doubles/triples
weight is an order of magnitude smaller
(\SI{0.7}{}/\SI{0.05}{\percent}) -- correlated configurations to which
VLUCJ assigns substantially less recovered weight under the present
circuit construction, consistent with the structural argument above.
UVCCSD, despite having by far the deepest circuit of the five
(Table~\ref{tab:ansatzlayout}), places \emph{less} weight on
doubles/triples at this qubit count (\SI{2.2}{}/\SI{0.1}{\percent})
than CHC, VIm-uCJ, or Vg-uCJ do -- its raw one-hot retention at this
parameter setting has already fallen to zero valid one-hot shots at
the applied shot budget (Appendix~\ref{app:fulltables-artifacts}), so its
recovered distribution reflects almost no real signal in the first
place, regardless of what its ideal circuit could reach.

This is exactly the divide this paper's IBM hardware runs show on the retention axis. Averaged over both bases and all
molecule--modal-count combinations (the 24 parameter settings evaluated
for each ansatz, Section~\ref{sec:results}), VLUCJ retains
\SI{25.6}{\percent} of raw shots on \code{ibm\_kingston} and
\SI{54.7}{\percent} on \code{ibm\_phoenix}, well above the individual-ansatz averages for UVCCSD, VIm-uCJ, Vg-uCJ, and CHC, which
range from \SI{4.6}{}--\SI{17.0}{\percent} on \code{ibm\_kingston} and
\SI{10.2}{}--\SI{27.5}{\percent} on \code{ibm\_phoenix}
(Appendix~\ref{app:fulltables-artifacts}, Figure~\ref{fig:onehotcliff}). This
structural distinction between VLUCJ's same-modal-index phase and the
other ansätze's population-transferring or cross-modal-index
correlators is worth keeping as a methodological point about VLUCJ's
own construction, even though it does not decide the practical utility described
below.

The retention gap this structural argument is consistent with is not,
however, the deciding factor for which ansatz to run on IBM hardware,
and it is not simply ``VLUCJ wins, everything else loses.''
Section~\ref{sec:results}'s quantum computer runs, on both backends,
finds VLUCJ -- the ansatz whose correlator structure most limits its
noiseless-limit expressivity for ground-state correlation -- retaining
more one-hot signal than every other ansatz, in both bases, across all
three molecules (Figure~\ref{fig:onehotcliff}), at a small fraction of
the CZ cost (Table~\ref{tab:ansatzlayout}). But UVCCSD and CHC are not
simply ``worse'' versions of the same tradeoff: at CH$_2$ClF's largest
qubit counts ($M{=}6/8$, 54/72 qubits) on \code{ibm\_phoenix},
both retain zero valid one-hot shots at the applied shot budget, while
VIm-uCJ and Vg-uCJ instead retain a visible but reduced signal (VIm-uCJ
\SI{5.5}{}/\SI{0.2}{\percent}, Vg-uCJ \SI{9.5}{}/\SI{1.6}{\percent} at
$M{=}6/8$ respectively), well
below VLUCJ's own \SI{19.7}{}/\SI{18.1}{\percent} at the same two
parameter settings, and clearly intermediate between VLUCJ's
shallower circuit and UVCCSD/CHC's deeper circuits, which fail to
retain any valid one-hot shots at these settings
(Appendix~\ref{app:fulltables-artifacts}). This pattern
holds because VIm-uCJ and Vg-uCJ's own $D_2$ cost is not fixed at
$\binom{N}{2}(M-1)^2$ the way UVCCSD's is: it is a chosen subset of $P$
mode-pair/excited-modal couplings, kept at the largest classically
predicted amplitudes (Section~\ref{sec:warmstart},
Appendix~\ref{app:vim_vg}, Table~\ref{tab:ansatzlayout}), so their
circuit depth can be tuned
between VLUCJ's shallow extreme and UVCCSD's full-expressivity extreme
by choosing $P$, which is why these two ansätze are the ones most
likely to remain usable, not merely accurate in principle, as qubit
count grows well past what this paper reaches. All five ansätze track
the classical VCC($M$) ladder to within a similar few \cmi{} once
retention is high enough for recovery to succeed at all
(Appendix~\ref{app:fulltables-artifacts}); the qubit counts at which that stops
being true, not the noiseless-limit expressivity, is what separates the
five ansätze in practice. Once retention is high enough for recovery to
converge, several different ansätze on the lower-noise
\code{ibm\_phoenix} backend (UVCCSD, VIm-uCJ, Vg-uCJ, CHC) all reach
comparable, sub-\SI{2}{\cmi}-median accuracy against VCC($M$) -- with
VLUCJ, the retention leader, trailing behind all four on accuracy
specifically (Figure~\ref{fig:energyvsMground}) -- a reminder that
``accurate once it runs'' and ``most likely to retain usable signal in
the first place'' are two different questions, and this paper's
cost/retention argument for VLUCJ answers the second, not the first.

The excited-state fundamental shows the same qualitative pattern at
CH$_2$ClF's largest qubit counts, though the middle ground is
noticeably thinner there: at $M{=}6/8$ on \code{ibm\_phoenix},
UVCCSD and CHC both retain \SI{0.0}{\percent} of raw shots for the
excited-state parameter setting, VLUCJ retains \SI{4.9}{}/\SI{8.3}{\percent},
and VIm-uCJ/Vg-uCJ retain a real but much smaller
\SI{0.16}{}/\SI{0.01}{\percent} and \SI{0.02}{}/\SI{0.004}{\percent}
respectively (Appendix~\ref{app:fulltables-artifacts}) -- the same three-tier
structure as the ground state, just compressed, since the excited-state
loop's max-overlap selection (Section~\ref{sec:excited}) has no
variational guarantee to fall back on once retention is this low. This
paper reports only the lowest-frequency fundamental per molecule
(Section~\ref{sec:excited}); extending this same retention/correlation
analysis to higher excited states, as would be needed to reconstruct a
fuller IR spectrum, is not attempted here and is identified as future
work in Section~\ref{sec:future}.

These observations distinguish ideal ansatz expressivity from hardware
utility: the former describes the states accessible without noise,
whereas the latter depends jointly on circuit cost, retained one-hot
sampling, and the composition of the recovered subspace.

\subsection{Ansatz selection on IBM hardware: retention--accuracy trade-offs}
\label{sec:recommendation}

Taken together, the noiseless-limit expressibility considerations and the
measured IBM-hardware performance indicate that ansatz choice is governed
primarily by (i) one-hot retention (the availability of a physically
valid sample for recovery) and (ii) circuit cost at the targeted qubit
count, with numerical accuracy conditional on retention being sufficient
for the recovery loop to converge to a stable subspace.

\begin{itemize}[leftmargin=1.4em]
  \item \textbf{VLUCJ maximizes one-hot retention over the tested regime.}
  Averaged over both bases and all 24 parameter settings per ansatz,
  VLUCJ retains \SI{25.6}{\percent} of raw shots on IBM \code{ibm\_kingston}
  and \SI{54.7}{\percent} on IBM \code{ibm\_phoenix}; the corresponding
  averages for the other four ansätze range from \SI{4.6}{}--\SI{17.0}{\percent}
  on \code{ibm\_kingston} and \SI{10.2}{}--\SI{27.5}{\percent} on
  \code{ibm\_phoenix} (Appendix~\ref{app:fulltables-artifacts},
  Figure~\ref{fig:onehotcliff}). At the highest qubit counts, where several
  alternative ansätze reach the empirical zero-retention floor at the
  available shot budgets, VLUCJ is therefore the most robust choice for
  obtaining a nontrivial recovered subspace.

  \item \textbf{UVCCSD and CHC become retention-limited at the largest CH$_2$ClF settings.}
  Both contain gate blocks whose cost scales as $\binom{N}{2}(M-1)^2$
  (UVCCSD's paired-double excitations and CHC's cross-modal-index phase;
  Table~\ref{tab:ansatzlayout}). For CH$_2$ClF at $M{=}6/8$ (54/72 qubits)
  on IBM \code{ibm\_phoenix}, both retain zero valid one-hot shots at the
  applied shot budgets (Appendix~\ref{app:fulltables-artifacts}), so their
  additional noiseless-limit expressivity cannot be leveraged in this
  regime.

  \item \textbf{VIm-uCJ and Vg-uCJ provide a cost-tunable alternative when off-diagonal correlation is required.}
  At the same CH$_2$ClF settings where UVCCSD and CHC retain zero valid
  shots, VIm-uCJ and Vg-uCJ retain nonzero one-hot signal (VIm-uCJ
  \SI{5.5}{}/\SI{0.2}{\percent}, Vg-uCJ \SI{9.5}{}/\SI{1.6}{\percent} at
  $M{=}6/8$), albeit below VLUCJ's \SI{19.7}{}/\SI{18.1}{\percent}
  (Appendix~\ref{app:fulltables-artifacts}). Their circuit depth is set by a
  selected subset of $P\le\binom{N}{2}(M-1)^2$ mode-pair/excited-modal
  couplings (Section~\ref{sec:warmstart}), enabling an explicit trade-off
  between circuit cost and access to population-transferring two-mode
  correlators. Consistent with this structure, their recovered
  distributions place appreciable weight on correlated configurations
  (\SI{8.5}{}--\SI{10.0}{\percent} on double excitations at CH$_2$ClF,
  $M{=}8$; Section~\ref{sec:ansatzdiscuss}).
\end{itemize}

These recommendations prioritize retention and circuit cost rather than
post-recovery accuracy in isolation. In the lower-noise regime of IBM
\code{ibm\_phoenix}, multiple ansätze (UVCCSD, VIm-uCJ, Vg-uCJ, CHC)
achieve comparable ground-state accuracy against the VCC($M$) ladder
once nonzero retention is maintained (Figure~\ref{fig:energyvsMground}),
so retention does not uniquely determine accuracy.

The present ranking is specific to the hardware regime studied here
(18--72 qubits on IBM \code{ibm\_kingston} and \code{ibm\_phoenix}, one layer
with a separate two-layer check in Section~\ref{sec:ansatze}) and should
not be extrapolated beyond this range without additional measurements.

\subsection{Modal-count and basis tradeoffs}

The modal and HO bases exhibit different raw-sampling behavior, but
neither basis is uniformly superior across all ansätze and molecules.
In principle, the modal basis provides a reference constructed from
VSCF-optimized one-mode functions and may therefore represent
anharmonic structure more compactly than a truncated harmonic-oscillator
basis. On hardware, however, this potential advantage is observable
only when the circuit retains a sufficiently informative one-hot
sample. At $M{=}4$, the deep UVCCSD circuit shows a pronounced
difference between the modal- and HO-basis raw energies
(Figures~\ref{fig:beforeafterbarsmodal}--\ref{fig:beforeafterbarsho}),
whereas the shallower VLUCJ circuit can produce comparably useful
samples in both bases, and the HO result is closer for some settings
(Section~\ref{sec:results}). The observed basis dependence must
therefore be interpreted jointly with circuit depth, one-hot
retention, and recovered-subspace composition rather than attributed
to basis choice alone. Figures~\ref{fig:beforeafterbarsmodal} and
\ref{fig:beforeafterbarsho} establish this comparison at $M{=}4$;
conclusions about its persistence at larger $M$ should be based on the
complete per-cell numbers (Appendix~\ref{app:fulltables-artifacts})
rather than extrapolated from the plotted $M{=}4$ data.

\section{Conclusion and Future Work}
\label{sec:future}

We extended sample-based quantum diagonalization from electronic to
vibrational structure and evaluated the resulting workflow on \code{ibm\_kingston} and \code{ibm\_phoenix} hardware. This work
covered H$_2$O, CH$_2$O, and CH$_2$ClF, spanning three to nine
vibrational modes and 6--72 qubits across the full campaign (H$_2$O at
$M{=}2$ uses 6 qubits; CH$_2$ClF at $M{=}8$ uses 72), with five
warm-started ansätze, four modal counts, and both the VSCF-modal and
harmonic-oscillator bases.

Three main conclusions emerge. First, the relative quality of the
modal- and HO-basis raw samples depends jointly on the ansatz, circuit
cost, and one-hot retention (Section~\ref{sec:results}): a modal-basis
advantage is visible for some deep-circuit settings, but neither basis
is uniformly superior across all ansätze and molecules. When a circuit
retains a sufficiently informative one-hot sample, the VSCF-modal basis
can provide a more compact representation of anharmonic structure;
when retention approaches the empirical sampling floor, apparent basis
differences cannot be interpreted independently of hardware noise and
finite-sampling effects.

Second, VLUCJ gives the highest one-hot retention and the lowest
CZ-gate cost over the measured hardware settings, with its advantage
largest at the higher-qubit settings, where several alternative
ansätze approach or reach the empirical retention floor
(Section~\ref{sec:recommendation}). Averaged over the full campaign
(both bases, all molecule--modal-count combinations), VLUCJ retains
approximately 2.6--2.9$\times$ the valid-shot fraction of the other
four ansätze combined, with substantially larger ratios at selected
high-qubit settings (Section~\ref{sec:ansatzdiscuss}). This does not
make VLUCJ the most accurate ansatz after recovery: on \code{ibm\_phoenix}, the post-recovery accuracy ranking differs from
the retention ranking. UVCCSD, VIm-uCJ, Vg-uCJ, and CHC give
ground-state median absolute deviations of \SI{0.27}{}--\SI{1.40}{\cmi}
and fundamental-frequency median deviations of
\SI{0.21}{}--\SI{1.20}{\cmi} (Figures~\ref{fig:energyvsMground},
\ref{fig:energyvsMexcited}); VLUCJ has a larger median deviation for
both, at \SI{2.0}{\cmi} (ground state) and \SI{1.38}{\cmi} (fundamental
frequency). On \code{ibm\_kingston} backend, VLUCJ's corresponding median
deviations are \SI{2.2}{\cmi} (ground state) and \SI{2.2}{\cmi}
(fundamental frequency), each averaged over the same 24
molecule--modal-count--basis combinations. Retention and final
numerical accuracy must therefore be treated as distinct metrics:
retention quantifies how much valid encoded sampling survives hardware
execution, whereas accuracy additionally depends on whether the
retained and repaired configurations span the physically important
part of the target-state subspace. This ranking differs from one based
solely on ideal ansatz expressivity (Section~\ref{sec:ansatzdiscuss})
-- VLUCJ's diagonal Jastrow correlator, acting alone, cannot change
computational-basis populations, although the complete
mixing--phase--mixing circuit can do so through interference -- which
is itself the paper's central methodological lesson: at the qubit
counts a near-term device actually reaches, the survival of a
sufficiently informative one-hot sample can be at least as important
as the ideal correlation capacity of the noiseless circuit.

Third, VIm-uCJ and Vg-uCJ provide a useful intermediate regime between
the low-cost, high-retention VLUCJ circuit and the more expensive
UVCCSD and CHC constructions. Their selected $D_2$ couplings allow the
circuit cost to be adjusted through the number of retained
mode-pair/excited-modal terms $P$, providing a tunable compromise
between hardware retention and recovered higher-excitation content.
The present data support this tradeoff over the measured 6--72-qubit
range, but do not establish that the same ranking will persist at
larger system sizes, greater circuit depths, or on other hardware
architectures.

Diagonalizing in the complete repaired subspace, rather than a smaller
fixed-dimensional truncation of it, is itself an important part of
this accuracy picture for CH$_2$ClF at intermediate modal counts,
where the repaired distribution is large enough that a small subspace
can miss collectively important configurations
(Section~\ref{sec:ch2fcl}). At the largest modal counts, however,
one-hot retention becomes the dominant limitation: when the raw sample
contains little recoverable physical signal, enlarging the subspace
cannot compensate for information that was never measured.

A uniform valid-configuration control, processed with the same
repair-and-resampling procedure as the hardware data
(Section~\ref{sec:uniform}, Appendix~\ref{app:uniformcontrol}), agrees
with the hardware-derived results at the smallest modal count for
every molecule, and throughout the modal-count range studied for
H$_2$O specifically, but diverges from hardware for CH$_2$O and
CH$_2$ClF once their configuration space grows large enough that
uniform sampling can no longer cover it completely (from $M{=}6$ and
$M{=}4$ respectively), reaching a \SI{63.8}{\cmi}--\SI{74.5}{\cmi}
deviation from hardware's best ansatz by $M{=}8$. Uniform
sampling of the valid one-hot space is therefore insufficient to
reproduce the hardware-derived energies for the two larger molecules.
This rejects a structureless valid-sampling baseline once the
configuration space is large enough for one to be meaningfully
tested, although it does not by itself distinguish chemically
meaningful circuit structure from structured hardware, initialization,
or readout bias.

The workflow was also extended to the lowest-frequency fundamental of
each molecule, executed on IBM hardware for all three molecules
(Section~\ref{sec:hardware}), with Vg-uCJ's HO-basis, $M{=}6$ result on
IBM \code{ibm\_kingston} matching the converged classical excitation
energy to within \SI{0.36}{\cmi}. Both the ground- and excited-state
energies are obtained from the same SQD-selected subspace, and the
target excited state is identified by maximum overlap with a
guaranteed reference configuration rather than by eigenvalue ordering.
This state-tracking procedure avoids a failure mode of selecting the
lowest Ritz eigenvalue, which is appropriate for the ground state but
can instead return a lower-energy state when the sampled subspace
contains configurations outside the intended excited-state manifold;
we identify the target state by maximum overlap with the
guaranteed-included excited reference instead, promoting consistent
identification of the targeted state across modal counts. Because the
reported fundamental frequency is an energy difference, its deviation
from the classical reference is not one-sided even when the
corresponding ordered Ritz energies individually satisfy variational
upper bounds (Section~\ref{sec:results}); the sign of the frequency
deviation must therefore be interpreted together with its magnitude,
modal-count stability, and the continuity of the selected state.

The principal directions for future work are:
\begin{itemize}[leftmargin=1.4em]
  \item \textbf{Extension to broader vibrational spectra and IR
  intensities.} The present work considers only the lowest-frequency
  fundamental of each molecule. Computing additional excited states and
  IR intensities will require systematic multi-state tracking and a
  dipole-moment surface in addition to the potential energy surface
  used here.
  \item \textbf{Improved recovery and uncertainty analysis.} The
  repair-and-resampling rule (Section~\ref{sec:recovery}) should be
  refined to reduce sensitivity to low-retention samples. Repeated
  hardware executions, bootstrap or resampling analyses, and explicit
  propagation of finite-shot uncertainty would help distinguish stable
  recovered energies from fortuitous agreement with a classical
  reference.
  \item \textbf{Larger and classically challenging molecules.} The
  present systems remain accessible, at least in part, to VCC, VHCI, or
  VFCI benchmarking, so no quantum-advantage claim is made here. A
  natural next step is to extend the workflow---particularly the
  cost-tunable VIm-uCJ and Vg-uCJ constructions---to larger vibrational
  active spaces beyond the reach of exact diagonalization, targeting
  both ground and low-lying excited states. Such calculations will
  require validation strategies that do not rely exclusively on exact
  classical reference energies (e.g.\ convergence with respect to
  recovered-subspace size, internal consistency checks, and
  cross-method comparisons where available).
  \item \textbf{Additional backends, qubit counts, and circuit depths.}
  Future hardware experiments should test whether the observed VLUCJ
  retention advantage and the intermediate VIm-uCJ/Vg-uCJ tradeoff
  persist beyond 72 qubits, on other hardware architectures, and at
  layer counts beyond the one- layer comparison already run
  here (Section~\ref{sec:ansatze}).
  \item \textbf{Iterative subspace diagonalization.} Dense
  diagonalization with \code{numpy.linalg.eigh}\cite{Harris2020NumPy} becomes a computational
  bottleneck for the largest repaired subspaces, which contain up to
  $\sim$27{,}000 configurations in this study (several hours per
  parameter setting on a CPU-GPU cluster, Section~\ref{sec:results}). A
  sparse block Davidson (SBD) or related iterative solver, combined
  with overlap-based root tracking for the excited states (rather than
  targeting only the globally lowest eigenpairs), would reduce the
  memory and computational cost of the recovery loop.
  \item \textbf{Acceleration of classical preprocessing.} The {\sc MidasCpp}
  coupled-cluster calculations used to generate the $t_1$/$t_2$
  warm-start amplitudes (Section~\ref{sec:warmstart}) may become a
  significant cost for larger active spaces. Algorithmic optimization
  and accelerator-supported or distributed implementations of this
  preprocessing stage should be investigated alongside the hardware and
  diagonalization improvements above.
\end{itemize}

\section{Acknowledgments}
\label{sec:Acknowledgments}
%
A portion of this work at RIKEN is based on results obtained from Project No.~JPNP20017, supported by the New Energy and Industrial Technology Development Organization (NEDO). 
This work at RIKEN was also supported by JSPS KAKENHI Grants No.~JP21H04446, No.~JP24K08336, and No.~JP26K06972; JST COI-NEXT (Grant No.~JPMJPF2221); the Program for Promoting Research on the Supercomputer Fugaku (Grant No.~MXP1020230411) by MEXT, Japan; and the COE research grant in computational science from Hyogo Prefecture and Kobe City through the Foundation for Computational Science. 
E.L.~and S.Y.~acknowledge support from the UTokyo Quantum Initiative and the RIKEN TRIP initiative (RIKEN Quantum).   
We also thank the JHPC-quantum GPU supercomputer \textit{ROQUO} for providing GPU resources used in the diagonalization calculations. We gratefully acknowledge our colleagues, Sarah Mostame, Tanvi Gujarati,  Alberto Baiardi, and Mario Motta, for sharing complementary insights from their concurrent research on vibrational SQD methods. A related study detailing their specific findings is being prepared/published in parallel with this work.

\FloatBarrier
\clearpage

\appendix
\renewcommand{\thefigure}{A\arabic{figure}}
\setcounter{figure}{0}
\section{Ansatz provenance and operator-level expressibility}
\label{app:ansatz_structure}

This Appendix summarizes the literature provenance of the five circuit
families and analyzes their operator-level structure as used in the
main-text calculations. Our goal is to disentangle three elementary
mechanisms that these ans\"atze combine in different ways: (i)
intramode population transfer among modals within a single vibrational
mode, (ii) configuration-dependent phase accumulation, and (iii)
explicit two-mode correlated excitations with independent
parameterization.

UVCCSD follows the systematically motivated unitary vibrational coupled-
cluster constructions developed in
Refs.~\cite{McArdle2019DigitalVibrations,Ollitrault2020VibrationalQC,JensenChristiansen2025UVCC}.
VLUCJ is the vibrational analogue of the local unitary cluster Jastrow
construction introduced for molecular electronic structure by Motta
\textit{et al.}~\cite{Motta2023LUCJ}. Its defining structure is a
diagonal Jastrow operator surrounded by basis-changing rotations,
conventionally written as $U e^{iJ} U^\dagger$ in the present
implementation as well, with $U$ the intramode Givens rotation and $J$
the diagonal Jastrow phase defined below.

VIm-uCJ and Vg-uCJ are vibrational adaptations of the electronic Im-uCJ
and g-uCJ correlators introduced by Tkachenko
\textit{et al.}~\cite{Tkachenko2025BeyondReal}. In the present
implementation, both contain selected paired two-mode rotations (the
$D_{ij}^{kl}(\theta)$ paired-excitation primitive of
Eq.~\eqref{eq:paired_double_pauli}), while
Vg-uCJ supplements each selected rotation with a diagonal phase acting
on the associated excited-modal pair. CHC is a compact comparison
circuit inspired by the hardware-efficient vibrational construction of
Ollitrault, Baiardi, Reiher, and
Tavernelli~\cite{Ollitrault2020VibrationalQC}. The CHC circuit used
here is not a literal reproduction of the phase-optimized blocks of
that work; rather, it preserves the relevant compact
singles--phase--singles organization used in the comparison.

\subsection{Role of diagonal phases in VLUCJ and CHC}
\label{app:diagonal_phases}

The controlled-phase primitive $\mathrm{cp}(\varphi)$ and its Hermitian
generator $G_{\rm cp}$ are defined in
Eq.~\eqref{eq:cpdiscuss} of Section~\ref{sec:ansatzdiscuss}, together
with the argument for why a controlled-phase gate cannot, by itself,
change computational-basis populations, and why the complete
rotation--phase--rotation circuit can nonetheless do so through
interference. This subsection gives the specific coupling patterns
that distinguish VLUCJ from CHC.

In the present implementation, VLUCJ and CHC use different phase
connectivity patterns. VLUCJ applies its \code{cp} phase to selected
same-modal-index qubit pairs across two modes -- the identical pair its
own surrounding Givens blocks already rotate -- whereas CHC's
\code{cp} block instead couples pairs of \emph{different} excited-modal
indices across two modes. These different coupling graphs define
different variational manifolds, even though both circuits share the
same singles--phase--singles organization
(Figure~\ref{fig:ansatzcircuits}). Across the molecules, modal
truncations, and one-mode bases examined here in noiseless simulation,
the optimized VLUCJ energy coincides with the reference-configuration
energy to numerical precision; CHC's noiseless-simulation optimum
plateaus at the same value in this particular implementation and
optimization protocol. These are empirical observations specific to the
present implementation and optimization protocol, not a universal
statement about all rotation-sandwiched Jastrow or CHC-type circuits:
they should not be interpreted as showing that the complete VLUCJ block
cannot, under any parameter choice or circuit realization, redistribute
population. Hardware execution, warm-started directly from
classical VCC amplitudes rather than optimized from a generic starting
point, tells a different story for CHC specifically: its recovered
distributions (Section~\ref{sec:ansatzdiscuss}) carry visible non-zero double- and
triple-excitation weight, an empirical comparison consistent with
CHC's cross-modal-index coupling pattern, though not itself a proof
that this pattern is what enables it.

\subsection{VIm-uCJ and Vg-uCJ: role of the additional phase}
\label{app:vim_vg}

VIm-uCJ introduces selected $D_2$ primitives as explicit two-mode
correlators. Consider two modes $i$ and $j$ with reference modals $r_i$
and $r_j$. We denote by $|{\rm ref}\rangle$ the corresponding
computational-basis configuration with occupations $(r_i,r_j)$ in these
modes (all other modes fixed to their own references), and by
$|ab\rangle$ the paired excitation in which mode $i$ occupies excited
modal $a\neq r_i$ and mode $j$ occupies excited modal $b\neq r_j$.
For such a selected reference--double pair, the anti-Hermitian generator
produces an $SO(2)$-type rotation in the resulting two-dimensional
subspace,
\begin{equation}
  |{\rm ref}\rangle
  \longrightarrow
  \cos\theta\,|{\rm ref}\rangle
  +
  \sin\theta\,|ab\rangle,
\end{equation}
up to the sign convention used for the generator. Unlike a diagonal
phase, this operation directly transfers population between the
reference and the paired doubly excited configuration.

Vg-uCJ retains the same population-generating $D_2$ rotation and adds a
controlled phase acting on the excited modals populated by that
rotation. Once both $|{\rm ref}\rangle$ and $|ab\rangle$ have nonzero
amplitudes, the phase can modify their relative complex phase,
schematically yielding
\begin{equation}
  \cos\theta\,|{\rm ref}\rangle
  +
  e^{i\varphi}\sin\theta\,|ab\rangle.
\end{equation}
The additional parameter therefore enlarges the variational manifold:
VIm-uCJ fixes the relative phase imposed by its rotation, whereas
Vg-uCJ can vary the relative amplitude and phase separately.

This additional freedom does not guarantee a lower optimized energy for
every vibrational Hamiltonian or finite optimization. In the
electronic-structure study motivating these circuit families, the
generalized phase structure improved the recovered correlation energy
relative to the imaginary-only construction.\cite{Tkachenko2025BeyondReal}
In the present vibrational calculations, VIm-uCJ and Vg-uCJ recover
energies within a few tenths of a \cmi{} of one another at most tested
parameter settings, with a handful of CH$_2$ClF cells differing by up
to $\sim$2--3\,\cmi{} (largest observed: 2.65\,\cmi, HO basis,
$M{=}8$, \code{ibm\_phoenix}). Neither ansatz is consistently
lower across these outlier cells. Consequently, the additional phase's
benefit is inconsistent and, where present, modest relative to the
other error sources (finite sampling, hardware noise, recovery) present
throughout this study.

The distinction between population-generating rotations and phase-only
operations also motivates the ansatz-dependent warm starts used in the
main text. Classical $t_1$ and $t_2$ amplitudes map naturally onto Givens
and paired-excitation angles, respectively, whereas the additional
excited-modal phase in Vg-uCJ has no direct counterpart in the classical VCC amplitudes and is therefore initialized independently.

\section{Harmonic-oscillator vs.\ modal basis: full derivation}
\label{app:ho-basis}

This Appendix gives the full derivation summarized in
Section~\ref{sec:basis}, together with the complete warm-start
procedure summarized in Section~\ref{sec:warmstart}.

\subsection{The VSCF rotation matrix}

\begin{figure*}[t]
  \centering
  \includegraphics[width=0.9\textwidth]{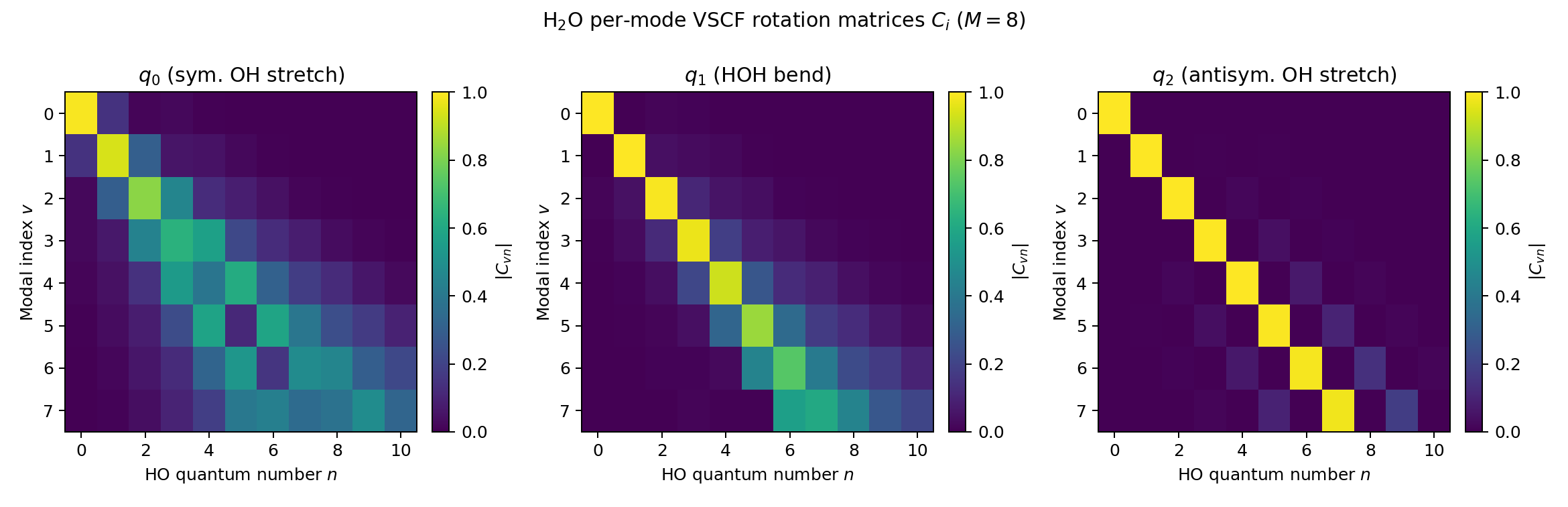}
  \caption{
  H$_2$O's three per-mode VSCF rotation matrices $C^{(i)}$
  ($M=8$), $|C^{(i)}_{kv}|$ as a heatmap. $q_2$ is close to diagonal;
  $q_0$ visibly mixes several HO quanta into each modal
  level; $q_1$ is intermediate. This ordering matches each mode's
  ground-state HO-basis probability spread in Figure~\ref{fig:ho-prob}
  exactly.}
  \label{fig:rotation}
\end{figure*}

$C^{(i)}$ is the same $M\times K$ matrix of Eq.~\eqref{eq:modal_expansion}
(Section~\ref{sec:encoding}): row $k$ is mode $i$'s $k$-th VSCF modal
eigenvector, expressed in the HO basis, $\phi_{ik}(q_i)=\sum_v
C^{(i)}_{kv}\chi_v(q_i)$, obtained by diagonalizing that mode's
single-mode mean-field (VSCF) Hamiltonian with
{\sc MidasCpp.}
\cite{MidasCpp,Bowman1978JCP,Bowman1986AccChemRes} Each row is
already normalized and mutually orthogonal by construction, since it
is an eigenvector of a Hermitian single-mode Hamiltonian; no separate
orthonormalization step is applied. Concretely, $C^{(i)}$ is the
change-of-basis matrix that rotates the $K$ primitive HO functions of
mode $i$ into that mode's $M$ VSCF modal functions: a row close to a
unit vector means that VSCF modal is essentially a single HO level
(the mode is close to harmonic), while a row spread across several HO
quanta means VSCF has mixed multiple HO levels together to build a
better-adapted, more anharmonic basis function. Figure~\ref{fig:rotation}
shows all three of H$_2$O's $C^{(i)}$ as heatmaps: $q_2$ (antisymmetric
OH stretch) is essentially the identity on its diagonal -- modal level
0 \emph{is} HO level 0 to 4 decimal places, i.e.\ this mode's potential
is close to harmonic at the order this force field retains -- while
$q_0$ (symmetric OH stretch) visibly spreads each modal row over
several adjacent HO quanta, and $q_1$ (HOH bend) sits between the two.
This trend is consistent with the reference-configuration energies
reported in the main text: modes with the largest off-diagonal spread in
$C^{(i)}$ are typically those for which the HO-basis reference is least
accurate, and for which switching to the modal basis yields the largest
improvement. This ordering is intended only as a qualitative guide based on
visual inspection of the rotation matrices shown here; we do not compute or
report a separate, quantitative per-mode error decomposition. The same rotation
matrices are what makes the modal-basis Hamiltonian denser than its
HO-basis counterpart, visualized directly in Figure~\ref{fig:ho-vs-modal}
below.

\begin{figure*}[t]
  \centering
  \includegraphics[width=0.9\textwidth]{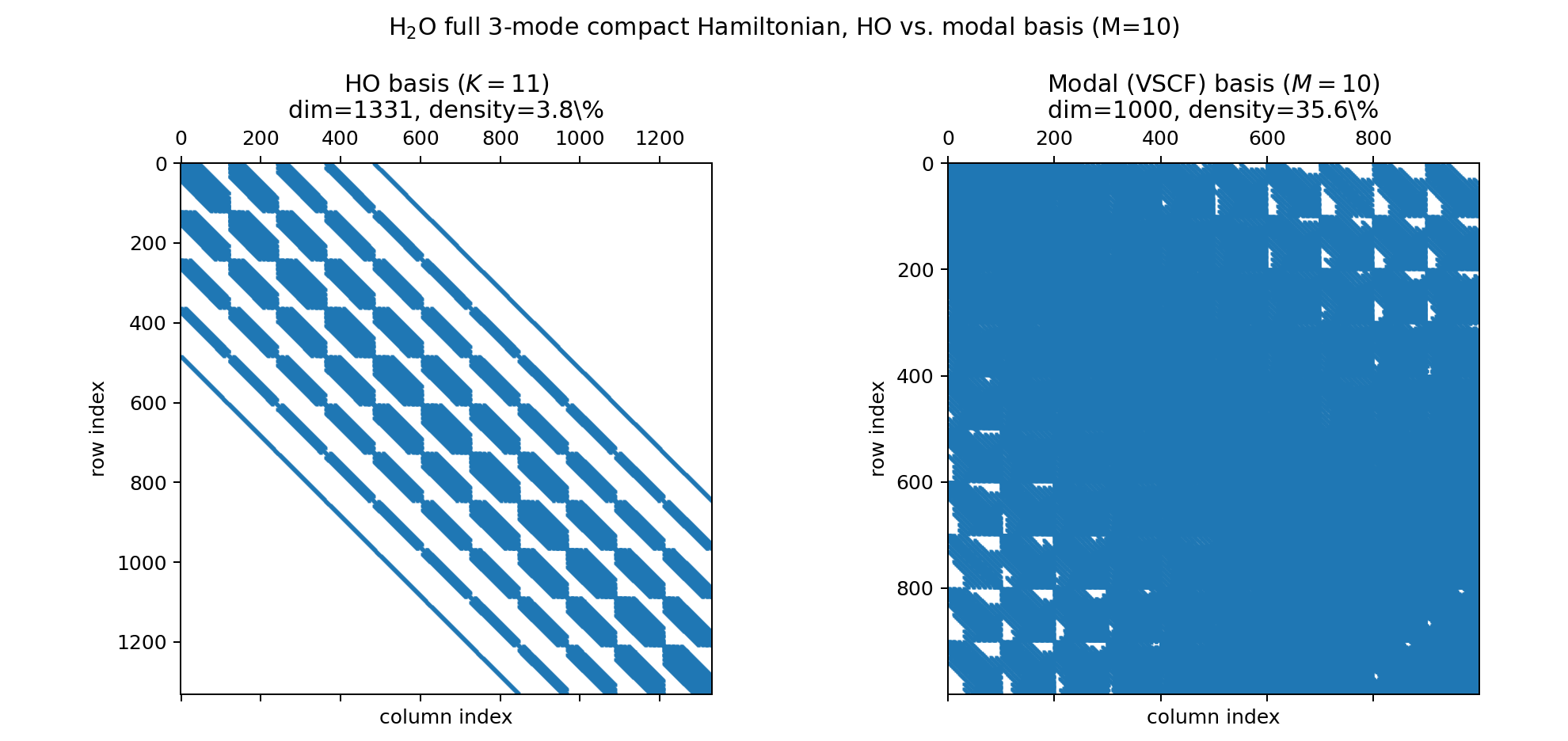}
  \caption{Full 3-mode compact Hamiltonian for H$_2$O, HO basis
  ($K=11$, dim 1331, left) vs.\ modal/VSCF basis ($M=10$, dim
  1000, right); a point is drawn for every matrix element above
  $10^{-10}$. The HO-basis matrix is confined to a small number of
  diagonal bands (density \SI{3.8}{\percent}): each anharmonic term in
  the force field is a bounded-degree polynomial in the normal
  coordinates, so it can only connect HO quanta a few levels apart,
  independent of how many quanta are kept overall. The modal-basis
  matrix (density \SI{35.6}{\percent}, almost ten times denser) has no
  such band structure, since the VSCF rotation $C^{(i)}$
  (Appendix~\ref{app:ho-basis}) mixes every HO quantum number into
  every modal level to varying degrees per mode. The density is
  \emph{not} uniform across the modal-basis matrix -- it is highest in
  the upper-left block (low modal levels for every mode), exactly the
  physically relevant sector this paper's warm-started ansätze and
  self-consistent recovery loop actually operate in
  (Section~\ref{sec:recovery}), and thins out toward the
  rarely-occupied high-modal sector in the lower right.}
  \label{fig:ho-vs-modal}
\end{figure*}

\subsection{Motivation for reporting both harmonic-oscillator and VSCF-modal bases}

\begin{figure*}[t]
  \centering
  \includegraphics[width=0.9\textwidth]{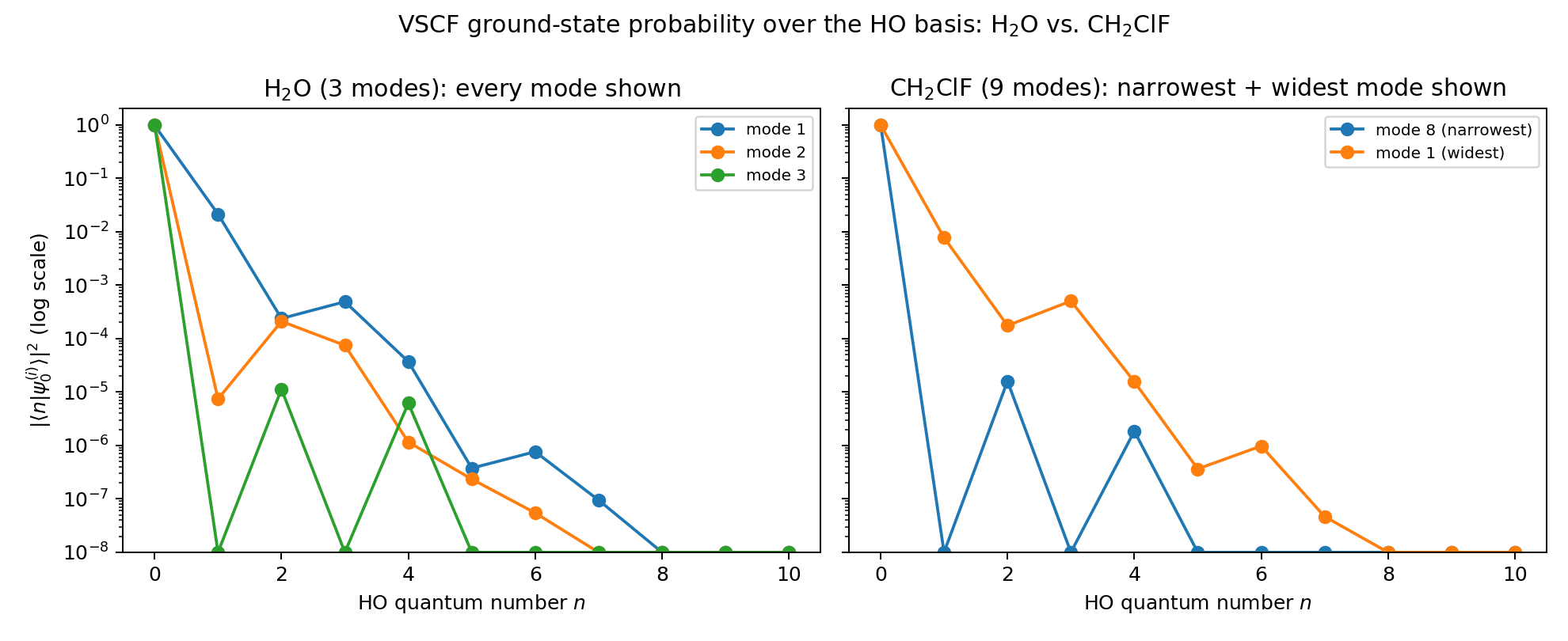}
  \caption{VSCF ground state's probability distribution over the HO
  basis, $|\langle n|\psi_0^{(i)}\rangle|^2 = |C^{(i)}_{0n}|^2$ (log
  scale; H$_2$O left, CH$_2$ClF right, $M=8$ for both). H$_2$O's most-
  and least-mixed modes bracket CH$_2$ClF's: CH$_2$ClF's narrowest mode
  (mode 8) is essentially pure HO level 0 ($|C_{00}|^2>0.9999$),
  matching H$_2$O's own narrowest mode; its widest mode (mode 1)
  spreads out over several quanta with a similar overall shape to
  H$_2$O's own widest mode, despite CH$_2$ClF having three times as
  many modes. The narrowest mode shown for CH$_2$ClF drops to
  essentially zero (below $10^{-10}$) at every odd $n$ -- an even/odd
  selection rule from a potential that is symmetric under $q_i \to
  -q_i$ for this specific mode at the order this force field retains,
  the same symmetry argument behind the even/odd checkerboard pattern
  visible within a single mode's own rotation matrix $C^{(i)}$ in
  Figure~\ref{fig:rotation} ($q_2$'s row 0 mixes only even HO levels).
  The widest mode
  shows no such suppression (odd-$n$ probabilities of order $10^{-3}$
  to $10^{-4}$), i.e.\ a genuine odd-order (cubic) anharmonic term is
  active for that specific mode. The lesson is that \emph{how much}
  a mode mixes is a property of that individual mode's own anharmonicity,
  not of how many other modes the molecule has -- what scales with mode
  count is the total Hilbert space dimension ($K^N$ or $M^N$), not
  the per-mode spread that dimension needs to represent.}
  \label{fig:ho-prob}
\end{figure*}

The motivation is computational rather than conceptual: basis compactness
(i.e., accuracy per retained one-mode function) and operator-assembly cost
are distinct metrics. By construction, the VSCF/modal basis represents each
mode's mean-field ground state with a single function (modal level~0),
whereas an HO expansion generally requires multiple quanta to reproduce the
same one-mode state (Figures~\ref{fig:rotation} and \ref{fig:ho-prob}).
This favors the modal basis for representing low-energy states.

However, the modal transformation mixes HO quanta and typically destroys the
banded sparsity of one-mode operators. As a result, the one-mode operator
factors entering the force-field Hamiltonian become significantly denser in
the modal basis, and the corresponding many-mode Hamiltonian assembled from
their products is more costly to construct and apply at large $(N,M)$.
Accordingly, we report HO-basis results alongside modal-basis results to
separate representational accuracy from practical operator-assembly
scalability. The resulting many-mode density does not follow from a single
universal multiplicative factor, because the full Hamiltonian is a sum
of such product terms across every mode, not one simple Kronecker
product, so density products do not compound by a fixed ratio.
Nevertheless, the modal representation generally loses the narrow band
structure present in the primitive HO basis, increasing the cost of
explicitly constructing the full many-mode matrix. In practice, this
cost renders explicit construction of the full modal-basis Hamiltonian
infeasible beyond the smallest test cases considered here. Accordingly,
our implementation evaluates only the matrix elements required within
the SQD-selected subspace (Section~\ref{sec:recovery}), assembling the
subspace Hamiltonian directly from the force field rather than forming
the full operator.

The concrete dimensions and sparsity/factorization argument for CH$_2$ClF
are given in Section~\ref{sec:basis}.

\subsection{Warm-start amplitude-to-angle mapping, in full}

Section~\ref{sec:warmstart} summarizes the warm-start rule for every
ansatz; the one piece not given there in closed form is the bounded
map used for VIm-uCJ's and Vg-uCJ's $D_2$-type paired-rotation angle,
\begin{equation}
  \theta = \arctan(s\, t_2), \qquad s = 1 \text{ by default},
  \label{eq:t2toangle}
\end{equation}
which -- at this default scale -- agrees with the raw copy to
first order for small amplitudes ($\arctan(t_2)\approx t_2$) while
smoothly bounding $\theta$ to $(-\pi/2,\pi/2)$ so no amplitude, however large, can
overshoot or wrap.

\subsection{Choosing the $P$ pairs}

For VIm-uCJ and Vg-uCJ, only the $P$ largest-$|t_2|$ (mode$_i$,
mode$_j$, exc$_a$, exc$_b$) couplings are kept per layer, ranked by
amplitude magnitude, out of the $\binom{N}{2}(M-1)^2$ possible pairs --
the same amplitude-ranking principle used to select which excitation
UVCCSD would need if it were run in a truncated/selected mode. Every VIm-uCJ/Vg-uCJ result in this
paper -- every molecule, modal count, and basis -- uses the same fixed
$P=8$ per layer; $P$ is not swept or tuned per parameter setting in the
present study.

\section{Self-consistent configuration recovery: full derivation}
\label{app:recovery}

This Appendix gives the full formal derivation of the self-consistent
configuration-recovery procedure summarized in
Section~\ref{sec:recovery}.

\subsection{Average modal occupancy}

Restating Eq.~\eqref{eq:occ-main} of Section~\ref{sec:occ}: given a set
of physical configurations $\{\mathbf v^{(p)}\}_{p=1}^{|\mathcal S|}$
(each $\mathbf v^{(p)} = (v_1^{(p)},\dots,v_N^{(p)})$) with amplitudes
$c_p$ from a diagonalization in the subspace $\mathcal S$
(Section~\ref{sec:subspace}), the average modal occupancy is
\begin{equation}
  \bar n_{i,k} \;=\; \sum_{p=1}^{|\mathcal S|} |c_p|^2\,
    \delta_{v_i^{(p)},\,k}
  \,, \qquad i=1,\dots,N,\;\; k=0,\dots,M-1,
  \label{eq:occ}
\end{equation}
with $\sum_k \bar n_{i,k} = 1$ for a normalized subspace eigenvector.
Before any diagonalization is performed, we initialize the iteration-0
prior as a uniform distribution,
\begin{equation}
  \bar n^{(0)}_{i,k} = \frac{1}{M} \quad \text{for all } i,k,
  \label{eq:occ0}
\end{equation}
which provides an intentionally unbiased starting point. Although the
target reference configuration (Section~\ref{sec:mainconfig}) is known a
priori, we do not bias the prior toward it, ensuring that the first
repair step is guided by the measured samples rather than trivially
reproducing the reference by construction.

\subsection{Repairing a single measurement outcome}

For each mode block $\beta_i$ of a raw (possibly non-physical)
bitstring, let $\mathrm{ones}(\beta_i) = \{k : \beta_{i,k}=1\}$ (every
mode block contains exactly $M$ qubits, so no additional range
condition on $k$ is needed). If the block is already physical
($|\mathrm{ones}(\beta_i)|=1$, i.e.\ $\mathrm{ones}(\beta_i)=\{k^\star\}$
for some $k^\star$) it is kept unchanged. Otherwise---whether no qubit
is set or multiple qubits are set---the block is repaired by resampling
from the current occupancy prior $\bar n_{i,k}$. The resulting repair
rule for mode $i$ is the conditional kernel
\begin{equation}
  r_i(k\mid\beta_i)=
  \begin{cases}
    \delta_{k,k^\star}, & \mathrm{ones}(\beta_i)=\{k^\star\},\\[4pt]
    \bar n_{i,k}, & \text{otherwise.}
  \end{cases}
  \label{eq:repair}
\end{equation}
Thus $v_i=k^\star$ deterministically in the physical case, whereas in
all non-physical cases $v_i$ is drawn once from the discrete
distribution over $\{0,\dots,M-1\}$ with probabilities
$(\bar n_{i,0},\dots,\bar n_{i,M-1})$. Each broken block is resampled
independently. Assuming mode blocks are repaired independently, the repair kernel for a full bitstring $b$ is
\begin{equation}
  R_{\bar n}(b'\mid b) = \prod_{i=1}^N r_i(v_i'\mid\beta_i).
\end{equation}
Because repair is stochastic rather than deterministic, the
distribution induced by an entire raw counts dictionary
$\{b_j\mapsto w_j\}$ is
\begin{equation}
  P_{\text{rec}}(b') \;=\; \frac{1}{Z}
  \sum_j w_j\, R_{\bar n}(b'\mid b_j) , \qquad
  Z = \sum_j w_j .
  \label{eq:recdist}
\end{equation}
In the numerical implementation, we estimate this expectation by
performing exactly one stochastic repair for each \emph{distinct} raw
bitstring key in the counts dictionary (i.e., one draw per key, not one
per shot), and by assigning that key's full shot weight to the single
repaired outcome obtained. We do not explicitly enumerate all possible
repair outcomes implied by the underlying discrete prior.

\subsection{Subspace Hamiltonian construction}

Every distinct configuration with nonzero weight in the realized
repaired distribution is included, together with the guaranteed
reference configuration (Section~\ref{sec:mainconfig}) whether or not
it was itself observed,
\begin{equation}
  \mathcal S^{(t)} = \operatorname{supp} P_{\text{rec}}^{(t)}
  \cup \{\mathbf v_{\rm ref}\}.
\end{equation}
For an excited-state calculation, $\mathbf v_{\rm ref}$ is the
guaranteed excited reference of Section~\ref{sec:excited}; the ground
reference is not separately guaranteed in that case. A subspace-size
cap -- retaining only the $N_{\rm cap}$ highest-weight repaired
configurations, discarding the rest before diagonalization -- bounds
the cost of the eigenvalue problem below as the repaired distribution
grows, at the risk of excluding configurations that carry real but
smaller weight. Rather than fix $N_{\rm cap}$ once, the production
calculations reported in this paper diagonalize at a schedule of
increasing subspace sizes (200 up to 8000 configurations, or the full
repaired support if smaller) and report the value at which the energy
has converged: for the ground state, whose energy is variational, this
is the global minimum over the schedule; for an excited state, it is
the point at which successive schedule steps agree to within
\SI{0.1}{\cmi} (Section~\ref{sec:results}). The repaired support itself
is never truncated before this schedule is applied -- every distinct
repaired configuration is eligible to enter $\mathcal S^{(t)}$ at some
subspace size in the schedule -- but for CH$_2$ClF's largest parameter
settings, whose repaired distributions can exceed 27{,}000
configurations, the 8000-configuration schedule ceiling is reached
before the energy has necessarily stopped changing at the level of
\SI{0.1}{\cmi}: this occurs for 19 of 240 ground-state and 1 of 192
excited-state (molecule, modals, basis, ansatz, backend) settings, all
CH$_2$ClF, where the reported energy is therefore the best value
found within the schedule rather than a value with a directly verified
plateau. Table~\ref{tab:configcounts} shows no cap near 8000 because it
reports only $M{=}4$, where every repaired distribution's support is
far smaller than the schedule's ceiling (at most 1613 configurations,
CH$_2$O/UVCCSD); the 8000 ceiling becomes relevant only at larger $M$,
for CH$_2$ClF's largest settings as stated above. For every other
parameter setting, $\mathcal S^{(t)}$ above
is exactly the full support of the repaired distribution, and
the vibrational Hamiltonian is restricted to this subspace,
\begin{equation}
  H^{\mathcal S}_{pq} \;=\; \langle \bm{v}^{(p)} | \hat H |
  \bm{v}^{(q)} \rangle , \qquad p,q = 1,\dots,|\mathcal S^{(t)}| ,
  \label{eq:hsub}
\end{equation}
never forming the full $M^N$-dimensional operator. $\hat H$ is the
\textcolor{cyan}{quartic force field Hamiltonian defined in Eq.~\eqref{eq:mop};} in the VSCF-modal basis, each
corresponding one-mode operator (harmonic number operator, or a
coordinate power $\hat q^{\,p}$ from an anharmonic term) is rotated
from the primitive HO basis via the per-mode rotation $C^{(i)}$ of
Appendix~\ref{app:ho-basis},
\begin{equation}
  \hat q^{\,p}_{i,\text{modal}} \;=\; C^{(i)}\, \hat q^{\,p}\, C^{(i)\dagger} ,
  \qquad
  \hat n_{i,\text{modal}} \;=\; C^{(i)}\, \hat n\, C^{(i)\dagger} ,
  \label{eq:modalop}
\end{equation}
whereas in the primitive HO basis the corresponding one-mode matrices
are evaluated directly, with no such rotation. $H^{\mathcal S}_{pq}$
is assembled term-by-term from Eq.~\eqref{eq:modalop} (modal basis) or
the direct HO one-mode matrices (HO basis), restricted to the sampled
configurations. Diagonalizing,
\begin{equation}
  H^{\mathcal S} \mathbf c \;=\; E\, \mathbf c ,
  \label{eq:diag}
\end{equation}
gives the current best energy estimate $E$ and a Ritz vector
$\mathbf c$. For a ground-state calculation, $\mathbf c^{(t)}$ is the
lowest eigenvector of $H^{\mathcal S^{(t)}}$. For an excited-state
calculation, $\mathbf c^{(t)}$ is instead the Ritz vector selected by
the state-tracking criterion of Section~\ref{sec:excited} -- maximum
overlap with the guaranteed excited reference -- which need not be the
lowest eigenvector once the sampled subspace contains a lower-energy
configuration. The selected vector then feeds back into
Eq.~\eqref{eq:occ} to produce the next iteration's occupancy prior
$\bar n^{(t+1)}$.

\subsection{The self-consistent loop}

\begin{equation}
\begin{aligned}
  \bar n^{(0)} \;&\xrightarrow{\text{Eq.~\eqref{eq:repair}}}\;
  P_{\text{rec}}^{(0)} \;\xrightarrow{\text{support}\,\cup\,\{\mathbf v_{\rm ref}\}}\;
  \mathcal S^{(0)} \\
  &\xrightarrow{\text{Eq.~\eqref{eq:hsub}--\eqref{eq:diag}}}\;
  E^{(0)}, \mathbf c^{(0)} \;\xrightarrow{\text{Eq.~\eqref{eq:occ}}}\;
  \bar n^{(1)} \;\to\; \cdots
\end{aligned}
  \label{eq:loop}
\end{equation}
This loop is iterated until the energy change satisfies
$\lvert E^{(t)}-E^{(t-1)}\rvert < \SI{0.1}{\cmi}$ for one consecutive
update, with a hard cap of 30 iterations. For reproducibility, the
stochastic repair draws (Eq.~\eqref{eq:repair}) use a fixed random seed
for each (molecule, modals, basis, ansatz) parameter setting. If the
stopping criterion is not met within the iteration cap, we report
$E^{(t)}$ from the final iteration.

This procedure is a fixed-point iteration: the occupancy estimate
$\bar n^{(t)}$ inferred from iteration $t$ determines the repair kernel
used to generate the repaired sample distribution at iteration $t{+}1$.
All ground-state parameter settings reported in this paper
(Appendix~\ref{app:fulltables-artifacts}) satisfy the convergence
criterion within 2--11 iterations; no max-iteration termination was
observed. Iteration counts were not stored separately for the
excited-state calculations.

\subsection{Logical relabeling for excited-state references}

All ansatz constructions and amplitude-indexing conventions used in
this work assume that, for each mode $i$, the reference occupation is
stored at modal index $0$. To treat an excited reference in which mode
$i$ is occupied in modal $r_i\neq 0$ (Section~\ref{sec:excited}), we do
not modify the circuits or permute physical qubits. Instead, we apply a
per-mode \emph{logical relabeling} that reinterprets indices.

For mode $i$ with reference modal $r_i$, we define a mapping from
logical indices to absolute modal indices,
\begin{equation}
\begin{aligned}
  \pi_i(0) &= r_i, \\
  \pi_i(1),\dots,\pi_i(M-1) &= \operatorname{sorted}\bigl(\{0,\dots,M-1\}\setminus\{r_i\}\bigr).
\end{aligned}
  \label{eq:logicalperm}
\end{equation}
so that logical index $0$ always denotes ``the chosen reference modal''
for that mode, and logical indices $1,\dots,M-1$ enumerate the
remaining modals in increasing absolute order.

This relabeling is required only in (i) classical amplitude
bookkeeping, to associate each excitation label with the correct
absolute modal, and (ii) Hamiltonian matrix-element evaluation for
recovered configurations, where decoded logical occupations are mapped
back to absolute occupations via $\pi_i$ prior to indexing. When
$r_i=0$, $\pi_i$ is the identity and the ground-state workflow is
recovered exactly. Importantly, the quantum circuit is unchanged: an
excited reference is incorporated purely through classical index
management, with no additional swap network or circuit modification.

\section{Complete per-parameter-setting results tables}
\label{app:fulltables}

This appendix reports every (molecule, modals, basis,
ansatz) parameter setting in this paper's IBM-hardware measurement
study, on both IBM \code{ibm\_kingston} and \code{ibm\_phoenix}
backends -- Section~\ref{sec:results}'s main-text figures show the
aggregate picture (which basis, which ansatz, how the answer evolves
with modal count); this appendix gives a dedicated pair of tables
reporting CZ count and circuit depth for the ground state
(Tables~\ref{tab:czdepthmodal}--\ref{tab:czdepthho}), the additional
ground- and excited-state comparison figures of
Section~\ref{app:fulltables-ground-others} for the three ansätze not
shown in the main-text Figures~\ref{fig:energyvsMground} and
\ref{fig:energyvsMexcited}, and pointers
to the complete per-parameter-setting numbers, which are available as
interactive online ledgers rather than printed tables
(Section~\ref{app:fulltables-artifacts}).

\subsection{Circuit preparation and transpilation, in full}
\label{app:transpilation}

All circuits in this study are compiled and executed using
Qiskit\cite{Qiskit2024} and Qiskit Runtime. To enable methodological
reproducibility, we report the transpiler settings and the
error-suppression configuration used for all runs.

Every CZ count and depth number in this appendix's
tables -- and in Figures~\ref{fig:beforeafterbarsmodal}--\ref{fig:beforeafterbarsho}'s
underlying data -- is measured directly on the actual circuit submitted
to \code{ibm\_kingston} and \code{ibm\_phoenix}, rather than from a
closed-form estimate: both the two-qubit (CZ) gate count and the
circuit depth are read from the transpiled
circuit immediately after transpilation, before any hardware-execution
step. Neither number includes any gates the dynamical-decoupling
pass inserts at execution time (Section~\ref{sec:suppression}), since
that pass acts on idle-time scheduling at execution rather than
modifying the transpiled circuit itself, so both are reported here as
gate-layer counts of the logical/routed circuit, not a
scheduled-duration or hardware-execution-time metric.
Each logical
ansatz circuit (Section~\ref{sec:ansatze}) is bound to its warm-start
parameters, then transpiled immediately before submission, once per
(molecule, modals, basis, ansatz, backend) setting, at the compiler's
highest optimization level and with no forced initial qubit layout. No
transpiler seed is fixed, so a repeated transpilation of the same
logical circuit could in principle return a different CZ count or
depth than the one reported here; every number in this paper's tables
is nonetheless the actual value from the one transpilation that
produced the circuit actually submitted and measured, not a resubmission-averaged or re-derived figure.

This same unseeded transpilation, together with the fact that each
ansatz circuit is bound to its warm-start parameters
(Section~\ref{sec:warmstart}) \emph{before} transpilation runs, is why
the modal- and HO-basis versions of an otherwise identical logical
circuit -- same ansatz, same $N$, same $M$, same reference-state
construction (Appendix~\ref{app:ho-basis}) -- do not always transpile
to the same CZ count and depth, even though the two bases share the
same circuit topology. The classical VCC amplitudes used to warm-start
the circuit differ between bases (they are fit to the modal- or
HO-basis Hamiltonian, respectively), so the two bases' circuits carry
different bound rotation angles; the compiler's numeric-aware
optimization passes (commutative-gate cancellation and single-qubit
gate fusion) can therefore find different cancellation opportunities
in the two cases, and the unseeded routing stage can independently
return a different layout for what is otherwise the same interaction
graph. The resulting basis-dependent depth difference is real but
modest and not one-sided: across all 60 matched (molecule, modals,
ansatz) ground-state settings on each backend, the modal-basis circuit
is shallower than its HO-basis counterpart in some of the experimental settings (\SI{56.7}{\percent} on \code{ibm\_kingston},
\SI{53.3}{\percent} on \code{ibm\_phoenix}), with most of the
remainder shallower in HO basis instead (\SI{40.0}{\percent} and
\SI{41.7}{\percent} respectively) and a handful tied; it should not be
read as a structural property of one basis over the other. Leaving the
initial layout unset (rather than pinning logical qubit $i$ to a fixed
physical qubit $i$, or to a hand-chosen straight line of physical
qubits) lets Qiskit's own preset layout stage run: \code{VF2Layout} searches for a subgraph
isomorphism between the circuit's actual two-qubit interaction graph
and the device's coupling graph, falling back to \code{SabreLayout}'s
iterative swap-based heuristic when no such isomorphism exists (the
common case at the qubit counts and connectivity used here); both
stages read the backend's live calibration data,
so the chosen layout is simultaneously \emph{interaction-aware} (it
tries to place logical qubits that actually interact next to each
other on the device) and \emph{error-aware} (it prefers lower-error
physical qubits and couplers when more than one placement is otherwise
equivalent) -- unlike a fixed straight-line layout, which maps logical
qubit $i$ to physical position $i$ purely positionally, with no
knowledge of which logical qubits the circuit actually entangles. This
choice is not a default left unexamined: a direct, controlled
comparison on ansatz circuits from this pipeline
(UVCCSD/VIm-uCJ/CHC, H$_2$O/CH$_2$O at several modal counts, verified
on \code{ibm\_kingston}) found the interaction-aware path giving
\SI{15}{}--\SI{48}{\percent} fewer CZ gates and \SI{15}{}--\SI{48}{\percent}
less depth than the same circuits transpiled onto a fixed, error-scored
straight line of physical qubits -- the CZ counts and depths reported
throughout this paper are therefore already the better of the two
layout strategies tested, not an arbitrary or unoptimized choice.

\subsubsection{Error suppression: dynamical decoupling and Pauli twirling}
\label{sec:suppression}

Every circuit in this work's hardware measurement study
is submitted through Qiskit Runtime with the same two
native suppression options enabled, set identically for every
(molecule, modals, basis, ansatz) parameter setting on both backends: dynamical
decoupling with an XX sequence, which inserts a pair of $X$ gates into
otherwise-idle qubit timeslots during the circuit so that
low-frequency dephasing noise is refocused rather than allowed to
accumulate unchecked while other qubits are still being acted on; and
gate-level Pauli twirling with eight Pauli-conjugated randomizations
per two-qubit gate, which re-randomizes each two-qubit
gate's implementation across those eight variants and averages
the results, converting a coherent, structured two-qubit error into an
incoherent Pauli-channel error without changing the circuit's intended
unitary action -- a standard mitigation for coherent errors, though
twirling does not reduce the total error rate itself and can add
sampling overhead from the extra randomizations. Neither option is tuned per parameter setting -- the same two
settings are used whether the circuit has 6 qubits and 10 CZ gates or
72 qubits and nearly 48{,}000 CZ gates -- so any
parameter-setting-to-parameter-setting difference in reported CZ count,
depth, or one-hot retention is attributable to the circuit and the
transpiler's own layout choice for it (Section~\ref{app:transpilation}),
not to a different error-suppression configuration being used for some
parameter settings and not others.

\paragraph{Cross-backend compilation outcomes.}
Across the 120 matched ground-state settings compiled and executed on
both \code{ibm\_kingston} and \code{ibm\_phoenix} (same warm-start
parameters, layers=1, and suppression settings), the circuits transpiled
for \code{ibm\_phoenix} are systematically cheaper: the median CZ count
is lower by 29.3\% and the median depth is lower by 28.0\%
(Tables~\ref{tab:czdepthmodal} and \ref{tab:czdepthho}). This reduction
is consistent with the higher connectivity of \code{ibm\_phoenix}'s
square-lattice coupling graph relative to \code{ibm\_kingston}'s
heavy-hex, which reduces SWAP-routing overhead for the predominantly
local interaction patterns of the vibrational circuits; however, we do
not log per-circuit SWAP counts, so this mechanism should be viewed as
a plausible explanation rather than a directly measured attribution.
In the same matched set, one-hot retention is higher on
\code{ibm\_phoenix} by a median factor of 5.1$\times$ over the 89
settings where \code{ibm\_kingston} retention is nonzero, and
\code{ibm\_phoenix} frequently retains nonzero signal in cases where
\code{ibm\_kingston} collapses to zero.

\begin{table*}[p]
\centering
\caption{Ground state, modal basis, CZ count and circuit depth for
every (molecule, modals, ansatz) parameter setting, \code{ibm\_phoenix} and \code{ibm\_kingston} hardware, layers=1
(\code{ibm\_phoenix} row listed first in each pair). Both
counts are measured directly on the actual transpiled, submitted
circuit (Section~\ref{app:transpilation}), not a closed-form estimate.
``--'' marks a (molecule, modals, ansatz, backend) setting with no
recorded CZ/depth entry in this paper's saved result files. Left:
H$_2$O and CH$_2$O; right: CH$_2$ClF.}
\label{tab:czdepthmodal}
\scriptsize
\setlength{\tabcolsep}{3pt}
\begin{minipage}[t]{0.48\textwidth}
\centering
\begin{tabular}{@{}llllrr@{}}
\toprule
Mol. & $M$ & Backend & Ansatz & CZ & Depth \\
\midrule
H$_2$O & 2 & Phoenix & UVCCSD & 41 & 103 \\
 &  & Kingston & UVCCSD & 50 & 133 \\
 &  & Phoenix & VLUCJ & 10 & 31 \\
 &  & Kingston & VLUCJ & 10 & 32 \\
 &  & Phoenix & VIm-uCJ & 39 & 84 \\
 &  & Kingston & VIm-uCJ & 40 & 96 \\
 &  & Phoenix & Vg-uCJ & 39 & 84 \\
 &  & Kingston & Vg-uCJ & 40 & 96 \\
 &  & Phoenix & CHC & 10 & 31 \\
 &  & Kingston & CHC & 10 & 32 \\
 & 4 & Phoenix & UVCCSD & 661 & 1475 \\
 &  & Kingston & UVCCSD & 798 & 1902 \\
 &  & Phoenix & VLUCJ & 34 & 66 \\
 &  & Kingston & VLUCJ & 53 & 93 \\
 &  & Phoenix & VIm-uCJ & 239 & 552 \\
 &  & Kingston & VIm-uCJ & 270 & 634 \\
 &  & Phoenix & Vg-uCJ & 244 & 571 \\
 &  & Kingston & Vg-uCJ & 270 & 635 \\
 &  & Phoenix & CHC & 92 & 188 \\
 &  & Kingston & CHC & 124 & 233 \\
 & 6 & Phoenix & UVCCSD & 1934 & 4000 \\
 &  & Kingston & UVCCSD & 2586 & 5473 \\
 &  & Phoenix & VLUCJ & 44 & 78 \\
 &  & Kingston & VLUCJ & 89 & 161 \\
 &  & Phoenix & VIm-uCJ & 240 & 571 \\
 &  & Kingston & VIm-uCJ & 317 & 789 \\
 &  & Phoenix & Vg-uCJ & 239 & 568 \\
 &  & Kingston & Vg-uCJ & 270 & 634 \\
 &  & Phoenix & CHC & 243 & 428 \\
 &  & Kingston & CHC & 330 & 456 \\
 & 8 & Phoenix & UVCCSD & 2679 & 5194 \\
 &  & Kingston & UVCCSD & 4050 & 7969 \\
 &  & Phoenix & VLUCJ & 44 & 78 \\
 &  & Kingston & VLUCJ & 85 & 168 \\
 &  & Phoenix & VIm-uCJ & 249 & 590 \\
 &  & Kingston & VIm-uCJ & 282 & 681 \\
 &  & Phoenix & Vg-uCJ & 244 & 582 \\
 &  & Kingston & Vg-uCJ & 310 & 723 \\
 &  & Phoenix & CHC & 224 & 336 \\
 &  & Kingston & CHC & 311 & 397 \\
\addlinespace
CH$_2$O & 2 & Phoenix & UVCCSD & 206 & 344 \\
 &  & Kingston & UVCCSD & 294 & 533 \\
 &  & Phoenix & VLUCJ & 8 & 31 \\
 &  & Kingston & VLUCJ & 8 & 30 \\
 &  & Phoenix & VIm-uCJ & 139 & 287 \\
 &  & Kingston & VIm-uCJ & 231 & 409 \\
 &  & Phoenix & Vg-uCJ & 136 & 259 \\
 &  & Kingston & Vg-uCJ & 266 & 515 \\
 &  & Phoenix & CHC & 15 & 42 \\
 &  & Kingston & CHC & 18 & 45 \\
 & 4 & Phoenix & UVCCSD & 2943 & 4615 \\
 &  & Kingston & UVCCSD & 3822 & 5604 \\
 &  & Phoenix & VLUCJ & 46 & 83 \\
 &  & Kingston & VLUCJ & 100 & 159 \\
 &  & Phoenix & VIm-uCJ & 224 & 416 \\
 &  & Kingston & VIm-uCJ & 319 & 665 \\
 &  & Phoenix & Vg-uCJ & 216 & 419 \\
 &  & Kingston & Vg-uCJ & 345 & 665 \\
 &  & Phoenix & CHC & 296 & 434 \\
 &  & Kingston & CHC & 386 & 509 \\
 & 6 & Phoenix & UVCCSD & 7708 & 12422 \\
 &  & Kingston & UVCCSD & 10824 & 16648 \\
 &  & Phoenix & VLUCJ & 56 & 87 \\
 &  & Kingston & VLUCJ & 166 & 227 \\
 &  & Phoenix & VIm-uCJ & 230 & 434 \\
 &  & Kingston & VIm-uCJ & 353 & 816 \\
 &  & Phoenix & Vg-uCJ & 231 & 420 \\
 &  & Kingston & Vg-uCJ & 355 & 789 \\
 &  & Phoenix & CHC & 652 & 804 \\
 &  & Kingston & CHC & 1009 & 886 \\
 & 8 & Phoenix & UVCCSD & 12341 & 16354 \\
 &  & Kingston & UVCCSD & 16948 & 23656 \\
 &  & Phoenix & VLUCJ & 56 & 87 \\
 &  & Kingston & VLUCJ & 139 & 157 \\
 &  & Phoenix & VIm-uCJ & 229 & 422 \\
 &  & Kingston & VIm-uCJ & 319 & 637 \\
 &  & Phoenix & Vg-uCJ & 234 & 404 \\
 &  & Kingston & Vg-uCJ & 334 & 652 \\
 &  & Phoenix & CHC & 695 & 850 \\
 &  & Kingston & CHC & 967 & 933 \\
\bottomrule
\end{tabular}
\end{minipage}%
\hfill
\begin{minipage}[t]{0.48\textwidth}
\centering
\begin{tabular}{@{}llllrr@{}}
\toprule
Mol. & $M$ & Backend & Ansatz & CZ & Depth \\
\midrule
CH$_2$ClF & 2 & Phoenix & UVCCSD & 794 & 1339 \\
 &  & Kingston & UVCCSD & 1206 & 2117 \\
 &  & Phoenix & VLUCJ & 32 & 66 \\
 &  & Kingston & VLUCJ & 41 & 94 \\
 &  & Phoenix & VIm-uCJ & 234 & 494 \\
 &  & Kingston & VIm-uCJ & 320 & 571 \\
 &  & Phoenix & Vg-uCJ & 222 & 492 \\
 &  & Kingston & Vg-uCJ & 343 & 692 \\
 &  & Phoenix & CHC & 84 & 199 \\
 &  & Kingston & CHC & 103 & 153 \\
 & 4 & Phoenix & UVCCSD & 8554 & 11470 \\
 &  & Kingston & UVCCSD & 11196 & 16856 \\
 &  & Phoenix & VLUCJ & 86 & 91 \\
 &  & Kingston & VLUCJ & 220 & 225 \\
 &  & Phoenix & VIm-uCJ & 230 & 405 \\
 &  & Kingston & VIm-uCJ & 337 & 650 \\
 &  & Phoenix & Vg-uCJ & 228 & 410 \\
 &  & Kingston & Vg-uCJ & 348 & 705 \\
 &  & Phoenix & CHC & 874 & 937 \\
 &  & Kingston & CHC & 1275 & 884 \\
 & 6 & Phoenix & UVCCSD & 19207 & 28387 \\
 &  & Kingston & UVCCSD & 26878 & 37695 \\
 &  & Phoenix & VLUCJ & 107 & 87 \\
 &  & Kingston & VLUCJ & 312 & 261 \\
 &  & Phoenix & VIm-uCJ & 229 & 421 \\
 &  & Kingston & VIm-uCJ & 323 & 649 \\
 &  & Phoenix & Vg-uCJ & 234 & 439 \\
 &  & Kingston & Vg-uCJ & 351 & 571 \\
 &  & Phoenix & CHC & 1925 & 1635 \\
 &  & Kingston & CHC & 2798 & 1929 \\
 & 8 & Phoenix & UVCCSD & 30264 & 40956 \\
 &  & Kingston & UVCCSD & 44331 & 59937 \\
 &  & Phoenix & VLUCJ & 106 & 87 \\
 &  & Kingston & VLUCJ & 347 & 204 \\
 &  & Phoenix & VIm-uCJ & 227 & 401 \\
 &  & Kingston & VIm-uCJ & 346 & 657 \\
 &  & Phoenix & Vg-uCJ & 219 & 420 \\
 &  & Kingston & Vg-uCJ & 310 & 612 \\
 &  & Phoenix & CHC & 1943 & 1533 \\
 &  & Kingston & CHC & 2871 & 1951 \\
\bottomrule
\end{tabular}
\end{minipage}
\end{table*}

\begin{table*}[p]
\centering
\caption{Ground state, HO basis, CZ count and circuit depth for every
(molecule, modals, ansatz) parameter setting, \code{ibm\_phoenix}
and code{ibm\_kingston} hardware, layers=1 (\code{ibm\_phoenix}
row listed first in each pair, matching Table~\ref{tab:czdepthmodal}'s
own convention). Both
counts are measured directly on the actual transpiled, submitted
circuit (Section~\ref{app:transpilation}), not a closed-form estimate.
``--'' marks a (molecule, modals, ansatz, backend) setting with no
recorded CZ/depth entry in this paper's saved result files. Left:
H$_2$O and CH$_2$O; right: CH$_2$ClF.}
\label{tab:czdepthho}
\scriptsize
\setlength{\tabcolsep}{3pt}
\begin{minipage}[t]{0.48\textwidth}
\centering
\begin{tabular}{@{}llllrr@{}}
\toprule
Mol. & $M$ & Backend & Ansatz & CZ & Depth \\
\midrule
H$_2$O & 2 & Phoenix & UVCCSD & 44 & 120 \\
 &  & Kingston & UVCCSD & 50 & 135 \\
 &  & Phoenix & VLUCJ & 10 & 33 \\
 &  & Kingston & VLUCJ & 10 & 31 \\
 &  & Phoenix & VIm-uCJ & 35 & 80 \\
 &  & Kingston & VIm-uCJ & 40 & 95 \\
 &  & Phoenix & Vg-uCJ & 35 & 88 \\
 &  & Kingston & Vg-uCJ & 40 & 95 \\
 &  & Phoenix & CHC & 10 & 33 \\
 &  & Kingston & CHC & 10 & 32 \\
 & 4 & Phoenix & UVCCSD & 613 & 1418 \\
 &  & Kingston & UVCCSD & 757 & 1705 \\
 &  & Phoenix & VLUCJ & 34 & 68 \\
 &  & Kingston & VLUCJ & 58 & 154 \\
 &  & Phoenix & VIm-uCJ & 238 & 585 \\
 &  & Kingston & VIm-uCJ & 355 & 888 \\
 &  & Phoenix & Vg-uCJ & 236 & 559 \\
 &  & Kingston & Vg-uCJ & 317 & 773 \\
 &  & Phoenix & CHC & 92 & 194 \\
 &  & Kingston & CHC & 116 & 180 \\
 & 6 & Phoenix & UVCCSD & 1778 & 3834 \\
 &  & Kingston & UVCCSD & 2468 & 5250 \\
 &  & Phoenix & VLUCJ & 52 & 92 \\
 &  & Kingston & VLUCJ & 98 & 152 \\
 &  & Phoenix & VIm-uCJ & 237 & 565 \\
 &  & Kingston & VIm-uCJ & 329 & 782 \\
 &  & Phoenix & Vg-uCJ & 242 & 578 \\
 &  & Kingston & Vg-uCJ & 338 & 826 \\
 &  & Phoenix & CHC & 240 & 443 \\
 &  & Kingston & CHC & 341 & 476 \\
 & 8 & Phoenix & UVCCSD & 3053 & 6211 \\
 &  & Kingston & UVCCSD & 4202 & 8638 \\
 &  & Phoenix & VLUCJ & 60 & 100 \\
 &  & Kingston & VLUCJ & 111 & 195 \\
 &  & Phoenix & VIm-uCJ & 246 & 587 \\
 &  & Kingston & VIm-uCJ & 348 & 865 \\
 &  & Phoenix & Vg-uCJ & 252 & 597 \\
 &  & Kingston & Vg-uCJ & 317 & 772 \\
 &  & Phoenix & CHC & 378 & 512 \\
 &  & Kingston & CHC & 556 & 679 \\
\addlinespace
CH$_2$O & 2 & Phoenix & UVCCSD & 181 & 370 \\
 &  & Kingston & UVCCSD & 380 & 719 \\
 &  & Phoenix & VLUCJ & 8 & 31 \\
 &  & Kingston & VLUCJ & 8 & 28 \\
 &  & Phoenix & VIm-uCJ & 138 & 241 \\
 &  & Kingston & VIm-uCJ & 172 & 400 \\
 &  & Phoenix & Vg-uCJ & 135 & 302 \\
 &  & Kingston & Vg-uCJ & 174 & 398 \\
 &  & Phoenix & CHC & 15 & 44 \\
 &  & Kingston & CHC & 16 & 49 \\
 & 4 & Phoenix & UVCCSD & 3082 & 5628 \\
 &  & Kingston & UVCCSD & 4086 & 7016 \\
 &  & Phoenix & VLUCJ & 46 & 79 \\
 &  & Kingston & VLUCJ & 103 & 122 \\
 &  & Phoenix & VIm-uCJ & 236 & 410 \\
 &  & Kingston & VIm-uCJ & 320 & 621 \\
 &  & Phoenix & Vg-uCJ & 219 & 412 \\
 &  & Kingston & Vg-uCJ & 386 & 717 \\
 &  & Phoenix & CHC & 293 & 413 \\
 &  & Kingston & CHC & 414 & 503 \\
 & 6 & Phoenix & UVCCSD & 7667 & 11736 \\
 &  & Kingston & UVCCSD & 10716 & 17260 \\
 &  & Phoenix & VLUCJ & 65 & 90 \\
 &  & Kingston & VLUCJ & 137 & 162 \\
 &  & Phoenix & VIm-uCJ & 226 & 421 \\
 &  & Kingston & VIm-uCJ & 336 & 643 \\
 &  & Phoenix & Vg-uCJ & 234 & 414 \\
 &  & Kingston & Vg-uCJ & 333 & 656 \\
 &  & Phoenix & CHC & 736 & 648 \\
 &  & Kingston & CHC & 1107 & 984 \\
 & 8 & Phoenix & UVCCSD & 13052 & 18143 \\
 &  & Kingston & UVCCSD & 18766 & 24728 \\
 &  & Phoenix & VLUCJ & 67 & 92 \\
 &  & Kingston & VLUCJ & 178 & 158 \\
 &  & Phoenix & VIm-uCJ & 229 & 427 \\
 &  & Kingston & VIm-uCJ & 332 & 601 \\
 &  & Phoenix & Vg-uCJ & 225 & 416 \\
 &  & Kingston & Vg-uCJ & 325 & 668 \\
 &  & Phoenix & CHC & 1070 & 998 \\
 &  & Kingston & CHC & 1550 & 1282 \\
\bottomrule
\end{tabular}
\end{minipage}%
\hfill
\begin{minipage}[t]{0.48\textwidth}
\centering
\begin{tabular}{@{}llllrr@{}}
\toprule
Mol. & $M$ & Backend & Ansatz & CZ & Depth \\
\midrule
CH$_2$ClF & 2 & Phoenix & UVCCSD & 737 & 1279 \\
 &  & Kingston & UVCCSD & 1197 & 2398 \\
 &  & Phoenix & VLUCJ & 32 & 66 \\
 &  & Kingston & VLUCJ & 41 & 88 \\
 &  & Phoenix & VIm-uCJ & 228 & 410 \\
 &  & Kingston & VIm-uCJ & 372 & 659 \\
 &  & Phoenix & Vg-uCJ & 218 & 482 \\
 &  & Kingston & Vg-uCJ & 331 & 656 \\
 &  & Phoenix & CHC & 81 & 187 \\
 &  & Kingston & CHC & 109 & 164 \\
 & 4 & Phoenix & UVCCSD & 8744 & 12478 \\
 &  & Kingston & UVCCSD & 11850 & 18778 \\
 &  & Phoenix & VLUCJ & 88 & 85 \\
 &  & Kingston & VLUCJ & 253 & 267 \\
 &  & Phoenix & VIm-uCJ & 232 & 491 \\
 &  & Kingston & VIm-uCJ & 293 & 518 \\
 &  & Phoenix & Vg-uCJ & 232 & 443 \\
 &  & Kingston & Vg-uCJ & 340 & 595 \\
 &  & Phoenix & CHC & 894 & 836 \\
 &  & Kingston & CHC & 1363 & 978 \\
 & 6 & Phoenix & UVCCSD & 20666 & 28608 \\
 &  & Kingston & UVCCSD & 28322 & 41093 \\
 &  & Phoenix & VLUCJ & 102 & 77 \\
 &  & Kingston & VLUCJ & 329 & 242 \\
 &  & Phoenix & VIm-uCJ & 228 & 425 \\
 &  & Kingston & VIm-uCJ & 318 & 733 \\
 &  & Phoenix & Vg-uCJ & 230 & 420 \\
 &  & Kingston & Vg-uCJ & 324 & 574 \\
 &  & Phoenix & CHC & 2134 & 1629 \\
 &  & Kingston & CHC & 3351 & 2352 \\
 & 8 & Phoenix & UVCCSD & 33809 & 45248 \\
 &  & Kingston & UVCCSD & 47992 & 62782 \\
 &  & Phoenix & VLUCJ & 107 & 76 \\
 &  & Kingston & VLUCJ & 386 & 204 \\
 &  & Phoenix & VIm-uCJ & 228 & 405 \\
 &  & Kingston & VIm-uCJ & 337 & 675 \\
 &  & Phoenix & Vg-uCJ & 233 & 420 \\
 &  & Kingston & Vg-uCJ & 313 & 573 \\
 &  & Phoenix & CHC & 3162 & 2467 \\
 &  & Kingston & CHC & 4849 & 2855 \\
\bottomrule
\end{tabular}
\end{minipage}
\end{table*}

\subsection{Backend calibration data}
\label{app:calibration}

Table~\ref{tab:calibration} (Section~\ref{app:calibration})
reports each backend's own live calibration snapshot, queried directly through
\code{BackendProperties} on two consecutive days
(2026-09-09/2026-09-10) rather than a nominal specification figure --
median across all qubits (T1, T2, readout error, single-qubit gate
error) or all couplers (two-qubit gate error) on each backend on each
day. \code{ibm\_phoenix}'s lower CZ error is consistent with this
paper's own retention/CZ-count results (Section~\ref{app:transpilation}):
a device with both a lower per-gate error rate and a circuit that
needs fewer of those gates compounds into the large retention gains
measured above. Day-to-day drift within a backend (e.g.
\code{ibm\_kingston}'s median T1 rising from 214.8 to 250.8\,\textmu s)
is visible and expected of superconducting hardware -- recalibrated
routinely -- and is reported here for transparency about which
snapshot this paper's hardware runs were executed against, not as a
claim that either backend's calibration is static.

\begin{table}[t]
\captionsetup{justification=raggedright,singlelinecheck=false}
\raggedright
\caption{Calibration snapshots, queried directly from each
backend's \code{BackendProperties} (not a nominal specification).
Median across all qubits (T1, T2, readout, 1Q gate error) or all
couplers (2Q gate error). 1Q gate is \code{sx}; 2Q gate is \code{cz}
on both backends.}
\label{tab:calibration}
\footnotesize
\setlength{\tabcolsep}{3pt}
\hspace*{-0.1em}
\begin{tabular}{@{}llrrrrr@{}}
\toprule
Backend & Date & \multicolumn{2}{c}{Coherence ($\mu$s)} & \multicolumn{3}{c}{Median error (\%)} \\
\cmidrule(lr){3-4}\cmidrule(lr){5-7}
 &  & $T_1$ & $T_2$ & Readout & 1Q & 2Q \\
\midrule
Kingston & 2026-09-09 & 214.78 & 143.21 & 0.928 & 0.0258 & 0.2025 \\
Kingston & 2026-09-10 & 250.83 & 149.85 & 0.891 & 0.0259 & 0.2077 \\
Phoenix  & 2026-09-09 & 172.87 & 128.46 & 0.793 & 0.0245 & 0.1968 \\
Phoenix  & 2026-09-10 & 168.99 & 135.53 & 0.757 & 0.0256 & 0.1968 \\
\bottomrule
\end{tabular}
\end{table}

\subsection{Before/after recovery,\code{ibm\_kingston} hardware}
\label{app:beforeafterkingston}

\textcolor{deepred}{Section~\ref{sec:results}'s
Figures~\ref{fig:beforeafterbarsmodal}--\ref{fig:beforeafterbarsho}
report this same before/after-recovery comparison on \code{ibm\_phoenix} hardware. The original \code{ibm\_kingston}
version of both figures, superseded in the main text but not
otherwise reported anywhere else in this paper, is given here in full
for completeness.}

\begin{figure*}[t]
  \centering
  \begin{subfigure}[b]{\textwidth}
    \centering
    \includegraphics[width=\textwidth]{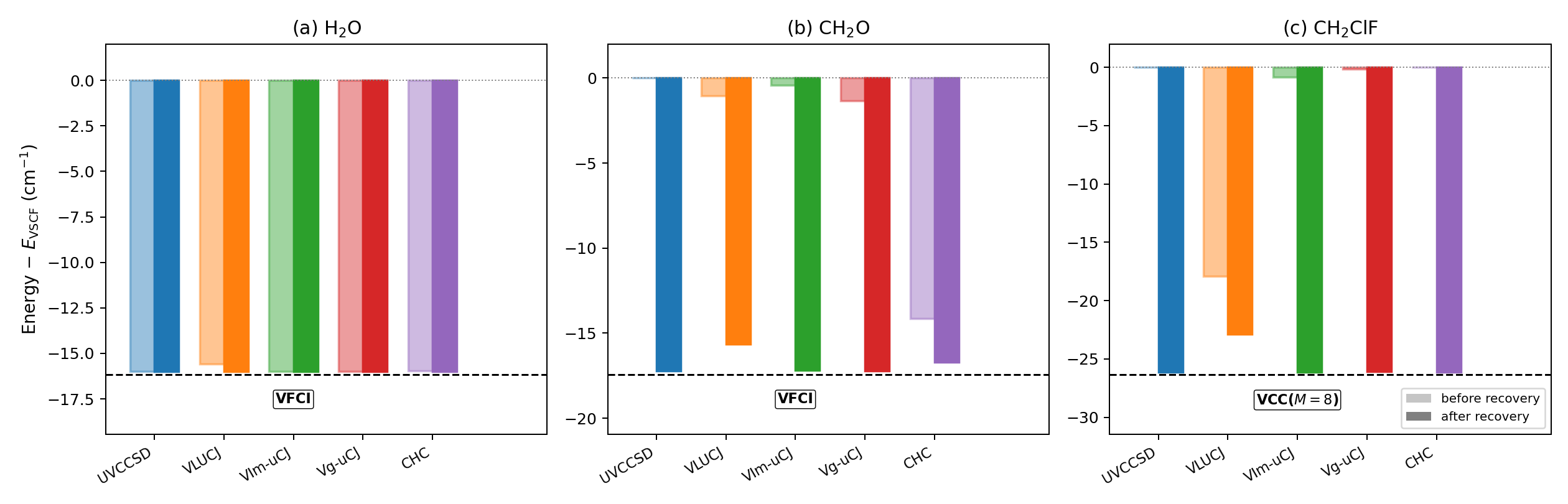}
    \caption{Modal basis. Modals=4 is chosen over a larger modal count
    specifically because pre-recovery signal is still visible here
    for multiple ansätze -- at $M{=}8$, one-hot retention has already
    collapsed far enough that every ``before'' bar is close to zero
    (Figure~\ref{fig:onehotcliff}), which would make this figure's
    actual point (recovery visibly working, not just returning
    the reference configuration) invisible.}
    \label{fig:beforeafterbarsmodalkingston}
  \end{subfigure}
  \\[1em]
  \begin{subfigure}[b]{\textwidth}
    \centering
    \includegraphics[width=\textwidth]{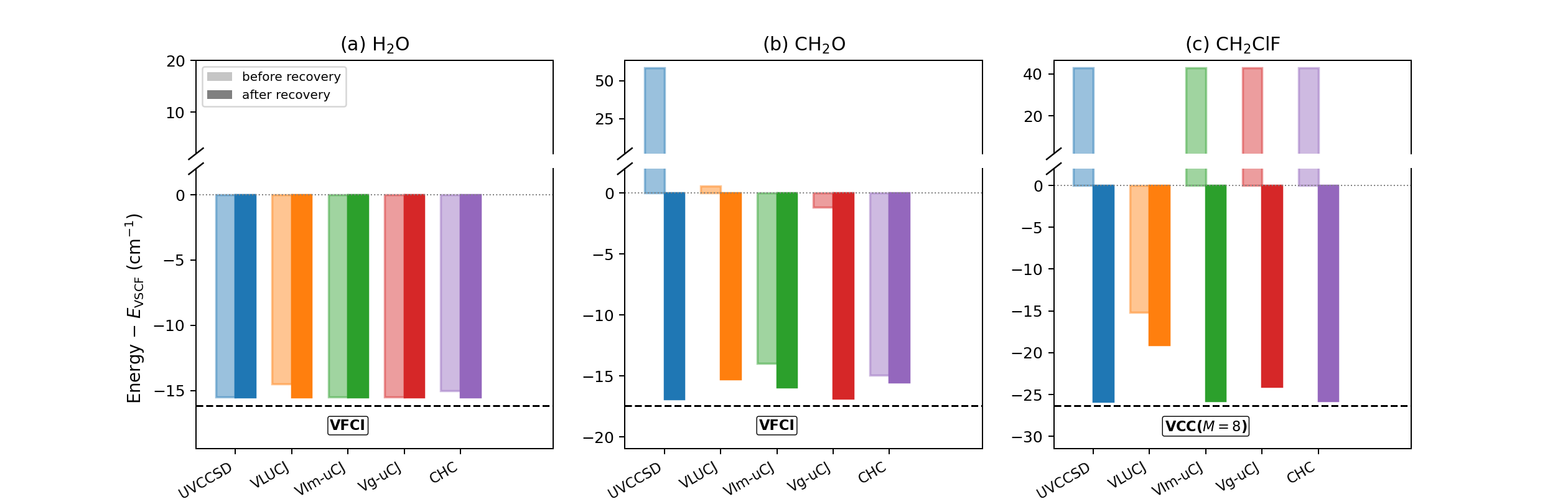}
    \caption{HO basis. Contrast with the modal-basis panel above: HO's
    ``before'' bars are far from converged (up to
    $+58.6\,\mathrm{cm}^{-1}$ for some ansätze/molecules) at the same
    modal count and qubit count -- modal basis starts the
    self-consistent loop from a much better raw hardware sample than
    HO basis does, at every molecule tested on \code{ibm\_kingston}. On
    \code{ibm\_phoenix} this pattern is less uniform
    (Figure~\ref{fig:beforeafterbarsho}'s caption).}
    \label{fig:beforeafterbarshokingston}
  \end{subfigure}
  \caption{\code{IBM\_Kingston} hardware, layers=1, energy relative
  to the VSCF reference ($E-E_{\rm VSCF}$) for each of the five
  ansätze, $M{=}4$ (12/24/36 qubits for H$_2$O/CH$_2$O/CH$_2$ClF
  respectively) -- faded bars are before self-consistent recovery (an
  explicit subspace diagonalization over whatever the raw sample
  actually contained), filled bars are after. The dashed line (labeled
  in-panel) is the converged classical target (VFCI for H$_2$O/CH$_2$O;
  VCC($M{=}8$) for CH$_2$ClF, whose Hamiltonian is too large for VFCI,
  Table~\ref{tab:classicalrefs}'s footnote).}
  \label{fig:beforeafterbarskingstonboth}
\end{figure*}

\subsection{Ground and excited state, remaining ansätze}
\label{app:fulltables-ground-others}

\begin{figure*}[!t]
  \centering
  \includegraphics[width=0.97\textwidth]{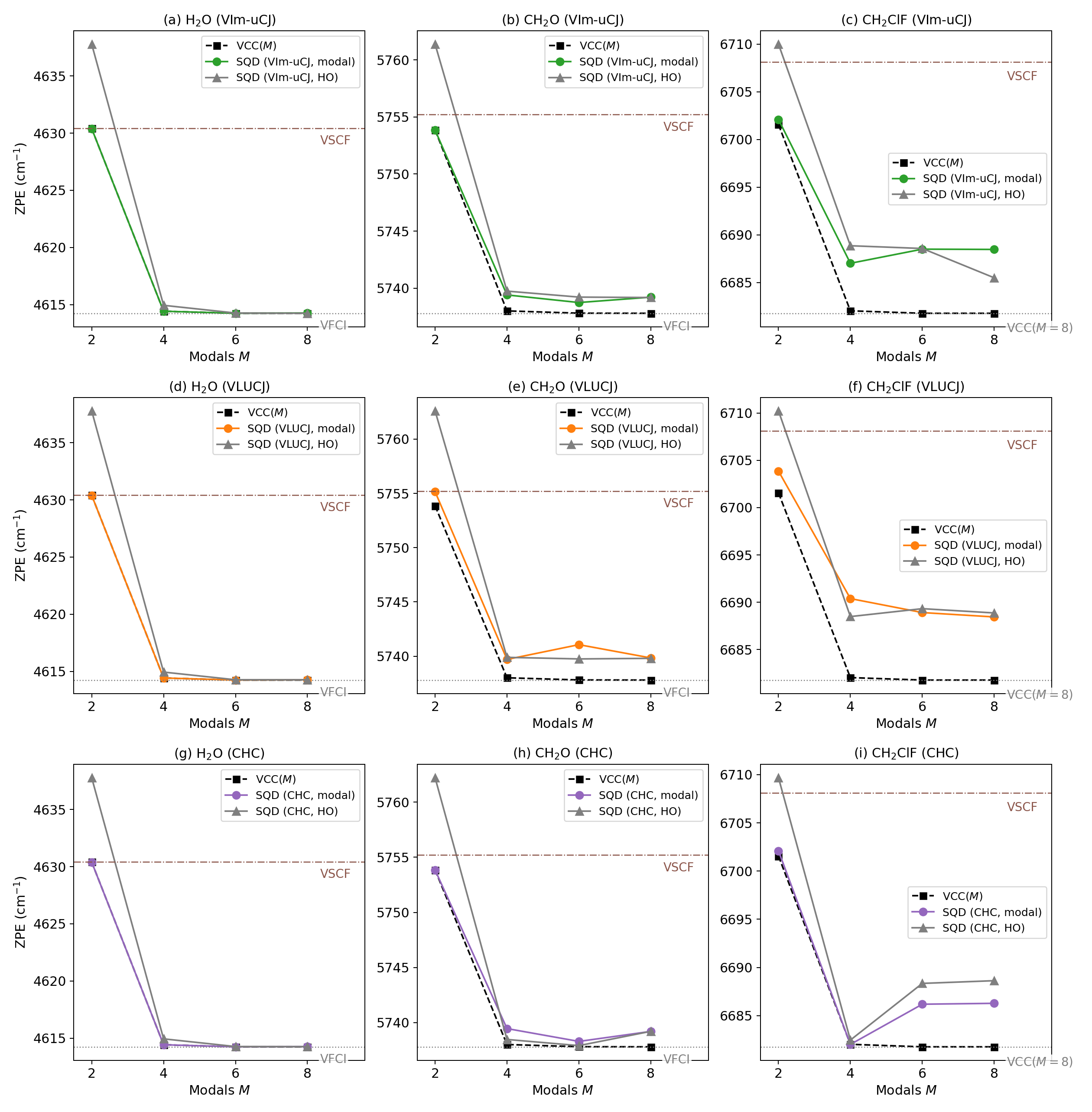}
  \caption{Same comparison as Figure~\ref{fig:energyvsMground}, for
  VIm-uCJ (top row, panels a--c), VLUCJ (middle row, panels d--f), and
  CHC (bottom row, panels g--i). All panels use the same axes,
  markers, and classical references.}
  \label{fig:energyvsMgroundothers}
\end{figure*}

The main text's Figure~\ref{fig:energyvsMground} compares UVCCSD and
Vg-uCJ's recovered ground-state energy against the classical VCC($M$)
hierarchy. Figure~\ref{fig:energyvsMgroundothers} gives the same
comparison for the three remaining ansätze -- VIm-uCJ, VLUCJ, and CHC --
on \code{ibm\_phoenix} hardware. Median absolute deviations from
VCC($M$) over all 24 (molecule, modals, basis) settings are
\SI{1.40}{\cmi} (VIm-uCJ), \SI{2.02}{\cmi} (VLUCJ, the largest of the
five ansätze), and \SI{0.50}{\cmi} (CHC), consistent with the
\SI{0.27}{}--\SI{1.40}{\cmi} range already reported for the other four
ansätze in Section~\ref{sec:results}.

\clearpage

The excited-state analogue of this comparison, for the same three
ansätze, is given in Figure~\ref{fig:energyvsMexcitedothers} below.

Figures~\ref{fig:energyvsMexcitedothers} (excited-state results for the
three remaining ansätze) and \ref{fig:uniformcontrol} (the uniform
random-sampler control) are laid out as full-width, multi-panel appendix
figures for readability. Figure~\ref{fig:energyvsMexcitedothers} follows
the same panel ordering and axis conventions as the ground-state grid
(Figure~\ref{fig:energyvsMgroundothers}), so that each molecule/ansatz
trend can be compared at a glance across ground and excited states.
Figure~\ref{fig:uniformcontrol} then provides a circuit-free baseline:
shots are drawn uniformly over valid one-hot configurations and passed
through the same recovery pipeline, isolating the role of genuine
circuit-induced structure in the hardware samples.

\clearpage
\begin{figure*}[!t]
  \centering
  \includegraphics[width=\textwidth]{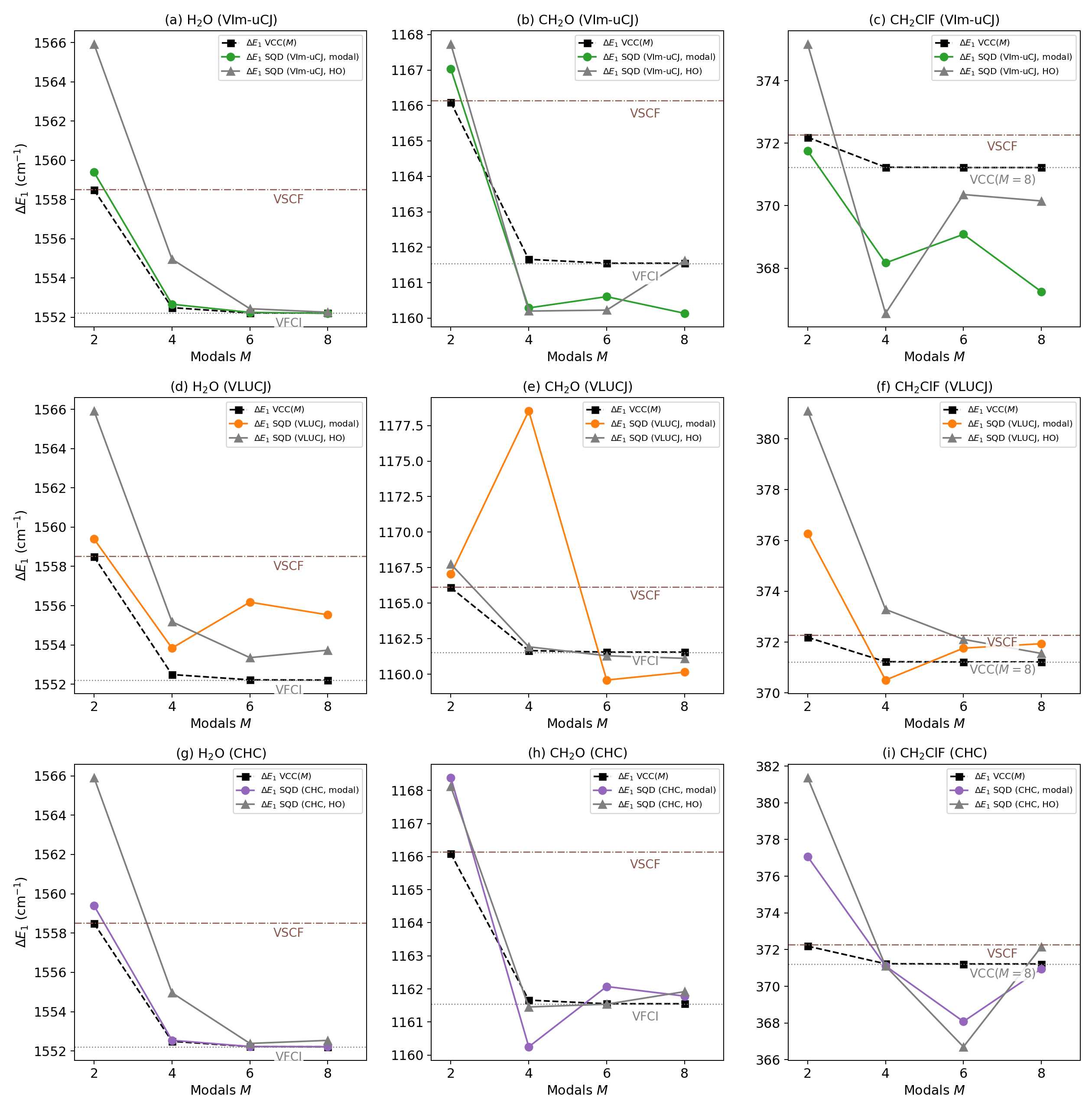}
  \caption{Same comparison as Figure~\ref{fig:energyvsMexcited}, for
  VIm-uCJ (top row, panels a--c), VLUCJ (middle row, panels d--f), and
  CHC (bottom row, panels g--i). All panels use the same axes,
  markers, and classical references.}
  \label{fig:energyvsMexcitedothers}
\end{figure*}

\clearpage
\twocolumn[{%
\begin{center}
  \includegraphics[width=\textwidth,height=0.45\textheight,keepaspectratio]{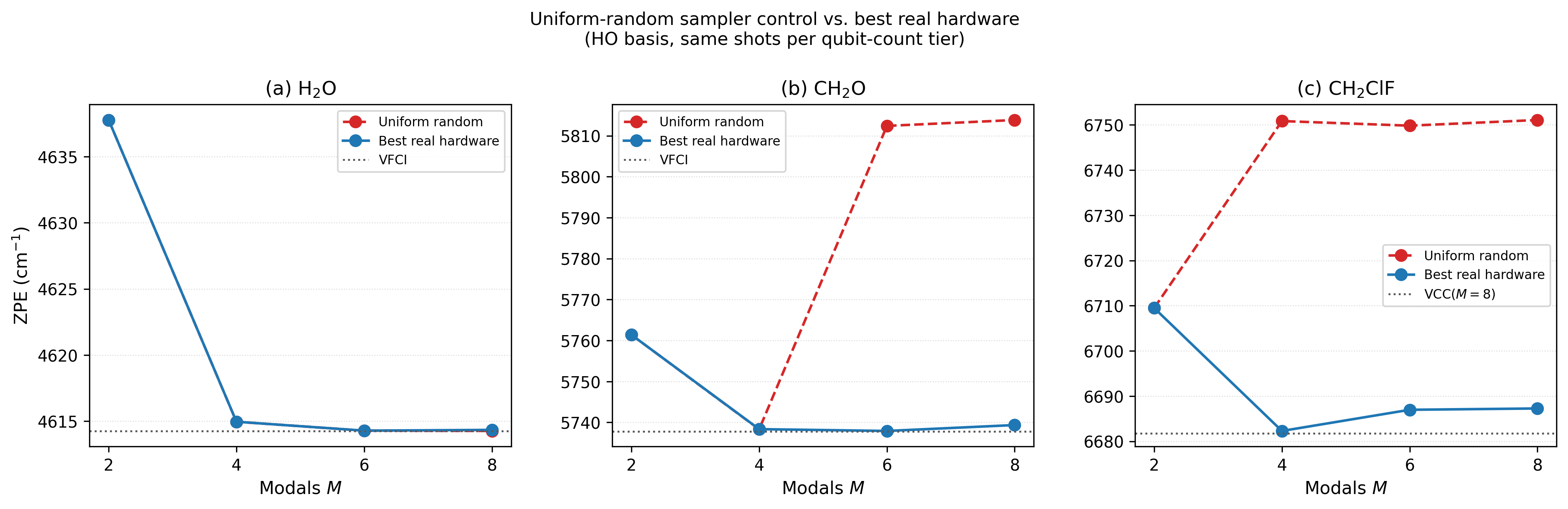}
  \captionof{figure}{Uniform-random-sampler control vs.\ the best \code{ibm\_kingston} result at each modal count, HO basis, one panel
  per molecule. At $M{=}2$ for every molecule, and throughout the full
  $M{=}2$--$8$ range for H$_2$O specifically, uniform guessing recovers
  as well as hardware (within a fraction of a
  \si{\cmi}) -- the valid-configuration space is small enough that
  random sampling covers it completely either way. For CH$_2$O and
  CH$_2$ClF the two diverge from $M{=}6$ and $M{=}4$ respectively: at
  $M{=}8$, the uniform control lands \SI{74.5}{\cmi} above hardware's best ansatz for CH$_2$O and \SI{63.8}{\cmi} above it for
  CH$_2$ClF. This divergence -- present for the two heavier molecules
  once the configuration space grows large enough that uniform sampling
  can no longer cover it completely, but not for H$_2$O at any modal
  count studied here -- is evidence that the circuits' structure is
  doing work once blind guessing alone is no longer sufficient.}
  \label{fig:uniformcontrol}
\end{center}
}]
\section{Uniform-random-sampler\\ control,\\ in full}
\label{app:uniformcontrol}

For every (molecule, modals) parameter setting in the current H$_2$O/CH$_2$O/CH$_2$ClF
measurement study, HO basis, the same qubit-count-tiered shot budget
(Section~\ref{sec:results}) was drawn \emph{uniformly at random} over
valid one-hot configurations -- no circuit, no backend, zero QPU quota --
and fed through the same self-consistent recovery loop used for the
IBM-hardware samples, including the same guaranteed inclusion of the
analytically-known reference configuration (Section~\ref{sec:mainconfig}).
Both the raw-sample diagonalization and the recovery loop's own final
diagonalization use the same basis-size convergence sweep described in
Appendix~\ref{app:recovery} (schedule 200/500/1000/2000/4000/8000
configurations, or the full unique-configuration count when smaller,
reporting the global minimum over that schedule): a single fixed
subspace cap is not well-defined for a uniform draw over a
combinatorially large configuration space, since it produces thousands
of configurations in an exact tie for sample count at any fixed cutoff,
so which configurations would be retained is decided by incidental
sort order rather than by physics. For the larger CH$_2$O and
CH$_2$ClF cells, the true valid-configuration count (up to
$\sim 10^6$) exceeds the schedule's 8000-configuration ceiling, so the
reported energy is the best value found within the schedule rather
than a value with an exhaustively verified global minimum.

At $M{=}2$ for every molecule, and throughout the full $M{=}2$--$8$
range for H$_2$O specifically, the uniform baseline is
indistinguishable from IBM hardware (within a fraction of a
\si{\cmi}), because the valid-configuration space is small enough that
random guessing covers it as completely as a structured quantum
sample does. For CH$_2$O and CH$_2$ClF, the two diverge once the
configuration space grows large enough that uniform sampling can no
longer cover it completely -- from $M{=}6$ for CH$_2$O and $M{=}4$ for
CH$_2$ClF -- and the gap remains roughly constant rather than growing
further at higher $M$: at $M{=}8$ the uniform control lands
\SI{74.5}{\cmi} above hardware's best ansatz for CH$_2$O and
\SI{63.8}{\cmi} above it for CH$_2$ClF (versus classical VCC($M{=}8$)
targets of \SI{5737.80}{\cmi} and \SI{6681.77}{\cmi} respectively). This
divergence, present for the two heavier molecules but not for H$_2$O
at any modal count studied here, is evidence that the ansatz's circuit
structure is doing most of the work once blind guessing alone is no longer
sufficient to cover the relevant configuration space: a
heavily-noise-corrupted quantum sample is not the same thing as a
uniform-random one, even once its one-hot retention has collapsed
toward the hardware floor (Figure~\ref{fig:onehotcliff}).

\subsection{Complete data for different parameter settings: interactive artifacts}
\label{app:fulltables-artifacts}

The complete per-(molecule, modals, basis, ansatz) numbers underlying
every figure in this paper -- CZ count, one-hot retention, before- and
after-recovery energy, recovery iterations, physical qubit mapping,
job identifiers, and per-stage timing where available -- were
previously printed here as four tables (ground state, modal and HO
basis; excited state, modal and HO basis). All
four ledgers are openly deposited at
\href{https://doi.org/10.5281/zenodo.23011464}{doi.org/10.5281/zenodo.23011464}:
the ground-state \code{ibm\_kingston} and \code{ibm\_phoenix}
comparisons (\code{recovery\_table\_artifact\_kingston.html},
\code{recovery\_table\_artifact\_phoenix.html}, one row per
(molecule, modals, basis, ansatz) parameter setting on each backend),
the excited-state comparison across both backends
(\code{excited\_comparison\_artifact.html}, including the classical
VCC($M$) target used for comparison, Table~\ref{tab:classicalrefs}),
and the uniform-random-sampler control
(\code{recovery\_table\_artifact\_uniform.html}, Section~\ref{sec:uniform},
Appendix~\ref{app:uniformcontrol}), each parameter setting shown
alongside its corresponding best hardware result for direct
comparison. Code to reproduce this work is available as open access in the Qiskit Community GitHub repository at https://github.com/qiskit-community/vib-sqd.

Both ground-state ledgers report a per-setting timing breakdown
(classical build time, transpile-and-submit time, and queue-wait time).
They also provide campaign-level aggregates and, where available, the
range of values observed across settings in their accompanying
descriptions.

\FloatBarrier
\clearpage
\bibliographystyle{rsc}
\bibliography{references}

\end{document}